\documentclass[%
superscriptaddress,
onecolumn,
nofootinbib,
 amsmath,amssymb,
 aps,
]{revtex4-2}

\usepackage{color}
\usepackage{graphicx}% Include figure files
\usepackage{dcolumn}% Align table columns on decimal point
\usepackage{bm}% bold math
\graphicspath{{./images/}}
\usepackage{enumerate}
\usepackage[normalem]{ulem}
\newcommand\redsout{\bgroup\markoverwith{\textcolor{red}{\rule[0.5ex]{2pt}{0.4pt}}}\ULon}

\newcommand{\araa}{Ann. Rev. Astron. Astrophys.}

\newcommand{\comment}[1]{}

\begin{document}

\preprint{APS/123-QED}

\title{Redshift-space streaming velocity effects on the Lyman-$\alpha$ forest baryon acoustic oscillation scale}% Force line breaks with \\

\author{Jahmour J. Givans}
 \email{Electronic address: givans.2@osu.edu}
  \affiliation{%
 Center for Cosmology and AstroParticle Physics, The Ohio State University, 191 West Woodruff Avenue, Columbus, Ohio 43210, USA
}%
 \affiliation{%
 Department of Physics, The Ohio State University, 191 West Woodruff Avenue, Columbus, Ohio 43210, USA
}%
\author{Christopher M. Hirata}%
 \affiliation{%
 Center for Cosmology and AstroParticle Physics, The Ohio State University, 191 West Woodruff Avenue, Columbus, Ohio 43210, USA
}%
 \affiliation{%
 Department of Physics, The Ohio State University, 191 West Woodruff Avenue, Columbus, Ohio 43210, USA
}%
 \affiliation{%
 Department of Astronomy, The Ohio State University, 140 West 18th Avenue, Columbus, Ohio 43210, USA
}%

\date{\today}% It is always \today, today,

% Current and upcoming BAO surveys such as the Dark Energy Spectroscopic Instrument (DESI), the Canadian Hydrogen Intensity Mapping Experiment (CHIME), and \textit{Euclid} are designed to have such great precision that it is imperative for researchers to properly understand and correct for systematics if we wish to make accurate cosmological measurements with BAO.

\begin{abstract}
The baryon acoustic oscillation (BAO) scale acts as a standard ruler for measuring cosmological distances and has therefore emerged as a leading probe of cosmic expansion history. However, any physical effect that alters the length of the ruler can lead to a bias in our determination of distance and expansion rate. One of these physical effects is the streaming velocity, the relative velocity between baryons and dark matter in the early Universe, which couples to the BAO scale due to their common origin in acoustic waves at recombination. In this work, we investigate the impact of streaming velocity on the BAO feature of the Lyman-$\alpha$ forest auto-power spectrum, one of the main tracers being used by the recently commissioned DESI spectrograph. To do this, we develop a  perturbative model for Lyman-$\alpha$ flux fluctuations which is complete to second order for a certain set of fields, and applicable to \textit{any} redshift-space tracer of structure since it is based only on symmetry considerations. We find that there are 8 biasing coefficients through second order. We find streaming velocity-induced shifts in the BAO scale of 0.081--0.149\% (transverse direction) and 0.053--0.058\% (radial direction), depending on the model for the biasing coefficients used. These are smaller than, but not negligible compared to, the DESI Lyman-$\alpha$ BAO error budget, which is 0.46\% on the overall scale. The sensitivity of these results to our choice of bias parameters underscores the need for future work to measure the higher-order biasing coefficients from simulations, especially for future experiments beyond DESI.
\end{abstract}

\pacs{Valid PACS appear here}% PACS, the Physics and Astronomy
\maketitle

%\tableofcontents

\section{\label{sec:intro}Introduction}

The standard $\Lambda$CDM model of cosmology presents a dynamical picture of the Universe dominated by two mysterious components: cold dark matter and dark energy. Cold dark matter (hereafter dark matter) forms the building blocks of large-scale structure and comprises 84\% of the total matter content \cite{Planck18}; its dynamics occur atop an expanding spacetime background. Just before the turn of the millennium, two independent research groups \cite{Riess98,Perlmutter99} discovered that spacetime expansion has accelerated over the last few billion years. This acceleration has been attributed to dark energy, a constant energy density component of the Universe frequently referred to as the cosmological constant $\Lambda$.

Baryon acoustic oscillations (BAOs) are one probe for determining the expansion history of the Universe (see \cite{obsprobes} for a review of this and other probes). During matter-radiation equality the Universe existed as an ionized plasma of cosmic microwave background photons, baryons, and electrons coupled via Thomson scattering and Coulomb interactions, with primordial dark matter inhomogeneities seeded throughout. These dark matter perturbations had an energy density roughly three times that of the photon-baryon fluid and thus governed its gravitational dynamics \cite{Ryden}. Gravity caused the fluid to infall toward dark matter overdensities until a critical point was reached and increased pressure pushed the fluid outward. Cycles of compression and rarefaction generating standing sound waves called BAOs continued as the Universe expanded. These could travel a comoving distance of $r_{d}=147.09 \pm 0.26$ Mpc before decoupling \cite{Planck18}, carrying along an excess of baryons and depositing them in a spherical shell of radius $r_{d}$ around the overdensities. Consequently all tracers of the matter density field show a distinct signature in their correlation functions and power spectra at a scale set by $r_d$. The BAO feature thus acts as a standard ruler we can use at a variety of known redshifts to determine the angular diameter distance $D_{A}(z)$ and Hubble parameter $H(z)$.

A preponderance of the literature (both theoretical and observational in nature) which focused on using BAOs to investigate dark energy addressed its applicability to galaxies \cite{Percival2001,Blake&Glazebrook,Hu&Haiman,Linder,Seo&Eisenstein,Cole2005,Eisenstein2005,Beutler2011,Blake2011,Beutler2017BOSS,Ross2017,Alam2017,Chen2019}. To search for the BAO signature at higher redshifts a different tracer is required. In the post-reionization Universe at redshifts $z \lesssim 6$, we can use fluctuations in Lyman-$\alpha$ forest flux to probe the intergalactic medium (IGM) \cite{1965ApJ...142.1677B,1998ApJ...495...44C,1999ApJ...520....1C,2000ApJ...543....1M}. Following recent detections of the BAO feature in Lyman-$\alpha$ forest autocorrelations and in Lyman-$\alpha$ forest-quasar cross-correlations \cite{Busca2013,Slosar2013,Font-Ribera2014,Delubac2015,Bautista2017,deSainteAgathe}, as measured by the Baryon Oscillation Spectroscopic Survey (BOSS) and extended BOSS (eBOSS), the Lyman-$\alpha$ forest now stands alongside other tools of precision cosmology \cite{hirataLya}. The Dark Energy Spectroscopic Instrument (DESI) will allow us to determine shifts in the BAO peak position down to 0.46\% precision when data from all redshift bins are combined \cite{DESI}. With such a fine measurement expected, even small systematics which can alter the BAO scale should be properly understood. 

Oscillations in the early Universe which lead to the BAO feature also give baryons a root-mean-square velocity of 33 km/s, coherent over the Silk damping scale \cite{Silk68} of several comoving Mpc, relative to dark matter at decoupling. This ``streaming velocity'' turns out to be supersonic since the sound speed in neutral hydrogen at decoupling is 6 km/s \cite{Tse10}. Streaming velocity directly affects small-scale structure $(\lesssim \textrm{few} \times 10^{6}M_{\odot})$ \cite{blazek} by setting the scale on which baryons become trapped in dark matter potential wells and changing the abundance of (and gas fraction within) early minihalos \cite{Tse_Bar_Hir2011}, but feedback processes during reionization could impact subsequent evolution of more massive galaxies \cite{Dalal}. If streaming velocity effects on early baryonic structure can leave an imprint on low-redshift tracers of the matter density field, then the BAO scale measured at low redshifts could be shifted \cite{hirataLya}. Previous works based on perturbation theory found that the BAO peak shifts to smaller (larger) physical scales when the streaming velocity bias $b_{v}>0$ ($b_{v}<0$) \cite{blazek,Yoo11,Yoo13,Slepian}.

The purpose of this work is to quantify the streaming velocity-induced BAO peak shift measured in the Lyman-$\alpha$ forest auto-power spectrum. This is the follow-up investigation to \cite{hirataLya} in which the author used a base power spectrum model from \cite{Arinyo2015} but included only the dominant term arising from streaming velocity bias, namely, the advection term. Our paper presents and accounts for many additional  streaming velocity contributions, up to one-loop in perturbation theory, to the Lyman-$\alpha$ forest auto-power spectrum in redshift space. We accomplish this by introducing a density contrast expansion applicable to \textit{any} redshift-space tracer, including all non-streaming velocity terms with total matter density, tidal fields, and total velocity divergence fields to second order (see Eqs.~\ref{first-order} and \ref{second-order}).

This paper is organized as follows: Section \ref{sec:fundamentals} reviews the basics of cosmological perturbation theory, and then presents bias models used for galaxies and the Lyman-$\alpha$ forest. In Section \ref{sec:galaxies}, we work through the more familiar case of calculating the galaxy power spectrum in redshift space, presenting terms arising from streaming velocity that were shown previously in \cite{Beutler}\footnote{Appendix \ref{sec: florian} demonstrates the equality between our power spectrum terms and theirs.}. In Section \ref{sec:lya}, we calculate the Lyman-$\alpha$ forest power spectrum, including streaming velocity contributions not previously noted in the literature. In Section \ref{sec:mapping}, we demonstrate how to rewrite the galaxy density contrast in a form that matches the generalized tracer density contrast. Here we also provide the corresponding mapping from generalized bias coefficients to those for galaxies. Section \ref{sec:shift} is where we outline our BAO model-fitting procedure and calculate the peak shift caused by a nonzero streaming velocity bias. In Section \ref{sec:conclusion} we summarize our findings and discuss their relevancy to current and future BAO surveys.  

\section{\label{sec:fundamentals}Principles and formalism}

We begin by introducing the general principles behind cosmological perturbation theory in Section \ref{sec:PT}. This provides the basis for our galaxy and Lyman-$\alpha$ forest biasing models in Section \ref{sec:galaxybias} and Section \ref{sec:lyabias}, respectively.

\subsection{\label{sec:PT}Perturbation theory basics}

The underpinnings of cosmological perturbation theory lie in the physics of the matter field. We assume this field consists exclusively of collisionless cold dark matter particles with such an enormous density that it can be treated as smooth. These particles obey the Vlasov equation for their phase space distribution function. At sufficiently large scales, the density contrast, peculiar velocity, and cosmological gravitational potential are small perturbations to the local cosmic density, particle velocity, and Newtonian potential, respectively. We can therefore linearize the Poisson, continuity, and Euler equations to solve for the evolution of the density and velocity divergence fields \cite{PT}.   

Nonlinear perturbation theory is built on the assumption that we can find higher order solutions to the density contrast $\delta$ (i.e. the deviation of a region from the mean density of the Universe) and peculiar velocity divergence $\theta={\boldsymbol\nabla}\cdot\mathbf v$, by expanding $\delta$ and $\theta$ about their linear solutions. Here ${\boldsymbol\nabla}$ is the comoving derivative operator. The full velocity field is determined by $\theta$ as long as the vorticity ${\boldsymbol\nabla}\times\mathbf v$ remains zero, which is guaranteed for CDM with gravitational forces only and in the perturbative limit (no stream crossing). The full fields can be written as
\begin{equation}\label{eq:1}
    \delta(\mathbf{r}) = \sum_{n=1}^{\infty} \delta^{(n)}(\mathbf{r}), \; \theta(\mathbf{r}) = \sum_{n=1}^{\infty} \theta^{(n)}(\mathbf{r}), 
\end{equation}
where $\delta^{(1)}$ (i.e. $\delta_{\textrm{lin}}$) and $\theta^{(1)}$ are linear in the initial density contrast, $\delta^{(2)}$ and $\theta^{(2)}$ are quadratic in the initial density contrast, and so on \cite{PT}. Our convention is that the absence of a subscript implies that the matter field itself, as opposed to a particular tracer, is being referenced.

To perform calculations in Sections \ref{sec:galaxies} and \ref{sec:lya} we will need 3D Fourier space representations of different configuration space fields. Our Fourier transform convention is
\begin{equation}
    \tilde{f}(\mathbf{k})=\int d^{3}\mathbf{r}\, e^{-i\mathbf{k}\cdot\mathbf{r}} f(\mathbf{r}) \leftrightarrow f(\mathbf{r})=\int \frac{d^{3}\mathbf{k}}{(2\pi)^{3}} \, e^{i\mathbf{k}\cdot\mathbf{r}} \tilde{f}(\mathbf{k}).
\end{equation}
Additionally we will need the convolution of two functions $f(\mathbf{k})$ and $g(\mathbf{k})$. Defining $\mathbf{k}_{2} \equiv \mathbf{k}-\mathbf{k}_{1}$ allows us to write the convolution as
\begin{equation}
    (f*g)(\mathbf{k})=\int \frac{d^{3}\mathbf{k}_{1}}{(2\pi)^{3}} \,f(\mathbf{k}_{1})g(\mathbf{k}_{2}) 
\end{equation}
Expressed in Fourier space, the fields in the summations of Eq. (\ref{eq:1}) are
\begin{equation}
    \tilde{\delta}^{(n)}(\mathbf{k}) = \int\frac{d^{3}\mathbf{k}_{1}\ldots d^{3}\mathbf{k}_{n}}{(2\pi)^{3(n-1)}}\delta_{D}^{(3)}(\mathbf{k}-\mathbf{k}_{1\ldots n})F_{n}(\mathbf{k}_{1},\ldots,\mathbf{k}_{n})\tilde{\delta}^{(1)}(\mathbf{k}_{1})\ldots \tilde{\delta}^{(1)}(\mathbf{k}_{n})
\end{equation}
and
\begin{equation}
    \tilde{\theta}^{(n)}(\mathbf{k}) = -aH\int\frac{d^{3}\mathbf{k}_{1}\ldots d^{3}\mathbf{k}_{n}}{(2\pi)^{3(n-1)}}\delta_{D}^{(3)}(\mathbf{k}-\mathbf{k}_{1\ldots n})G_{n}(\mathbf{k}_{1},\ldots,\mathbf{k}_{n})\tilde{\delta}^{(1)}(\mathbf{k}_{1})\ldots \tilde{\delta}^{(1)}(\mathbf{k}_{n}),
\end{equation}
where $\mathbf{k}_{1\ldots n} = \mathbf{k}_{1} + \cdots + \mathbf{k}_{n}$, and $F_{n}$ and $G_{n}$ are kernels built from mode coupling functions in such a way that $F_{1}=G_{1}=1$. Note that these equations are slightly modified from their forms in \cite{PT}. With these definitions $\theta^{(1)} = -aH\delta^{(1)}$, which differs from the more common convention where $\theta$ and $\delta$ are constructed to be equal at first order. The kernels we will use explicitly are
\begin{eqnarray}
 S_{2}(\mathbf{k}_{1},\mathbf{k}_{2})&\equiv&\mu_{12}^{2}-\frac{1}{3},
 \nonumber \\
    F_{2}(\mathbf{k}_{1},\mathbf{k}_{2}) &=& \frac{5}{7} + \frac{1}{2}\mu_{12}\left(\frac{k_{1}}{k_{2}}+\frac{k_{2}}{k_{1}}\right) + \frac{2}{7}\mu^{2}_{12},\; \textrm{ ~~~and}
    \nonumber \\
    G_{2}(\mathbf{k}_{1},\mathbf{k}_{2}) &=& \frac{3}{7} + \frac{1}{2}\mu_{12}\left(\frac{k_{1}}{k_{2}}+\frac{k_{2}}{k_{1}}\right) + \frac{4}{7}\mu^{2}_{12},
\end{eqnarray}
where $\mu_{12} = \widehat{\mathbf{k}}_{1}\cdot \widehat{\mathbf{k}}_{2}$. $S_{2}$ is a kernel that arises when working with tidal fields.

\subsection{\label{sec:galaxybias}Galaxy biasing model}
We cannot directly probe $\delta$ or $\theta$ since the majority of matter responsible for the dynamics of these fields is dark matter. To overcome this issue we use nonlinear biasing \cite{mcdonald} in which some luminous cosmological tracer is written as an expansion of terms all dependent on $\delta$, where each term in the expansion comes with an unknown bias coefficient.

Consider what terms can be present in the expansion of $\delta_{g}(\mathbf{r})$.  Nonlinear galaxy biasing assumes galaxy formation is an exclusively local phenomenon, with gravity being the only long-range physics at play. At distances large compared to galaxy formation scales, the galaxy density contrast should depend only on the matter density field, the tidal field, the local streaming velocity, and their respective histories. The traceless-symmetric tidal tensor is given by 
\begin{equation} \label{eq:2}
 s_{ij}(\mathbf{r})=\left(\nabla_{i}\nabla_{j}\nabla^{-2}-\frac{1}{3}\delta_{ij}\right)\delta(\mathbf{r}). 
\end{equation}
If galaxy formation is local and depends only on gravitational clustering, then the biasing model will be based on $\delta$, $s_{ij}$, and (at higher orders) other combinations of the density and velocity fields. If galaxy formation also remembers the streaming velocity, we must include streaming velocity terms. The streaming velocity $\mathbf{v}_{bc} \equiv \mathbf{v}_{b}-\mathbf{v}_{c}$ is normalized to be
\begin{equation}
 \mathbf{v}_{s}(\mathbf{r})=\frac{\mathbf{v}_{bc}(\mathbf{r},a)}{\sigma_{bc}},
\end{equation}
where $\sigma_{bc}$ is the root-mean-square value of $\mathbf{v}_{bc}$ \cite{blazek,Slepian}.

Density contrast is a scalar under rotations in real space. This implies that the tidal tensor and streaming velocity fields may only enter the expansion in combinations that make each term a scalar. Our goal in later sections will be to calculate streaming velocity corrections, $\Delta P$, to a base power spectrum. Streaming velocity is a second-order contribution (at lowest order) to the density contrast. Our correlations will be of two types: (i) between second-order streaming velocity terms and second-order non-streaming velocity terms, and (ii) between third-order streaming velocity terms and first-order non-streaming velocity terms. With these restrictions on order of perturbation, we build the galaxy density contrast 
\begin{equation} \label{eq:4}
\begin{aligned}
  \delta_{g}(\mathbf{r}) =\, &b_{1}\delta(\mathbf{r})+\frac{b_{2}}{2}\left[\delta^{2}(\mathbf{r})-\left<\delta^{2}\right>\right] + \frac{b_{s}}{2}\left[s^{2}(\mathbf{r})-\left<s^{2}\right>\right] + \cdots \\ &+b_{v}\left[v_{s}^{2}(\mathbf{x})-1\right] + b_{1v}\delta(\mathbf{r})\left[v_{s}^{2}(\mathbf{x})-1\right] \\ &+b_{sv}s_{ij}(\mathbf{r})v_{s,i}(\mathbf{x})v_{s,j}(\mathbf{x}) + \cdots, 
\end{aligned}
\end{equation}
where \textbf{r} is the Eulerian position and \textbf{x} is the Lagrangian position. At a given conformal time $\eta$ the two coordinates are related by $\mathbf{r}(\mathbf{x},\eta)=\mathbf{x}+\Psi(\mathbf{x},\eta),$ where the Lagrangian displacement is $\Psi=-\nabla \nabla^{-2}\delta_{\textrm{lin}}(\mathbf{r},\eta)$ to leading order. Streaming velocities most naturally enter as functions of \textbf{x} since they are an early Universe effect -- at decoupling, the gas is near its Lagrangian position $\textbf{x}$. (A fully Eulerian treatment and investigation of Eulerian vs.\ Lagrangian approaches can be found in Appendix~C of Ref.~\cite{blazek}\footnote{The \textit{Physical Review Letters} version of this work has appendices as part of the supplemental material, not the main text.}.) Any field $\varphi$ of order $N$ given in Lagrangian coordinates can be mapped to order $N+1$ and expressed in Eulerian coordinates by 
\begin{equation}\label{zeldovich}
 \varphi(\mathbf{x})=\varphi(\mathbf{r})+\nabla\varphi(\mathbf{r})\cdot \nabla \nabla^{-2}\delta_{\textrm{lin}}(\mathbf{r},\eta)+\cdots,
\end{equation}
from which it follows 
\begin{equation}
    v_{s}^{2}(\mathbf{x})=v_{s}^{2}(\mathbf{r})+ [\nabla_{i}\nabla^{-2}\delta_{\textrm{lin}}(\mathbf{r})][\nabla_{i}v_{s}^{2}(\mathbf{r})].
\label{eq:vs-adv}
\end{equation}
This allows us to perform our analysis using Eulerian perturbation theory \cite{blazek}. In Eq. (\ref{eq:4}) we have included terms up to $\mathcal{O}(\delta_{\textrm{lin}}^{3})$ since leading corrections to the galaxy 2-point correlation function arise from terms of $\mathcal{O}(\delta_{\textrm{lin}}^{4})$. 

Our job is to now determine $\delta_{g}(\mathbf{r})$ as a function of redshift-space coordinate \textbf{s} instead of real-space coordinate \textbf{r}. The total number of galaxies in a given region is conserved in transformations between real and redshift space. Mathematically this is described by $n_{s}(\mathbf{s})d^{3}\mathbf{s}=n(\mathbf{r})d^{3}\mathbf{r}$, where $n_{s}$ is the redshift-space number density and $n$ is the real-space number density. From this equation we immediately find the Jacobian $J\equiv |d^{3}\mathbf{r}/d^{3}\mathbf{s}|$ which gives the relationship between \textbf{s} and \textbf{r},
\begin{equation}\label{redtoreal}
 \mathbf{s}=\mathbf{r}+\frac{v_{z}}{aH}\hat{\mathbf{e}}_{z},
\end{equation}
where the subscript $z$ indicates the line-of-sight direction, $a$ is the scale factor, and $H$ is the Hubble parameter. Combining the preceding equations and using the arguments of \cite{Kaiser} yields
\begin{equation} \label{delta_s}
 1+\delta_{g}(\mathbf{s})=\frac{1}{1+\frac{1}{aH}\frac{\partial v_{z}}{\partial r_{z}}}[1+\delta_{g}(\mathbf{r})].
\end{equation}
The line-of-sight velocity is given in terms of the velocity divergence by
\begin{equation}
    \tilde{v}_{z}(\mathbf{k})= -\frac{ik_{z}}{k^{2}}\tilde{\theta}(\mathbf{k}).
    \label{eq:vz}
\end{equation}

\subsection{\label{sec:lyabias}Lyman-alpha forest biasing model}
Neutral hydrogen ({H\,\sc{i}}) gas clouds abound throughout the IGM after reionization ends. Background quasars emit radiation which can interact with {H\,\sc{i}} in all foreground clouds along a given line of sight. When this radiation has a wavelength of $1216$ \AA \, in the rest frame of a hydrogen atom it will be absorbed and prompt a Lyman-$\alpha$ transition. Since several {H\,\sc{i}} gas clouds lie at different redshifts along the line of sight between us and a given quasar, we see an entire ``Lyman-$\alpha$ forest'' of absorption features in the fraction of transmitted flux for each quasar spectrum. This probes a 1D skewer through the Universe and allows us to investigate its matter distribution \cite{1998ARA&A..36..267R}.

The relation between observed flux $F$ and the matter field is not as straightforward as connection between galaxies and the matter field. In the Lyman-$\alpha$ forest case, hydrogen atoms track the underlying dark matter distribution through some nonlinear relation. Those atoms then undergo a redshift-space distortion (RSD) transformation akin to Eq. (\ref{delta_s}). Optical depth $\tau$ is proportional to this {H\,\sc{i}} density, but the quantity we measure is $F=\exp (-\tau)$ \cite{seljak}, a nonlinear function of $\tau$. In contrast, galaxies trace the dark matter distribution and undergo an RSD transformation; there are no further transformations required to obtain an observable. This is why our treatment of the Lyman-$\alpha$ forest must differ from that of galaxies (see Section 3.2 of \cite{mcdonald} and Section 9.3.2 of \cite{Desjacques2018}). Note that Ref.~\cite{seljak} derived a set of biasing formulas for the Lyman-$\alpha$ forest, but with additional physical assumptions (e.g., analytic approximations for the peak-background split) beyond the symmetries.

Once again we consider which terms can be present in $\delta_{F}(\mathbf{s})$ and then apply symmetry arguments to build the final expansion. Similarly to Eq. (\ref{eq:2}) we introduce the traceless-symmetric tensor \cite{mcdonald}
\begin{equation}
 t_{ij}(\mathbf{r})= \left(\nabla_{i}\nabla_{j}\nabla^{-2}-\frac{1}{3}\delta_{ij}\right)\left[\frac{-1}{aH}\theta(\mathbf{r})-\delta(\mathbf{r})\right],
\end{equation}
which is constructed to be zero at first order. With this field and others constructed from those given in Eqs. (\ref{eq:2})--(\ref{eq:4}), we now have several quantities which can contribute to the density contrast. The redshift-space transmitted flux density contrast has as a requirement rotational invariance along the line of sight (i.e. azimuthal symmetry); each term need not possess spherical symmetry. The complete expansion in these fields is
\begin{equation} \label{deltaF}
\begin{aligned}
 \delta_{F}(\mathbf{s})= &c_{1}\delta(\mathbf{r})+c_{2}s_{zz}(\mathbf{r})+c_{3}\delta^{2}(\mathbf{r})+c_{4}s^{2}(\mathbf{r})\\ &+c_{5}\delta(\mathbf{r})s_{zz}(\mathbf{r})+c_{6}t_{zz}(\mathbf{r})+c_{7}s_{zz}^{2}(\mathbf{r})+c_{8}[s_{xz}^{2}(\mathbf{r})+s_{yz}^{2}(\mathbf{r})]\\ &+c_{0} + b_{v}[v_{s}^{2}(\mathbf{x})-1]+b_{1v}\delta(\mathbf{r})[v_{s}^{2}(\mathbf{x})-1] \\ &+b_{sv}s_{ij}(\mathbf{r})v_{s,i}(\mathbf{x})v_{s,j}(\mathbf{x})+b_{vz}\left[v_{s,z}^{2}(\mathbf{x})-\frac{1}{3}v_{s}^{2}(\mathbf{x})\right] + \cdots,
\end{aligned}
\end{equation}
where $c_{0}$ is a constant counterterm chosen to ensure $\delta_{F}(\mathbf{s})$ has zero average value (see Section \ref{sec:counterterm}). Table \ref{term_equivalence} gives the correspondence between our first-order coefficients and fields and those used in the traditional description of Lyman-$\alpha$ forest perturbation theory, such as seen in Ref.~\cite{Arinyo2015}. The overdensities, velocity gradients (which are absorbed into other terms), and tidal fields are native to real space whereas the flux fluctuation field exists only in redshift space -- the optical depth to flux transformation $F=\exp(-\tau)$ makes sense only in redshift space. Equation (\ref{deltaF}) is the most general second-order expansion possible for the fields we considered, subject to the symmetries of the Lyman-$\alpha$ forest, and is applicable to \textit{any} redshift-space tracer (see Appendix \ref{sec:group theory} for a proof). We note that the first appearance of a similar expansion without streaming velocity was given in \cite{Desjacques2018jcap}. Just as we do here, the authors had eight terms through second order, as shown in their Eqs. (2.6) and (2.14).

This expansion is intended to include all terms which satisfy our criterion of contributing to the streaming velocity correction $\Delta P(k,\mu)$ at 1-loop order. This requirement eliminates from our consideration all third-order non-streaming velocity terms. Excluded also are streaming velocity gradients and the relative density perturbation between baryons and dark matter. We view these as distinct physical effects which must be handled in simulations separately from a ``simple'' (constant across the box) implementation of streaming velocity. A complete treatment, to third order in perturbation theory, of streaming velocity effects for galaxies in real space is given in Ref. \cite{Schmidt2016}. Therein, in addition to the fields we consider, the author considers spatial derivatives of the streaming velocity, the relative density perturbations between baryons and dark matter, and the initial amplitude of the constant relative density perturbation (see Section 8.2 of \cite{Desjacques2018} for a detailed review of these). All of these considerations are important for the Lyman-$\alpha$ forest in redshift space, but they are beyond the scope of this paper.

\begin{table}[t]
\caption{\label{term_equivalence}Comparison of first-order fields and biasing coefficients in Ref.~\cite{Arinyo2015} to those in this work.}
\begin{tabular}{|l|l|}\hline
Quantity appearing in Ref.~\cite{Arinyo2015} & Equivalent quantity in this work \\ \hline
$\delta$                                                 & $\delta$                         \\
$\eta$                                                   & $f(s_{zz}+\frac{1}{3}\delta)$    \\
$b_{F_{\delta}}$                                         & $c_{1}-\frac{1}{3}c_{2}$         \\
$b_{F_{\eta}}$                                           & $c_{2}/f$                        \\
$b_{F_{\delta}}+\frac{1}{3}fb_{F_{\eta}}$                & $c_{1}$                          \\
$\beta b_{F_{\delta}}=fb_{F_{\eta}}$                     & $c_{2}$             \\ \hline        
\end{tabular}
\end{table}

The above biasing analysis applies to total matter. The difference between baryonic and dark matter perturbations could be important in two ways. At small scales, the baryons are smoothed by pressure effects that can be parameterized by the filtering scale $k_{\rm F}$ \cite{1998MNRAS.296...44G} and in redshift space they are smeared by thermal broadening; this is very important at $k\sim k_{\rm F}$ (in 3D; in the 1D power spectrum the effect extends to all $k$ because of the projection integrals, e.g., \cite{2011MNRAS.415.2257M}), but should have little effect on the large scale correlation function or BAOs. The other difference is that the BAO feature is not present with the same amplitude and morphology in both the dark matter and baryons. This is because the BAO feature is present in the baryons at decoupling, and then appears in the dark matter because there is a combined growing mode of the baryons and dark matter. The difference in BAO feature amplitudes decays slowly and is still significant at low $z$ (see Fig.~1 of Ref.~\cite{2007ApJ...664..660E}).

\section{\label{sec:galaxies}Galaxies analysis}

Define the power spectrum by
\begin{equation}
    \left<\tilde{\delta}(\mathbf{k})\tilde{\delta}^{*}(\mathbf{k}') \right> = (2\pi)^{3}P(k)\delta_{D}^{(3)}(\mathbf{k}-\mathbf{k}')\,,
    \label{eq:powerspectrum}
\end{equation}
where $\left<\vphantom{Y}\ldots\vphantom{Y}\right>$ denotes an ensemble average. When necessary we will specify whether our ensemble average is taken in real space or redshift space with a subscript $\mathbf{r}$ or $\mathbf{s}$, respectively.
The Fourier transform of Eq. (\ref{delta_s}) is, to second order in perturbation theory,
\begin{equation}
\begin{aligned}
    \tilde{\delta}_{g}^{(s)}(\mathbf{k})&=\int d^{3}\mathbf{s}\,\left|\frac{d^{3}\mathbf{r}}{d^{3} \mathbf{s}}\right|[1+\delta_{g}^{(r)}(\mathbf{r})]e^{-i\mathbf{k}\cdot\mathbf{s}}-\int d^{3}\mathbf{s}\,e^{-i\mathbf{k}\cdot\mathbf{s}} \\  &= \int d^{3}\mathbf{r}\,\delta^{(r)}_{g}(\mathbf{r})\,e^{-i\mathbf{k}\cdot\mathbf{r}}\,e^{-ik_{z}v_{z}/aH} + \int d^{3}\mathbf{r}\, e^{-i\mathbf{k}\cdot\mathbf{r}}\,e^{-ik_{z}v_{z}/aH} - (2\pi)^{3}\delta^{(3)}_{D}(\mathbf{k}) \\ &= \int d^{3}\mathbf{r}\,\delta^{(r)}_{g}(\mathbf{r}) \left[1-\frac{ik_{z}v_{z}}{aH}-\frac{(k_{z}v_{z})^{2}}{2(aH)^{2}}\right]e^{-i\mathbf{k}\cdot\mathbf{r}}+\int d^{3}\mathbf{r}\, \left[-\frac{ik_{z}v_{z}}{aH}-\frac{(k_{z}v_{z})^{2}}{2(aH)^{2}}+\frac{i(k_{z}v_{z})^{3}}{6(aH)^{3}}\right]e^{-i\mathbf{k}\cdot\mathbf{r}}. 
\end{aligned}
\end{equation}
Taking its autocorrelation generates power spectrum components of $P_{g}$. In this section we list only those pieces with a single power of $b_{v}$; if $b_{v}$ is expected to be small then $b_{v}^{2}$ ought to be negligible. Any terms containing $\mathbf{k}=0$, i.e. those with $\delta_{D}^{(3)}(\mathbf{k})$, will be disregarded since those modes feed back into the mean number density of galaxies in the Universe; they contribute to overall shot noise \cite{mcdonald} which doesn't affect shifts in BAO peak position. As is true in real space, correlations proportional to $b_{1v}$ or $b_{sv}$ vanish by parity considerations in redshift space \cite{blazek}. What remains at one-loop is
\begin{equation} \label{Pg}
\begin{aligned}
 P_{g,b_{v}\textrm{ only}}(k,\mu)=&2b_{1}b_{v}\left[P_{\delta |v^{2}}(k)+P_{\textrm{adv}|\delta}(k)+f\mu^{2}P_{\delta v_{z}|v^{2}}(k) \right. \\ &\left.+f\mu^{2}P_{v_{z}v^{2}|\delta}(k)\right]+2b_{2}b_{v}P_{\delta^{2}|v^{2}}(k) \\ &+2b_{s}b_{v}P_{s^{2}|v^{2}}(k)+2f\mu^{2}b_{v}\left[P_{{\textrm{adv}}|v_{z}}(k)\right. \\ & \left.+P_{v_{z}|v^{2}}(k) +fP_{v_{z}^{2}|v^{2}}(k)\right.\\ &\left.+f\mu^{2}P_{v_{z}|v^{2}v_{z}}(k)\right].
\end{aligned}
\end{equation}
Here the subscripts indicate which contributions to $\delta_g$ give rise to each term in the power spectrum, and ``adv'' indicates the advection term (second term of Eq.~(\ref{eq:vs-adv}), see \cite{blazek} for further discussion).
As examples, we will detail how to calculate three of the power spectra in Eq.~(\ref{Pg}) and put them in a format consistent for use in FAST-PT, an algorithm to calculate convolution integrals of scalar and tensor quantities in cosmological perturbation theory \cite{fastpt1,fastpt2}. These examples are chosen to illustrate the range of mathematical techniques that are used.

\subsection{Galaxy example term I: $P_{\delta^{2}| v^{2}}$}

We begin with the correlation function
\begin{equation}\label{eq:20}
    \left<(\tilde{\mathbf{v}}_{s} * \tilde{\mathbf{v}}_{s})^{*}(\mathbf{k})(\tilde{\delta} * \tilde{\delta})(\mathbf{k}')\right>.
\end{equation}
Using the equation 
\begin{equation}
    \mathbf{v}_{s}(\mathbf{k})= -iT_{v}(k)\delta_{\textrm{lin}}(\mathbf{k})\widehat{\mathbf{k}},
\end{equation}
where $T_v$ is the velocity transfer function, and defining $\mathbf{k}_{2}\equiv\mathbf{k}-\mathbf{k}_{1}$ and $\mathbf{k}'_{2}\equiv\mathbf{k}'-\mathbf{k}'_{1}$, we can recast Eq. (\ref{eq:20}):
 \begin{equation}
 \begin{aligned}
     \left<(\tilde{\mathbf{v}}_{s} * \tilde{\mathbf{v}}_{s})^{*}(\mathbf{k})(\tilde{\delta} * \tilde{\delta})(\mathbf{k}')\right> &= \left<-\int \frac{d^{3}\mathbf{k}_{1}}{(2\pi)^{3}}\mu_{12}T_{v}(k_{1})T_{v}(k_{2})\tilde{\delta}^{*}_{\textrm{lin}}(\mathbf{k}_{1})\tilde{\delta}^{*}_{\textrm{lin}}(\mathbf{k}_{2})\int \frac{d^{3}\mathbf{k}'_{1}}{(2\pi)^{3}}\tilde{\delta}_{\textrm{lin}}(\mathbf{k}'_{1})\tilde{\delta}_{\textrm{lin}}(\mathbf{k}'_{2})\right> \\ &= -\int \frac{d^{3}\mathbf{k}_{1}}{(2\pi)^{3}}\int \frac{d^{3}\mathbf{k}'_{1}}{(2\pi)^{3}}\mu_{12}T_{v}(k_{1})T_{v}(k_{2})\left<\tilde{\delta}^{*}_{\textrm{lin}}(\mathbf{k}_{1})\tilde{\delta}^{*}_{\textrm{lin}}(\mathbf{k}_{2})\tilde{\delta}_{\textrm{lin}}(\mathbf{k}'_{1})\tilde{\delta}_{\textrm{lin}}(\mathbf{k}'_{2})\right>.
 \end{aligned}
 \end{equation}
 We apply Wick's theorem to the four-point function:
 \begin{equation}
 \begin{aligned}
    \left<\tilde{\delta}^{*}_{\textrm{lin}}(\mathbf{k}_{1})\tilde{\delta}^{*}_{\textrm{lin}}(\mathbf{k}_{2})\tilde{\delta}_{\textrm{lin}}(\mathbf{k}'_{1})\tilde{\delta}_{\textrm{lin}}(\mathbf{k}'_{2})\right> &= (2\pi)^{6}P_{\textrm{lin}}(k_{1})P_{\textrm{lin}}(k_{2})\delta_{D}^{(3)}(\mathbf{k})\delta_{D}^{(3)}(\mathbf{k}') \\ &+ (2\pi)^{6}P_{\textrm{lin}}(k_{1})P_{\textrm{lin}}(k_{2})\delta_{D}^{(3)}(\mathbf{k}_{1}-\mathbf{k}'_{1})\delta_{D}^{(3)}(\mathbf{k}_{2}-\mathbf{k}'_{2}) \\ &+ (2\pi)^{6}P_{\textrm{lin}}(k_{1})P_{\textrm{lin}}(k_{2})\delta_{D}^{(3)}(\mathbf{k}_{1}-\mathbf{k}'_{2})\delta_{D}^{(3)}(\mathbf{k}_{2}-\mathbf{k}'_{1}).
 \end{aligned}
 \end{equation}
As mentioned previously we disregard the first term on the right-hand side. The second and third terms are identical under the exchange $\mathbf{k}'_{1} \leftrightarrow \mathbf{k}'_{2}$ so we combine these. Eq. (\ref{eq:20}) may then be written as
 \begin{equation}
     \begin{aligned}
     \left<(\tilde{\mathbf{v}}_{s} * \tilde{\mathbf{v}}_{s})^{*}(\mathbf{k})(\tilde{\delta} * \tilde{\delta})(\mathbf{k}')\right> &= -2(2\pi)^{6}\int \frac{d^{3}\mathbf{k}_{1}}{(2\pi)^{3}}\int \frac{d^{3}\mathbf{k}'_{1}}{(2\pi)^{3}}\mu_{12}T_{v}(k_{1})T_{v}(k_{2})P_{\textrm{lin}}(k_{1})P_{\textrm{lin}}(k_{2})\delta_{D}^{(3)}(\mathbf{k}_{1}-\mathbf{k}'_{2})\delta_{D}^{(3)}(\mathbf{k}_{2}-\mathbf{k}'_{1}) \\ &= -2\int d^{3}\mathbf{k}_{1} \mu_{12}T_{v}(k_{1})T_{v}(k_{2})P_{\textrm{lin}}(k_{1})P_{\textrm{lin}}(k_{2})\delta_{D}^{(3)}(\mathbf{k}-\mathbf{k}'), 
     \end{aligned}
 \end{equation}
 from which we read off the non-renormalized (NR) power spectrum
 \begin{equation}
     P_{\delta^{2}|v^{2}, \textrm{NR}}(k)=-2\int\frac{d^{3}\mathbf{k}_{1}}{(2\pi)^{3}}\mu_{12}T_{v}(k_{1})T_{v}(k_{2})P_{\textrm{lin}}(k_{1})P_{\textrm{lin}}(k_{2}).
 \end{equation}

To renormalize the power spectrum we take the $\mathbf{k} \rightarrow 0$ limit and subtract off that constant piece. This is sensible since there should be no power on arbitrarily large physical scales. The constant power contribution looks like shot noise and can be absorbed into an overall shot noise term \cite{mcdonald}. The final answer is  
\begin{equation}
 P_{\delta^{2}|v^{2}}(k)= -2\int \frac{d^{3}\mathbf{k}_{1}}{(2\pi)^{3}}T_{v}(k_{1})P_{\textrm{lin}}(k_{1})[\mu_{12}T_{v}(k_{2})P_{\textrm{lin}}(k_{2}) +T_{v}(k_{1})P_{\textrm{lin}}(k_{1})].
\end{equation}

\subsection{\label{sec3B}Galaxy example term II: $P_{v_{z}^{2}|v^{2}}$}

Determining $P_{v_{z}^{2}|v^{2}}$ proceeds in similar fashion. Note that $v_{z}$ at linear order is given by the linear term in Eq.~(\ref{eq:vz}):
\begin{equation}
    \tilde{v}_{z}(\mathbf{k})=\frac{iaHk_{z}}{k^{2}}\tilde{\delta}(\mathbf{k}).
    \label{eq:vz-lin}
\end{equation}
We write the correlation function
\begin{equation}
    \frac{-k'^{2}}{2(aH)^{2}}\left<(\tilde{\mathbf{v}}_{s} * \tilde{\mathbf{v}}_{s})^{*}(\mathbf{k})(\tilde{v}_{z}*\tilde{v}_{z})(\mathbf{k}')\right> =  \frac{-k'^{2}}{2(aH)^{2}} \left<-\int \frac{d^{3}\mathbf{k}_{1}}{(2\pi)^{3}}\mu_{12}T_{v}(k_{1})T_{v}(k_{2})\tilde{\delta}^{*}_{\textrm{lin}}(\mathbf{k}_{1})\tilde{\delta}^{*}_{\textrm{lin}}(\mathbf{k}_{2})\int \frac{d^{3}\mathbf{k}'_{1}}{(2\pi)^{3}}\tilde{v}_{z}(\mathbf{k}_{1}')\tilde{v}_{z}(\mathbf{k}_{2}')\right>  
\end{equation}
and read off the power spectrum
\begin{equation}\label{eq:29}
    P_{v_{z}^{2}|v^{2}} = -k^{2}\int \frac{d^{3}\mathbf{k}_{1}}{(2\pi)^{3}}\frac{\mu_{12}}{k_{1}k_{2}}\hat{k}_{1z}\hat{k}_{2z}T_{v}(k_{1})T_{v}(k_{2})P_{\textrm{lin}}(k_{1})P_{\textrm{lin}}(k_{2}).
\end{equation}
As written, this equation is not in the form of Eq. (1.1) in Ref.~\cite{fastpt2} since it contains specific components of wave vectors. We will resolve this issue below.

A general symmetric tensor may be written as
\begin{equation}
\begin{aligned}
    A_{ij}(\mathbf{k})=\int \frac{d^{3}\mathbf{k}_{1}}{(2\pi)^{3}}\hat{k}_{1i}\hat{k}_{2j}f(k_{1})f(k_{2})g(k_{1},k_{2},\mu_{12})&=\int \frac{d^{3}\mathbf{k}_{1}}{(2\pi)^{3}}\frac{\mu_{12}}{3}\delta_{ij}f(k_{1})f(k_{2})g(k_{1},k_{2},\mu_{12}) \\ &+\int \frac{d^{3}\mathbf{k}_{1}}{(2\pi)^{3}}\left(\hat{k}_{1i}\hat{k}_{2j}-\frac{\mu_{12}}{3}\delta_{ij}\right)f(k_{1})f(k_{2})g(k_{1},k_{2},\mu_{12})
\end{aligned}    
\end{equation}
to explicitly show its spin-0 component $A_{ij}^{(0)}$ and spin-2 component $A_{ij}^{(2)}$, given by the first and second integral on the right-hand side, respectively. Note that $f$ and $g$ depend on the size and shape of the $\mathbf k_1,\mathbf k_2,\mathbf k$ triangle but not its 3D orientation. In any coordinate system the spin-0 component takes the form
\begin{equation}
    A_{zz}^{(0)}(\mathbf{k})=\int \frac{d^{3}\mathbf{k}_{1}}{(2\pi)^{3}}\frac{\mu_{12}}{3}f(k_{1})f(k_{2})g(k_{1},k_{2},\mu_{12}).
\end{equation}
We now build a new coordinate system where $\mathbf{k}$ lies on the $\overline{z}$-axis and consider the spin-2 component. Since $A_{ij}^{(2)}$ is traceless by construction and there is rotational symmetry of the integral around the $\overline{z}$ axis, it follows that $A_{\overline{x}\overline{x}}^{(2)}=A_{\overline{y}\overline{y}}^{(2)}=-\frac{1}{2}A_{\overline{z}\overline{z}}^{(2)}$ and $A_{\overline i \overline j}^{(2)}=0$ for $\overline i\neq \overline j$. With these conditions we find
\begin{equation}
\begin{aligned}
 A_{zz}^{(2)}(\mathbf{k}) &= \sum_{\overline{ij}}R_{\overline{i}z}R_{\overline{j}z}A_{\overline{ij}}^{(2)}(\mathbf{k}) \\
 &= A_{\overline{zz}}^{(2)}(\mathbf{k})\left[-\frac{1}{2}R_{\overline{x}z}^{2}-\frac{1}{2}R_{\overline{y}z}^{2}+R_{\overline{z}z}^{2}\right] \\ &= A_{\overline{zz}}^{(2)}(\mathbf{k})\left[-\frac{1}{2}+\frac{3}{2}R_{\overline{z}z}^{2}\right] \\ &= A_{\overline{zz}}^{(2)}(\mathbf{k})\mathcal{P}_{2}(\mu),
\end{aligned}
\end{equation}
where $R_{ij}$ is the rotation matrix between axes $i$ and $j$, $\mathcal{P}_{n}$ is the Legendre polynomial of order $n$, and $\mu$ is the cosine of the angle between the wave vector $\mathbf k$ and the line of sight $\hat{\mathbf e}_z$. In going from the second to third line we used the relation $R_{\overline{x}z}^{2}+R_{\overline{y}z}^{2}+R_{\overline{z}z}^{2}=1$ and in going from the third to fourth line we identified $R_{\overline{z}z}=\mu$. Then Eq. (\ref{eq:29}) simplifies to 
\begin{equation}\label{comparison_ex}
\begin{aligned}
    P_{v_{z}^{2}|v^{2}} = &-k^{2}\left[\int \frac{d^{3}\mathbf{k}_{1}}{(2\pi)^{3}}\frac{\mu_{12}^{2}}{3k_{1}k_{2}}T_{v}(k_{1})T_{v}(k_{2})P_{\textrm{lin}}(k_{1})P_{\textrm{lin}}(k_{2}) \right. \\ &+ \left. \mathcal{P}_{2}(\mu)\int \frac{d^{3}\mathbf{k}_{1}}{(2\pi)^{3}}\left(\mu_{1}\mu_{2}-\frac{\mu_{12}}{3}\right)\frac{\mu_{12}}{k_{1}k_{2}}T_{v}(k_{1})T_{v}(k_{2})P_{\textrm{lin}}(k_{1})P_{\textrm{lin}}(k_{2})\right],
\end{aligned}    
\end{equation}
where $\mu_{1} = \widehat{\mathbf{k}}_{1}\cdot \widehat{\mathbf{k}}$ and $\mu_{2} = \widehat{\mathbf{k}}_{2}\cdot \widehat{\mathbf{k}}$.

\subsection{Galaxy example term III: $P_{\textrm{adv}|\delta}$}

An advection term arises from the real-space correlation 
\begin{equation}
 \left<\delta|\nabla_{i}[v_{s,j}v_{s,j}][\nabla_{i}\nabla^{-2}\delta]\right> = 2\left<\delta|v_{s,j}(\nabla_{i}v_{s,j})\nabla_{i}\nabla^{-2}\delta\right>.
\end{equation}
Using the definition $\mathbf{k}_{3} \equiv \mathbf{k}-\mathbf{k}_{1}-\mathbf{k}_{2}$ and Fourier transforming the above equation gives
\begin{equation}\label{advterm}
 \begin{aligned}
  \left<\tilde{\delta}^{*}(\mathbf{k}')|\textrm{adv}(\mathbf{k})\right> &=
  2\left<\int \frac{d^{3}\mathbf{k}_{1}}{(2\pi)^{3}}\frac{d^{3}\mathbf{k}_{2}}{(2\pi)^{3}}\tilde{\delta}^{*}_{\textrm{lin}}(\mathbf{k}')v_{s,j}(\mathbf{k}_{1})ik_{2,i}v_{s,j}(\mathbf{k}_{2})\frac{ik_{3,i}}{-k_{3}^{2}}\tilde{\delta}_{\textrm{lin}}(\mathbf{k}_{3})\right> \\ &= -2\int \frac{d^{3}\mathbf{k}_{1}}{(2\pi)^{3}}\frac{d^{3}\mathbf{k}_{2}}{(2\pi)^{3}}\left(\widehat{\mathbf{k}}_{1}\cdot\widehat{\mathbf{k}}_{2} \right)\frac{\mathbf{k}_{2}\cdot\mathbf{k}_{3}}{k_{3}^{2}}T_{v}(k_{1})T_{v}(k_{2})\left<\tilde{\delta}^{*}_{\textrm{lin}}(\mathbf{k}')\tilde{\delta}_{\textrm{lin}}(\mathbf{k}_{1})\tilde{\delta}_{\textrm{lin}}(\mathbf{k}_{2})\tilde{\delta}_{\textrm{lin}}(\mathbf{k}_{3})\right>
 \end{aligned}
\end{equation}
The four-point function can be simplified using Wick's theorem:
\begin{equation}
 \begin{aligned}
  \left<\tilde{\delta}^{*}(\mathbf{k}')\tilde{\delta}_{\textrm{lin}}(\mathbf{k}_{1})\tilde{\delta}_{\textrm{lin}}(\mathbf{k}_{2})\tilde{\delta}_{\textrm{lin}}(\mathbf{k}_{3})\right> &= (2\pi)^{6}P_{\textrm{lin}}(k')P_{\textrm{lin}}(k_{3})\delta_{D}^{(3)}(\mathbf{k}'-\mathbf{k}_{1})\delta_{D}^{(3)}(\mathbf{k}_{2}+\mathbf{k}_{3}) \\ &+ (2\pi)^{6}P_{\textrm{lin}}(k')P_{\textrm{lin}}(k_{3})\delta_{D}^{(3)}(\mathbf{k}'-\mathbf{k}_{2})\delta_{D}^{(3)}(\mathbf{k}_{1}+\mathbf{k}_{3}) \\ &+ (2\pi)^{6}P_{\textrm{lin}}(k')P_{\textrm{lin}}(k_{1})\delta_{D}^{(3)}(\mathbf{k}'-\mathbf{k}_{3})\delta_{D}^{(3)}(\mathbf{k}_{1}+\mathbf{k}_{2}). 
 \end{aligned}
\end{equation}
Substituting these results into Eq. (\ref{advterm}) and integrating gives
\begin{equation}\label{advcorr}
 \begin{aligned}
  \left<\tilde{\delta}^{*}(\mathbf{k}')|\textrm{adv}(\mathbf{k})\right> = &-2 \int d^{3}\mathbf{k}_{3} \left(\widehat{\mathbf{k}}\cdot\widehat{\mathbf{k}}_{3}\right)T_{v}(k)T_{v}(k_{3})P_{\textrm{lin}}(k)P_{\textrm{lin}}(k_{3})\delta_{D}^{(3)}(\mathbf{k}-\mathbf{k}') \\ &+ 2\int d^{3}\mathbf{k}_{1}\left(\widehat{\mathbf{k}}_{1}\cdot\widehat{\mathbf{k}}\right)\frac{\mathbf{k}\cdot \mathbf{k}_{1}}{k_{1}^{2}}T_{v}(k_{1})T_{v}(k)P_{\textrm{lin}}(k_{1})P_{\textrm{lin}}(k)\delta_{D}^{(3)}(\mathbf{k}-\mathbf{k}') \\ &- 2\int d^{3}\mathbf{k}_{1} \frac{\mathbf{k}_{1}\cdot \mathbf{k}}{k^{2}}T_{v}(k_{1})^{2}P_{\textrm{lin}}(k)P_{\textrm{lin}}(k_{1})\delta_{D}^{(3)}(\mathbf{k}-\mathbf{k}') .
 \end{aligned}
\end{equation}
To proceed we will simplify the integrals above. Each integral is a 3D volume integral over either $\mathbf k_3$ (or $\mathbf k_1$), which depends on both the magnitude and direction (on $S^2$) of that wave vector. Because the dependence on the direction is simple (involving only dot products), we can simplify by taking an angular average; the integral of the angular average is equal to the integral of the original function: $\int d^3{\mathbf k}_3 \,f({\mathbf k}_3) = \int d^3{\mathbf k}_3 \,\langle f({\mathbf k}_3) \rangle_{S^2}$. We denote the average over direction $\hat{\mathbf k}_3$ (or $\hat{\mathbf k}_1$) using the symbol $\langle \rangle_{S^2}$, where $S^2$ is the unit sphere. Letting $\widehat{\mathbf{k}}\cdot\widehat{\mathbf{k}}_{3} = \mu_{3}$, the dot product in the first term on the right-hand-side of Eq. (\ref{advcorr}) simplifies to
\begin{equation}
 \left<\mu_{3}\right>_{S^{2}} = \frac{1}{4\pi}\int_{0}^{2\pi} d\zeta \int_{-1}^{1}\mu_{3}\, d\mu_{3} = 0,
\end{equation}
where $\zeta$ is the angular rotation variable in the plane perpendicular to $\mathbf{k}$. The same is true for $\mu_{1}$ appearing in the third term. From the second term we get
\begin{equation}
 \left<\left(\widehat{\mathbf{k}}_{1}\cdot\widehat{\mathbf{k}}\right)\frac{\mathbf{k}\cdot \mathbf{k}_{1}}{k_{1}^{2}}\right>_{S^{2}} = \left<\frac{k}{k_{1}}\mu_{1}^{2}\right>_{S^{2}} = \frac{k}{4\pi k_{1}}\int_{0}^{2\pi} d\zeta \int_{-1}^{1}\mu_{1}^{2}\, d\mu_{1} = \frac{k}{3k_{1}}.
\end{equation}
The power spectrum can now be read from Eq. (\ref{advcorr}) as
\begin{equation}
 P_{\textrm{adv}|\delta}(k) = \frac{2}{3}kT_{v}(k)P(k)\int \frac{d^{3}\mathbf{k}_{1}}{(2\pi)^{3}}\frac{T_{v}(k_{1})}{k_{1}}P(k_{1}).
\end{equation}

\subsection{Remaining terms}
Each remaining power spectrum in Eq.~(\ref{Pg}) may be computed similarly. We find:
\begin{equation}
\begin{aligned}
 P_{\delta |v^{2}}(k)= &-2\int \frac{d^{3}\mathbf{k}_{1}}{(2\pi)^{3}}\mu_{12}F_{2}(\mathbf{k}_{1},\mathbf{k}_{2})T_{v}(k_{1})T_{v}(k_{2})  P_{\textrm{lin}}(k_{1})P_{\textrm{lin}}(k_{2}),
\nonumber \\
P_{s^{2}|v^{2}}(k)=&-2\int \frac{d^{3}\mathbf{k}_{1}}{(2\pi)^{3}}T_{v}(k_{1})P_{\textrm{lin}}(k_{1})\left[\vphantom{\int}\mu_{12}S_{2}(\mathbf{k}_{1},\mathbf{k}_{2})
T_{v}(k_{2}) P_{\textrm{lin}}(k_{2}) + \frac{2}{3}T_{v}(k_{1})P_{\textrm{lin}}(k_{1})\vphantom{\int}\right],
\nonumber \\
 P_{\delta v_{z}|v^{2}}(k)=&-2k\int \frac{d^{3}\mathbf{k}_{1}}{(2\pi)^{3}}\frac{\mu_{12}\mu_{2}}{k_{2}}T_{v}(k_{1})T_{v}(k_{2}) P_{\textrm{lin}}(k_{1})P_{\textrm{lin}}(k_{2}),
\nonumber \\
 P_{v_{z}|v^{2}}(k)=&-2\int \frac{d^{3}\mathbf{k}_{1}}{(2\pi)^{3}}\mu_{12}G_{2}(\mathbf{k}_{1},\mathbf{k}_{2})T_{v}(k_{1})T_{v}(k_{2})  P_{\textrm{lin}}(k_{1})P_{\textrm{lin}}(k_{2}),
\nonumber \\
 P_{\textrm{adv}|v_{z}}(k)=&\,P_{\textrm{adv}|\delta} (k),\,\textrm{ and}
\nonumber \\
 P_{v_{z}v^{2}|\delta}(k)=&\,P_{v_{z}v^{2}|v_{z}}(k)=-P_{\textrm{adv}|\delta}(k).
\end{aligned}
\end{equation}

\section{\label{sec:lya}Lyman-$\alpha$ forest analysis}

We now turn our attention to the Lyman-$\alpha$ forest. The determination of the power spectrum proceeds similarly to the case of galaxies, with the exception that we must do our Fourier integrals in $\mathbf s$-space, rather than writing the Fourier integrals in $\mathbf r$-space and using conversion factors of $e^{-ik_zv_z/aH}$.
To find the flux fluctuation power spectrum we need the Fourier transform of Eq. (\ref{deltaF}), given by
\begin{equation}
    \tilde{\delta}_{F}^{s}(\mathbf k) = \int d^{3}\mathbf{s}\, \delta_{F}^{s}(\mathbf{s})e^{-i\mathbf{k}\cdot\mathbf{s}}.
\end{equation}
The right-hand-side of Eq. (\ref{deltaF}) is written in terms of \textbf{r}, which we rewrite in terms of \textbf{s} via Eq. (\ref{redtoreal}). Each field transforms according to a Taylor series. Since we are restricting our analysis to streaming velocity correlations that are net fourth-order in the matter field, only the $c_{1}, c_{2},$ and $b_{v}$ terms pick up an additional contribution. For all others we may make the substitution \textbf{r} $\rightarrow$ \textbf{s}.

We now work through the detailed transformation for the three terms specified. The $c_{1}$ term is
\begin{equation}
    \delta\left(\mathbf{s}-\frac{v_{z}}{aH}\hat{\mathbf{e}}_{z}\right) = \delta(\mathbf{s}) - \frac{v_{z}}{aH}[\partial_{z}\delta(\mathbf{s})].
\end{equation}
Similarly the $c_{2}$ term becomes
\begin{equation}
    s_{zz}\left(\mathbf{s}-\frac{v_{z}}{aH}\hat{\mathbf{e}}_{z}\right) = s_{zz}(\mathbf{s})-\frac{v_{z}}{aH}[\partial_{z}s_{zz}(\mathbf{s})].
\end{equation}
The $b_{v}$ term first maps according to Eq. (\ref{eq:vs-adv}) and then can be Taylor expanded,
\begin{equation}
 \begin{aligned}
 v_{s}^{2}(\mathbf{x}) &= v_{s}^{2}\left(\mathbf{s}-\frac{v_{z}}{aH}\hat{\mathbf{e}}_{z}\right) + \left[\nabla v_{s}^{2}\left(\mathbf{s}-\frac{v_{z}}{aH}\hat{\mathbf{e}}_{z}\right)\right]\cdot\left[\nabla \nabla^{-2}\delta\left(\mathbf{s}-\frac{v_{z}}{aH}\hat{\mathbf{e}}_{z}\right)\right] \\ &= v_{s}^{2}(\mathbf{s}) - \frac{v_{z}}{aH}[\partial_{z}v_{s}^{2}(\mathbf{s})] + [\nabla v_{s}^{2}(\mathbf{s})]\cdot[\nabla\nabla^{-2}\delta(\mathbf{s})].
 \end{aligned}
\end{equation}

The full power spectrum containing a base term and pieces with a contribution from $b_{v}$ is
\begin{equation} \label{PF}
\begin{aligned}
 P_{F}(k,\mu) &= P_{\textrm{base}}(k,\mu)+2b_{v}\left[c_{1}+c_{2}f\left(\mu^{2}-\frac{1}{3}\right)\right]\left[P_{\delta|v^{2}}(k)+(1+f\mu^{2})P_{\textrm{adv}|\delta}(k)\right] \\ &+2b_{v}f\left(c_{1}-\frac{1}{3}c_{2}\right)\left[P_{v^{2}|-\frac{v_{z}}{aH}\partial_{z}\delta,\textrm{I}}(k)+\mathcal{P}_{2}(\mu)P_{v^{2}|-\frac{v_{z}}{aH}\partial_{z}\delta,\textrm{II}}(k)\right] + 2c_{2}b_{v}fP_{v^{2}|-\frac{v_{z}}{aH}\partial_{z}s_{zz},\textrm{ other}}(k,\mu) \\ &+2c_{3}b_{v}P_{\delta^{2}|v^{2}}(k)+2c_{4}b_{v}P_{s^{2}|v^{2}}(k) +2c_{5}b_{v}fP_{\delta s_{zz}|v^{2}}(k,\mu)+2c_{6}b_{v}fP_{t_{zz}|v^{2}}(k,\mu) \\ &+2c_{7}b_{v}f^{2}P_{s_{zz}^{2}|v^{2}}(k,\mu)+2c_{8}b_{v}P_{s_{xz}^{2}+s_{yz}^{2}|v^{2}}(k,\mu) + b_{v}^{2}P_{v^{2}|v^{2}}(k),
\end{aligned}
\end{equation}
where $P_{\textrm{base}}$ is given by Eqs. (3.3) and (3.6) in \cite{Arinyo2015}. (Note that the power spectrum in Ref.~\cite{Arinyo2015} is based on fits to simulations and not to perturbation theory; however it still serves the purpose of a $P_{\textrm{base}}$ that does not contain streaming velocity contributions.) Subscripts I and II denote the spin-0 and spin-2 contributions, respectively, to $P_{v^{2}|-\frac{v_{z}}{aH}\partial_{z}\delta}$. The term $P_{v^{2}|-\frac{v_{z}}{aH}\partial_{z}s_{zz}}$ has three components, two of which are identical to I and II (up to a constant prefactor) and a third term which we label ``other.'' Streaming velocity corrections are built from the autocorrelation of Eq. (\ref{deltaF}). In this expansion, we disregard quadrupolar streaming velocity corrections (i.e., those containing $b_{vz}$) since the simulations carried out in \cite{hirataLya} show $b_{vz} = (1.6\pm 7.5)\times 10^{-5}$ and $b_v = (-3.7\pm 0.4) \times 10^{-4}$ at $z=2.5$. 

\subsection{Lyman-$\alpha$ flux example term: $P_{s_{xz}^{2}+s_{yz}^{2}|v^{2}}$}

Here we will detail how to calculate the $P_{s_{xz}^{2}+s_{yz}^{2}|v^{2}}$ term listed above. The procedure followed applies to all kernels with directionally-dependent wave vectors.

We start with the correlation functions
\begin{widetext}
\begin{equation}
    \left<(\tilde{\mathbf{v}}_{s} * \tilde{\mathbf{v}}_{s})^{*}(\mathbf{k})(\tilde{s}_{xz}*\tilde{s}_{xz})(\mathbf{k}')\right> + \left<(\tilde{\mathbf{v}}_{s} * \tilde{\mathbf{v}}_{s})^{*}(\mathbf{k})(\tilde{s}_{yz}*\tilde{s}_{yz})(\mathbf{k}')\right>
\end{equation}
which by the methods used in Section \ref{sec:galaxies} can have its power spectrum put into the form
\begin{equation}
    P_{c_{8}b_{v}}(k)= -2\int \frac{d^{3}\mathbf{k}_{1}}{(2\pi)^{3}}\mu_{12}\left(\hat{k}_{1x}\hat{k}_{1z}\hat{k}_{2x}\hat{k}_{2z} + \hat{k}_{1y}\hat{k}_{1z}\hat{k}_{2y}\hat{k}_{2z}\right)T_{v}(k_{1})T_{v}(k_{2})P_{\textrm{lin}}(k_{1})P_{\textrm{lin}}(k_{2}).
\end{equation}
As before, we want to simplify these wave vectors by taking an angular average. Unlike the case of Eq.~(\ref{advcorr}), here both ${\mathbf k}_1$ and ${\mathbf k}_2$ appear non-trivially in the integrand, so a full angular average $\langle \rangle_{S^2}$ will not result in a simplification. However, the integrand does behave in a straightforward way if we rotate around the vector ${\mathbf k}$; therefore, we consider an angular average $\langle \rangle_{S^1}$ over the ring swept out as we rotate the ${\mathbf k}_1, {\mathbf k}_2, {\mathbf k}$ triangle around ${\mathbf k}$.
Writing the wave vectors in terms of the spherical harmonics yields
\begin{equation}
\begin{aligned}
    \left<\hat{k}_{1x}\hat{k}_{1z}\hat{k}_{2x}\hat{k}_{2z} + \hat{k}_{1y}\hat{k}_{1z}\hat{k}_{2y}\hat{k}_{2z} \right>_{S^{1}} = &\hphantom{+}\left<\frac{8\pi}{15}\left[\frac{Y_{21}(\hat{\mathbf{k}}_{1})-Y_{2,-1}(\hat{\mathbf{k}}_{1})}{2}\times\frac{Y_{21}(\hat{\mathbf{k}}_{2})-Y_{2,-1}(\hat{\mathbf{k}}_{2})}{2} \right. \right. \\ &+ \left. \left.\frac{Y_{21}(\hat{\mathbf{k}}_{1})+Y_{2,-1}(\hat{\mathbf{k}}_{1})}{2i}\times\frac{Y_{21}(\hat{\mathbf{k}}_{2})+Y_{2,-1}(\hat{\mathbf{k}}_{2})}{2i}\right] \right>_{S^{1}}\\ = &-\frac{8\pi}{15}\operatorname{Re}\left[\left<Y_{21}(\hat{\mathbf{k}}_{1})Y_{2,-1}(\hat{\mathbf{k}}_{2})\right>_{S^{1}}\right].
\end{aligned}
\end{equation}
Just as we did in Section \ref{sec3B} we need to represent each unit wave vector, currently expressed in the observer's coordinate system, in the barred coordinate system where the $\overline z$-axis points in the direction $\hat{\mathbf k}$. We do this using Eulerian angles, following the convention outlined in Section 11.8 of \cite{Thornton&Marion}. The first rotation is done counterclockwise through an angle $\phi$ about the $z$-axis. Next, rotate counterclockwise through an angle $\theta$ about the new ``$y$-axis.'' Last, rotate counterclockwise through an angle $\psi$ about the new ``$z$-axis.'' This series of rotations from unbarred to barred coordinates is given by the rotation matrix 
\begin{equation}
    \mathbf{R}_{\textrm{Euler}} = 
    \begin{bmatrix}
    \cos\psi\, \cos\phi - \cos\theta\, \sin\phi\, \sin\psi & \cos\psi\, \sin\phi + \cos\theta\, \cos\phi\, \sin\psi & \sin\psi\, \sin\theta \\
    -\sin\psi\, \cos\phi - \cos\theta\, \sin\phi\, \cos\psi & -\sin\psi\, \sin\phi + \cos\theta\, \cos\phi\, \cos\psi & \cos\psi\, \sin\theta \\
    \sin\theta\, \sin\phi & -\sin\theta\, \cos\phi & \cos\theta
    \end{bmatrix}.
\end{equation}
Rotation of spherical harmonics from the unbarred to barred coordinate system is carried out via the relation
\begin{equation}
    Y_{lm}(\hat{\mathbf{k}})=\sum_{\overline{m}}{D}^{\dagger(l)}_{m\overline{m}}(\phi,\theta,\psi)Y_{l\overline{m}}(\overline{\hat{\mathbf{k}}})
\end{equation}
where $D^{\dagger(l)}_{m\overline{m}}$ is the (adjoint) Wigner \textit{D}-matrix defined by 
\begin{equation}
   D^{\dagger(l)}_{m,-s}(\phi,\theta,\psi)=(-1)^{s}\sqrt{\frac{4\pi}{2l+1}}\hphantom{|}_{s}Y_{lm}(\theta,\phi)e^{-is\psi}. 
\end{equation}
Here $\overline{m}=-s$ and $\hphantom{|}_{s}Y_{lm}$ are the spin-weighted spherical harmonics\footnote{We made use of the SpinWeightedSpheroidalHarmonics package, part of the Black Hole Perturbation Toolkit \cite{BHPToolkit}, to calculate these functions.}. With these relations we simplify the product of spherical harmonics: 
\begin{equation}
    \begin{aligned}
     \left<Y_{21}(\hat{\mathbf{k}}_{1})Y_{2,-1}(\hat{\mathbf{k}}_{2})\right>_{S^{1}} &= \left<\sum_{\overline{{m_{1}}},\overline{m_{2}}} D^{\dagger(2)}_{1,\overline{m_{1}}} D^{\dagger(2)}_{-1,\overline{m_{2}}}Y_{2\overline{m_{1}}}(\overline{\hat{\mathbf{k}}}_{1})Y_{2\overline{m_{2}}}(\overline{\hat{\mathbf{k}}}_{2})\right>_{S^{1}} \\
     &= \sum_{\overline{m_{1}}} D^{\dagger(2)}_{1,\overline{m_{1}}} D^{\dagger(2)}_{-1,-\overline{m_{1}}}Y_{2\overline{m_{1}}}(\overline{\hat{\mathbf{k}}}_{1})Y_{2,-\overline{m_{1}}}(\overline{\hat{\mathbf{k}}}_{2}) \\
     &= -\frac{15}{64\pi}(1-\mu^{4})(1-\mu_{1}^{2})(1-\mu_{2}^{2})-\frac{15}{16\pi}\frac{k_{2}}{k_{1}}(3\mu^{2}-4\mu^{4}-1)(1-\mu^{2}_{2})\mu_{1}\mu_{2} \\ &\hphantom{=}-\frac{15}{32\pi}\mu^{2}(1-\mu^{2})(3\mu_{1}^{2}-1)(3\mu_{2}^{2}-1).
     \end{aligned}
\end{equation}
 In going from the first to second line we used the fact that averaging forces all terms for which $m_{1}' \neq -m_{2}'$ to equal zero. Substituting the above equations into the kernel and renormalizing the resultant integral (i.e., subtracting the contribution at $\mathbf k=0$) yields the power spectrum
\begin{equation}
    \begin{aligned}
        P_{s_{xz}^{2}+s_{yz}^{2}|v^{2}}(k,\mu)= &-\int \frac{d^{3}\mathbf{k}_{1}}{(2\pi)^{3}}T_{v}(k_{1})P_{\textrm{lin}}(k_{1})\left\{\mu_{12}\left[\frac{(1-\mu^{4})(1-\mu_{1}^{2})(1-\mu_{2}^{2})}{4}+(3\mu^{2}-4\mu^{4}-1)(1-\mu^{2}_{2})\frac{k_{2}}{k_{1}}\mu_{1}\mu_{2} \right. \right. \\  &+ \left. \left. \frac{\mu^{2}(1-\mu^{2})(3\mu_{1}^{2}-1)(3\mu_{2}^{2}-1)}{2}\right]T_{v}(k_{2})P_{\textrm{lin}}(k_{2}) + \left[ \frac{(1-\mu^{4})(1-\mu_{1}^{2})^{2}}{4} \right. \right. \\ &+ \left. \left. (3\mu^{2}-4\mu^{4}-1)(1-\mu_{1}^{2})\mu_{1}^{2}+\frac{\mu^{2}(1-\mu^{2})(3\mu_{1}^{2}-1)^{2}}{2}\right]T_{v}(k_{1})P_{\textrm{lin}}(k_{1})\right\}.
    \end{aligned}
\end{equation}
\end{widetext}
\subsection{Remaining terms}
The other new contributions are 
\begin{equation}
\begin{aligned}
 P_{v^{2}|-\frac{v_{z}}{aH}\partial_{z}\delta,\textrm{I}}(k) = 2\int \frac{d^{3}\mathbf{k}_{1}}{(2\pi)^{3}}T_{v}(k_{1})P_{\textrm{lin}}(k_{1})\left[\frac{k_{2}}{k_{1}}\frac{\mu_{12}^{2}}{3}T_{v}(k_{2})P_{\textrm{lin}}(k_{2})+\frac{1}{3}T_{v}(k_{1})P_{\textrm{lin}}(k_{1})\right],
\end{aligned}
\end{equation}

\begin{equation}
\begin{aligned}
 P_{v^{2}|-\frac{v_{z}}{aH}\partial_{z}\delta,\textrm{II}}(k) = 2\int\frac{d^{3}\mathbf{k}_{1}}{(2\pi)^{3}}\frac{k_{2}}{k_{1}}\mu_{12}\left(\mu_{1}\mu_{2}-\frac{\mu_{12}}{3}\right)T_{v}(k_{1})T_{v}(k_{2})P_{\textrm{lin}}(k_{1})P_{\textrm{lin}}(k_{2}),
\end{aligned}
\end{equation}

\begin{equation}
\begin{aligned}
P_{v^{2}|-\frac{v_{z}}{aH}\partial_{z}s_{zz},\textrm{ other}}(k,\mu) = &-\int\frac{d^{3}\mathbf{k}_{1}}{(2\pi)^{3}}\frac{k_{2}}{k_{1}}\mathcal{P}_{1}(\mu_{12})\left[\frac{6}{35}(1-\mu^{2})(5\mu^{2}-1)\frac{k_{2}}{k_{1}}[\mathcal{P}_{2}(\mu_{2})-\mathcal{P}_{4}(\mu_{2})]\right. \\ &\left. + \frac{2}{5}(5\mu^{2}-3)\mathcal{P}_{1}(\mu_{1})\mathcal{P}_{3}(\mu_{2}) + \frac{2}{5}\frac{k_{2}}{k_{1}}(1-\mu^{2})(1-\mathcal{P}_{2}(\mu_{2})) \right. \\ &\left. +\frac{3}{10}\mu^{2}\mathcal{P}_{1}(\mu_{1}){P}_{1}(\mu_{2}) \right]T_{v}(k_{1})T_{v}(k_{2})P_{\textrm{lin}}(k_{1})P_{\textrm{lin}}(k_{2}) \\ &- \frac{8\pi}{5}(1-\mu^{2})\int\frac{dk_{1}}{(2\pi)^{3}}k_{1}^{2}T_{v}^{2}(k_{1})P_{\textrm{lin}}^{2}(k_{1}) -  \frac{2\pi}{5}\mu^{2}\int\frac{dk_{1}}{(2\pi)^{3}}k_{1}^{2}T_{v}^{2}(k_{1})P_{\textrm{lin}}^{2}(k_{1}),
\end{aligned}
\end{equation}

\begin{equation}
\begin{aligned}
P_{\delta s_{zz}|v^{2}}(k,\mu) = &-2\left(\mu^{2}-\frac{1}{3}\right)\int \frac{d^{3}\mathbf{k}_{1}}{(2\pi)^{3}}T_{v}(k_{1})P_{\textrm{lin}}(k_{1})[\mu_{12}\mathcal{P}_{2}(\mu_{2})T_{v}(k_{2})P_{\textrm{lin}}(k_{2}) + \mathcal{P}_{2}(\mu_{1})T_{v}(k_{1})P_{\textrm{lin}}(k_{1})], \end{aligned}    
\end{equation}

\begin{equation}
\begin{aligned}
 P_{t_{zz}|v^{2}}(k,\mu)= -4\mathcal{P}_{2}(\mu)\int \frac{d^{3}\mathbf{k}_{1}}{(2\pi)^{3}}[G_{2}(\mathbf{k}_{1},\mathbf{k}_{2})-F_{2}(\mathbf{k}_{1},\mathbf{k}_{2})] \mu_{12}T_{v}(k_{1})T_{v}(k_{2})P_{\textrm{lin}}(k_{1})P_{\textrm{lin}}(k_{2}), 
\end{aligned}
\end{equation}

\begin{equation}
\begin{aligned}
 P_{s_{zz}|v^{2}}(k, \mu)= &-2\int \frac{d^{3}\mathbf{k}_{1}}{(2\pi)^{3}}T_{v}(k_{1})P_{\textrm{lin}}(k_{1})\left\{\left[\frac{1}{8}(1-\mu^{2})^{2}(1-\mu_{1}^2)(1-\mu_{2}^2)-2\mu^{2}(1-\mu^{2})(1-\mu_{2}^{2})\mu_{1}\mu_{2}\frac{k_{2}}{k_{1}}\right. \right. \\ &\left. \left.  + \frac{4}{9}\mathcal{P}_{2}(\mu)^{2}\mathcal{P}_{2}(\mu_{1})\mathcal{P}_{2}(\mu_{2})\vphantom{\frac{1}{1}}\right] \mu_{12}T_{v}(k_{2})P_{\textrm{lin}}(k_{2}) + \left[\frac{1}{8}(1-\mu^{2})^{2}(1-\mu_{1}^{2})^{2}-2\mu^{2}(1-\mu^{2})(1-\mu_{1}^{2})\mu_{1}^{2} \right. \right. \\ &\left. \left. +\frac{4}{9}\mathcal{P}_{2}(\mu)^{2}\mathcal{P}_{2}(\mu_{1})^{2}\right] T_{v}(k_{1})P_{\textrm{lin}}(k_{1})\vphantom{\int}\right\}, \textrm{ and}
\end{aligned}
\end{equation}
\begin{equation}
 \begin{aligned}
  P_{v^{2}|v^{2}}(k) = 2\int \frac{d^{3}\mathbf{k}_{1}}{(2\pi)^{3}}T_{v}^{2}(k_{1})P_{\textrm{lin}}(k_{1}) \left[\mu_{12}^{2}T_{v}^{2}(k_{2})P_{\textrm{lin}}(k_{2}) - T_{v}^{2}(k_{1})P_{\textrm{lin}}(k_{1})\right].
 \end{aligned}
\end{equation}

\subsection{Bias coefficient estimation}\label{bias_est}

Different methods exist for determining the bias parameters $\{c_{i},b_{v}\}$. The most direct way is by comparing measured auto- and cross-correlations of appropriate tracers. This method is excellent at leading order but becomes increasingly difficult to implement in practice as more unique cross-correlations are needed. Alternatively, given physical model parameters and an observed flux one point distribution function, there are analytic expressions for certain coefficients \cite{seljak}. Unfortunately these expressions exist in literature only for $c_{1}$ and $c_{2}$.  

For the analysis in Section \ref{sec:shift} we use smoothed particle hydrodynamics (SPH) simulation results from \cite{Arinyo2015} for $c_{1}$ and $c_{2}$ and from \cite{hirataLya} for $b_{v}$, both of which ran {\sc{GADGET}}-2 code \cite{Gadget2}. Additionally, we use the fluctuating Gunn-Peterson approximation (FGPA) \cite{FGPA,FGPAexplained} as a guide for determining bias coefficients. The FGPA is only a crude approximation, but should be a useful qualitative guide to what to expect for the 2nd order coefficients. Later in Section \ref{sec:shift} we will show how much the choice of these coefficients matters for determining the BAO peak shift. In the FGPA (without thermal broadening) the optical depth in redshift space is
\begin{equation}\label{tau}
\tau(\mathbf{s})=\frac{A(1+\delta)^{2-0.7(\gamma-1)}}{1+\frac{1}{aH}\frac{\partial v_{z}}{\partial r_{z}}},
\end{equation}
where $A\approx 0.3$ at $z=2.4$ \cite{seljak}, $\gamma\approx 1.6$, and $\gamma -1$ is the slope of the IGM temperature-density relation. We Taylor expand the denominator
\begin{equation}
    \left(1+\frac{1}{aH}\frac{\partial v_{z}}{\partial r_{z}}\right)^{-1}=1-\frac{1}{aH}\frac{\partial v_{z}}{\partial r_{z}}+\left(\frac{1}{aH}\frac{\partial v_{z}}{\partial r_{z}}\right)^{2},
\end{equation}
and make use of the relation
\begin{equation}
    -\frac{1}{aH}\frac{\partial v_{z}}{\partial r_{z}} = f\left(t_{zz}+s_{zz}-\frac{1}{3aH}\theta\right),
\end{equation}
where $f$ is the dimensionless linear growth rate which is approximately unity at this redshift. At first and second order $\theta$ is not an independent variable \cite{mcdonald}. Instead it can be expressed as
\begin{equation}\label{eq:mcdonald}
    -\frac{1}{aH}\theta = \delta + \frac{2}{7}s^{2} -\frac{4}{21}\delta^{2}.
\end{equation}
With the preceding results Eq. (\ref{tau}) can be expressed as
\begin{equation}
    \tau(\mathbf{s}) = 0.32\delta^{2}+0.68\delta s_{zz}+0.58\delta+0.029s^{2} + 0.3\,(t_{zz}+s_{zz}+s_{zz}^{2}+1).
\end{equation}

The transmitted flux is nonlinearly related to the optical depth via
\begin{equation}
    F(\mathbf{s})=\exp[-\tau(\mathbf{s})],
\end{equation}
which we Taylor expand to second order. Ultimately the statistical quantity we care about is transmitted flux density contrast 
\begin{equation}
    \delta_{F}(\mathbf{s}) = \frac{F(\mathbf{s})}{\left<F(\mathbf{s})\right>_{\mathbf{s}}}-1.
\end{equation}
Ref. \cite{Arinyo2015} gives  $\left<F(\mathbf{s})\right>_{\mathbf{s}}=0.8185$ at $z=2.4$, therefore 
\begin{equation}\label{delta_fgpa}
    \delta_{F}(\mathbf{s})_{\textrm{FGPA}} = -0.137\delta^{2}-0.46\delta  s_{zz}-0.52\delta-0.026s^{2}-0.27t_{zz}-0.27s_{zz}-0.23s_{zz}^{2}-0.09. 
\end{equation}
  Our plan is to use FGPA estimations for unknown bias parameters while relying on simulation results for the others. Simulations performed at $z=2.4$ using a first-order bias expansion found $c_{1}=-0.183$ and $c_{2}=-0.1714$ \cite{Arinyo2015}. In that work, $b_{F_{\delta}}=b_{\tau_{\delta}}\textrm{ln}\,\overline{F}$, $b_{F_{\eta}}=b_{\tau_{\eta}}\textrm{ln}\,\overline{F}$, and parameter values come from Table 4. Ref. \cite{hirataLya} found $b_{v}=-3.7\times10^{-4}$ at $z=2.5$. More recent work which included a single second-order term in the expansion of $\delta_{F}$ found $c_{3} \simeq 0.05$ at $z=2.3$ \cite{Tie}. For our purposes we can ignore differences in the redshifts and treat them all as occurring at $z=2.4$.
Therefore the values we choose for bias parameters are
\begin{align}\label{bias_values}
    \begin{bmatrix}
           c_{1,\textrm{ sim}} \\
           c_{2,\textrm{ sim}} \\
           c_{3,\textrm{ sim}} \\
           c_{4,\textrm{ FGPA}} \\
           c_{5,\textrm{ FGPA}} \\
           c_{6,\textrm{ FGPA}} \\
           c_{7,\textrm{ FGPA}} \\
           c_{8,\textrm{ FGPA}} \\
           b_{v}
       \end{bmatrix}
       =
       \begin{bmatrix}
           -0.183 \\
           -0.1714 \\
           0.05 \\
           -0.026 \\
           -0.46 \\
           -0.27 \\
           -0.23 \\
           0 \\
           -3.7\times 10^{-4}
       \end{bmatrix}.
  \end{align}

\subsection{\label{sec:counterterm}Counterterm expression}

  In this section we provide an analytic expression for the components of the counterterm $c_{0}$ introduced in Eq.~(\ref{deltaF}). The key idea is that $c_0$ is a constant that must be added to ensure the mean tracer fluctuation ($\delta_F$ in the case of the Lyman-$\alpha$ forest) is zero. This is a simple exercise in real space to second order (e.g., $\delta^2$ has a counterterm $-\sigma^2$, where $\sigma^2$ is the variance of the density field). In redshift space, it is more complicated. It is generally true that
  \begin{equation}
      \langle \delta_{F}(\mathbf{s}) \rangle_{\mathbf{s}} = \left<\delta_{F}(\mathbf{s})\left|\frac{\partial\mathbf{s}}{\partial \mathbf{r}}\right|\right>_{\mathbf{r}} = \left<\delta_{F}(\mathbf{s})\left[1+\frac{1}{aH}\frac{\partial v_{z}}{\partial r_{z}} \right]\right>_{\mathbf{r}}.
  \end{equation}
 Proceeding term-by-term in the expansion above we find the following, keeping only second-order terms at the end: 
 \begin{equation}
  \langle c_{1}\delta\rangle_{\mathbf{s}} = c_{1}\langle\delta\rangle_{\mathbf{r}} - c_{1}f \left<\delta\left(s_{zz}+\frac{1}{3}\delta\right)\right>_{\mathbf{r}} = -\frac{1}{3}fc_{1}\sigma^{2},
 \end{equation}
 \begin{equation}
     \langle c_{2}s_{zz}\rangle_{\mathbf{s}} = -c_{2}f\langle s_{zz}^{2}\rangle_{\mathbf{r}},
 \end{equation}
 \begin{equation}
     \langle c_{3}\delta^{2}\rangle_{\mathbf{s}} = c_{3}\langle\delta^{2}\rangle_{\mathbf{r}} + c_{3}f\left<\delta^{2}\left(-t_{zz}-s_{zz}+\frac{1}{3aH}\theta\right)\right>_{\mathbf{r}} = c_{3}\sigma^{2},
 \end{equation}
 \begin{equation}
    \langle c_{4}s^{2}\rangle_{\mathbf{s}} = c_{4}\langle s^{2}\rangle_{\mathbf{r}},  
 \end{equation}
 \begin{equation}
     \langle c_{5}\delta s_{zz}\rangle_{\mathbf{s}} = \mathcal{O}(\delta^{3}),
 \end{equation}
 \begin{equation}
     \langle c_{6}t_{zz}\rangle_{\mathbf{s}} = \mathcal{O}(\delta^{3}),
 \end{equation}
 \begin{equation}
     \langle c_{7}s_{zz}^{2}\rangle_{\mathbf{s}} = c_{7}\langle s_{zz}^{2}\rangle_{\mathbf{r}},\textrm{ and}
 \end{equation}
 \begin{equation}
     \langle c_{8}(s_{xz}^{2}+s_{yz}^{2})\rangle_{\mathbf{s}} = c_{8}\langle(s_{xz}^{2}+s_{yz}^{2})\rangle_{\mathbf{r}}.
 \end{equation}
 The sum of each of these terms equals $-c_{0}$ since the streaming velocity contributions in Eq. (\ref{deltaF}) already have zero mean.
 
\section{\label{sec:mapping}bias coefficient mapping}

In Section \ref{sec:lyabias} we introduced a generalized density contrast applicable to any redshift-space tracer of the matter field. As an example of its utility, we will show how the redshift-space version of Eq. (\ref{eq:4}) can be made to look like Eq. (\ref{deltaF}) given some algebraic manipulation and the correct choice of $c$ coefficients.

Expanding Eq. (\ref{delta_s}) gives 
\begin{equation}
\begin{aligned}
\delta_{g}(\mathbf{s}) &= \delta_{g}(\mathbf{r}) - \frac{1}{aH}\frac{\partial v_{z}}{\partial r_{z}} + \left(\frac{1}{aH}\frac{\partial v_{z}}{\partial r_{z}}\right)^{2} - \frac{1}{aH}\frac{\partial v_{z}}{\partial r_{z}}\delta_{g}(\mathbf{r}) + \left(\frac{1}{aH}\frac{\partial v_{z}}{\partial r_{z}}\right)^{2}\delta_{g}(\mathbf{r}) \\ &= \delta_{g}(\mathbf{r}) + f\left[t_{zz}+s_{zz}+\frac{1}{3}\left(\delta+\frac{2}{7}s^{2}-\frac{4}{21}\delta^{2}\right)\right] + f^{2}\left[t_{zz}+s_{zz}+\frac{1}{3}\left(\delta+\frac{2}{7}s^{2}-\frac{4}{21}\delta^{2}\right)\right]^{2} \\ &+  f\left[t_{zz}+s_{zz}+\frac{1}{3}\left(\delta+\frac{2}{7}s^{2}-\frac{4}{21}\delta^{2}\right)\right]\delta_{g}(\mathbf{r}) + f^{2}\left[t_{zz}+s_{zz}+\frac{1}{3}\left(\delta+\frac{2}{7}s^{2}-\frac{4}{21}\delta^{2}\right)\right]^{2}\delta_{g}(\mathbf{r}) \\ &= \left(b_{1}+\frac{1}{3}f\right)\delta + fs_{zz}+\left(\frac{1}{2}b_{2}-\frac{4}{63}f+\frac{1}{9}f^{2}+ \frac{1}{3}b_{1}f\right)\delta^{2} + \left(\frac{1}{2}b_{s}+\frac{2}{21}f\right)s^{2}+ \left(\frac{2}{3}f^{2}+b_{1}f\right)\delta s_{zz} \\ &+ ft_{zz} +f^{2}s_{zz}^{2}- \frac{1}{2}b_{2}\langle\delta^{2}\rangle- \frac{1}{2}b_{s}\langle s^{2}\rangle + b_{v}\left[v_{s}^{2}(\mathbf{x})-1\right] + b_{1v}\delta(\mathbf{r})\left[v_{s}^{2}(\mathbf{x})-1\right] +b_{sv}s_{ij}(\mathbf{r})v_{s,i}(\mathbf{x})v_{s,j}(\mathbf{x}).
\end{aligned}
\end{equation}
The prefactors of each field ($\delta$, $s_{zz}$, etc.) can be identified as $c_1$ ... $c_8$. Their values, written in terms of $b$, are provided in Table \ref{galaxymap}.
\begin{table}[t]
\begin{tabular}{|l|l|}
\hline
General coefficient & Galaxy bias equivalent                                                \\
\hline 
$c_{1}$             & $b_{1}+\frac{1}{3}f$                                                  \\[2pt]
$c_{2}$             & $f$                                                                   \\[2pt]
$c_{3}$             & $\frac{1}{2}b_{2}-\frac{4}{63}f+\frac{1}{9}f^{2} + \frac{1}{3}b_{1}f$ \\[2pt]
$c_{4}$             & $\frac{1}{2}b_{s}+\frac{2}{21}f$                                      \\[2pt]
$c_{5}$             & $\frac{2}{3}f^{2}+b_{1}f$                                             \\[2pt]
$c_{6}$             & $f$                                                                   \\[2pt]
$c_{7}$             & $f^{2}$                                                               \\[2pt]
$c_{8}$             & 0        \\ \hline                                                                                                                
\end{tabular}
\caption{Mapping between generalized coefficients and galaxy bias parameters. These are obtained by rewriting Eq. (\ref{delta_s}) to match Eq. (\ref{deltaF}) and making a term-by-term comparison.}
\label{galaxymap}
\end{table}

\section{\label{sec:shift}BAO peak shift}

\begin{figure}
    \centering
    \includegraphics[width=0.7\textwidth]{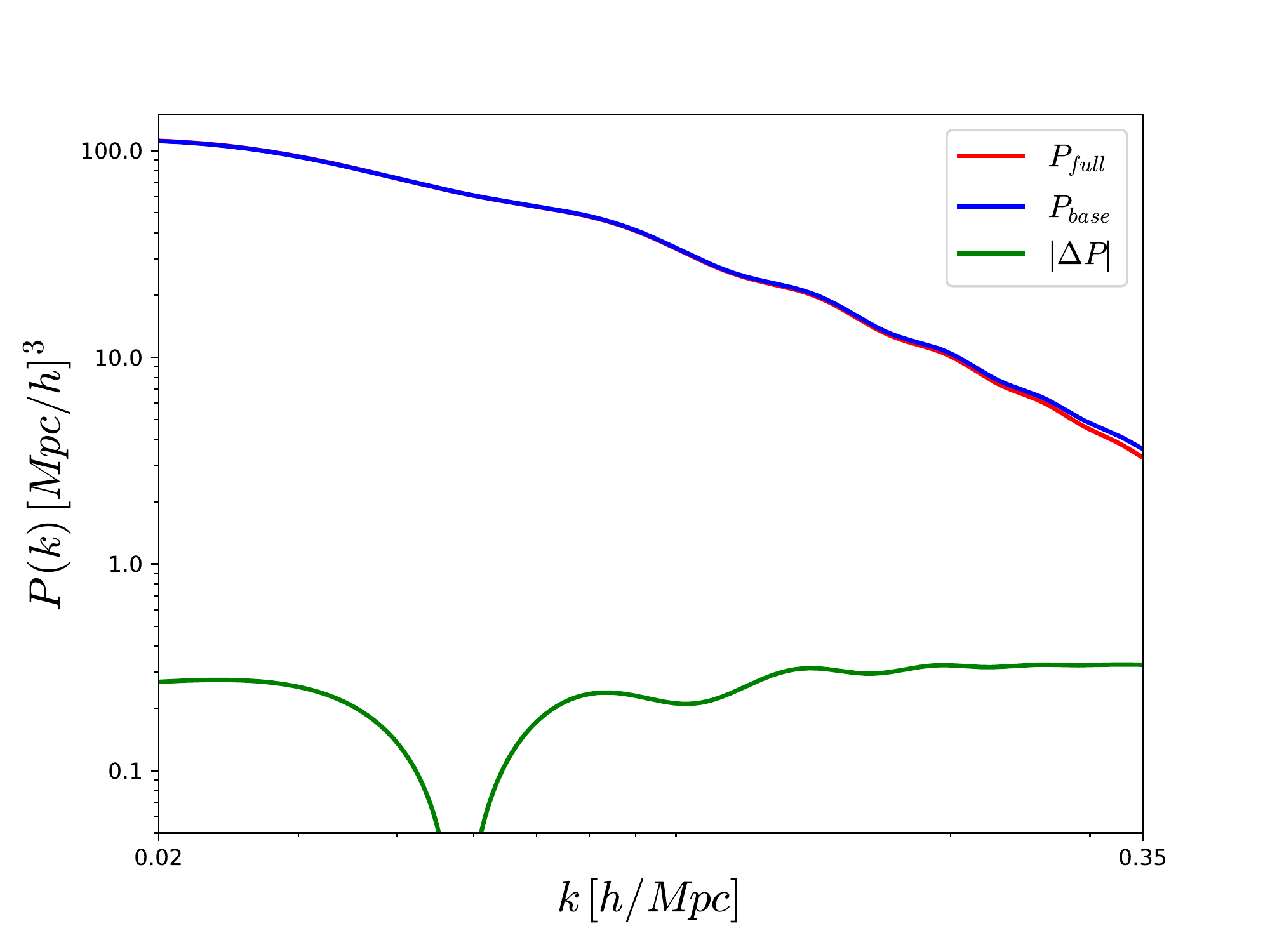}
    \caption{Shown here are the full power spectrum (red) of Eq. \ref{PF}, the base model (blue) from \cite{Arinyo2015}, and the absolute value of the difference between them (green). $\Delta P$ changed sign from positive at small $k$ to negative at $k \gtrsim 0.04 \,h/\textrm{Mpc}$. Both here and in Figure \ref{delta_alpha} we took $c_1$ and $c_2$ from \cite{Arinyo2015} and the remaining $c$ coefficients from FGPA. These were evaluated at $\mu=1/\sqrt{3}$.}
    \label{BAO_shift_zoom}
\end{figure}

To determine the BAO peak shift induced by streaming velocity we fit a model power spectrum template to Eq. (45) using $\chi^{2}$ minimization, following the method used in Appendix B of \cite{blazek} which draws from \cite{Yoo13,Seo2008}. This template is given by 
\begin{equation}
 P_{\textrm{model}}(k) = \sum_{j=0}^{2}a_{j}k^{j}P_{\textrm{evo}}
(k/\alpha) + \sum_{j=0}^{5}b_{j}k^{j},
\end{equation}
where $\alpha$ sets the BAO scale and the coefficients $a_{j}$ and $b_{j}$ are marginalized over. We account for nonlinear BAO damping with the evolved power spectrum
\begin{equation}
P_{\textrm{evo}}(k)=[P_{\textrm{ww}}(k)-P_{\textrm{nw}}(k)]e^{-k^{2}\Sigma^{2}/2}+P_{\textrm{nw}},
\end{equation}
where $P_{\textrm{ww}}$ is the baryonic linear power spectrum ``with wiggles,'' $P_{\textrm{nw}}$ is the no-wiggle baryonic linear power spectrum from \cite{EH1998}, and $\Sigma$ is a damping parameter.  
The $\chi^{2}$ integral is
\begin{equation}
\chi^{2} = V\int_{k_{\textrm{min}}}^{k_{\textrm{max}}}\frac{d^{3}k}{(2\pi)^{3}}\frac{[P_{F}(k)-P_{\textrm{model}}(k)]^{2}}{2[P_{F}(k)+1/\overline{n}_{\textrm{eff}}]^{2}},
\end{equation}
where the effective density of quasar sight lines is $\overline{n}_{\textrm{eff}}=5.6\times 10^{-5}\, \textrm{Mpc}^{-3},$ quoting the value in \cite{hirataLya} which is based on DESI forecasts. The integration range we fit over is $0.02\,h\,\textrm{Mpc}^{-1}< k < 0.35\,h\,\textrm{Mpc}^{-1}$. $P_{F}(k)$ is given by Eq. (\ref{PF}) and plotted in Fig. \ref{BAO_shift_zoom}. The choice of volume $V$ is inconsequential since it cancels out with volume in the effective number density. 

\begin{figure}
    \centering
    \includegraphics[width=0.7\textwidth]{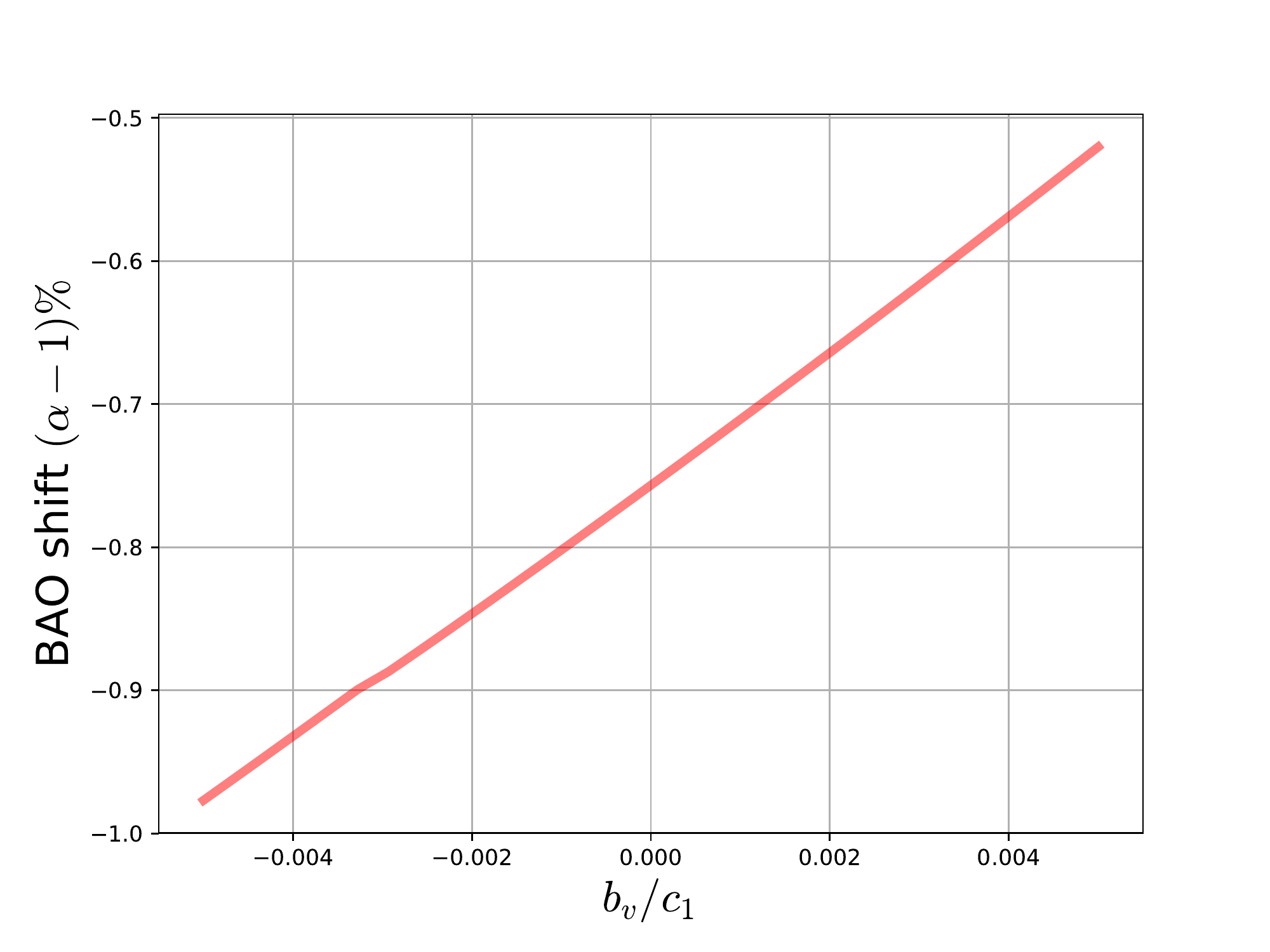}
    \caption{A plot of BAO scale shift as a function of $b_v/c_1$, evaluated at $\mu=1/\sqrt{3}$. One finds a percentage BAO shift, $\Delta \alpha\,\%$, by evaluating the function when streaming velocity is turned on (i.e. $b_v/c_1 = 0.002$), and subtracting from that the function evaluated when streaming velocity is turned off (i.e. $b_v/c_1 =0$). The scale shift is not zero when $b_v=0$ due to the inclusion of $P_{13}$ and $P_{22}$ terms from standard perturbation theory; if the base power spectrum were simply $P_{\textrm{lin}}$ there would be no offset. Similar plots were used to calculate all values in Table \ref{shift_percents}.}
    \label{delta_alpha}
\end{figure}

Our minimizer performs the required fits in the $\chi^{2}$ integral using a Nelder-Mead optimizer \cite{NelderMead} on $[\,\alpha,\Sigma\,]$. It starts by searching parameter space for $[\,\alpha,\Sigma\,]$ and for a given pair uses least squares fitting to find $a_{j}$ and $b_{j}$. If the minimizer ventures outside an acceptable region, it forces the integral to return a divergent answer. This kicks the minimizer back to an allowed region of parameter space. We ran this fit for different values of $\mu$ to account for anisotropic damping. The difference in $\alpha$ calculated when $b_v=0$ and $b_v = -3.7\times 10^{-4}$ defines the shift in BAO peak position; see Fig. \ref{delta_alpha} for additional details.

Table \ref{shift_percents} displays the percentage shift in BAO position for three different choices of $\mu$. We calculated these shifts using the bias coefficients from Eq. (\ref{bias_values}). To see how much the choice of coefficients affects the results, we also tested two cases where only simulational values were used and all unknown coefficients were set to zero. Our results show a noticeable dependence of BAO peak shift magnitude on choice of bias coefficients; this dependence is strongest when the power spectrum is evaluated perpendicular to the line of sight. This is also the case for which we see the greatest BAO peak shift. In other words, RSD has the effect of diminishing the BAO peak shift relative to what one would expect in real space. 

It is important to note that the FGPA overestimated, in terms of absolute value, the bias coefficients $c_{1}$ and $c_{2}$ relative to simulations in \cite{Arinyo2015}. These simulations included thermal broadening---an effect which FGPA ignores---which decreases the magnitude of leading order bias coefficients. This gives us reason to believe that other coefficients may be similarly overestimated, causing greater values for $\Delta\alpha$ in the last column of Table \ref{shift_percents} than we see in the other columns. 

As mentioned in Section \ref{sec:intro}, DESI will achieve Lyman-$\alpha$ forest BAO peak measurements down to a precision of 0.46\% over $2<z<2.7$, the range of redshifts used in forecasting. Results in Table \ref{shift_percents} make clear that streaming velocity effects alone can account for anywhere from $\sim 11-32$\% of the total error budget allotted to this measurement. While a $<1\sigma$ effect, future work should be done to better quantify the BAO peak shift and account for this effect in the DESI analysis pipeline. This will be even more true if a more ambitious Lyman-$\alpha$ forest mapping experiment takes place in the future (e.g., \cite{2019BAAS...51g.229S}).

\begin{table}[h]
\begin{tabular}{|l|l|l|l|}
\hline
$\Delta \alpha$ in \% & \begin{tabular}[c]{@{}l@{}}$c_{1}$, $c_{2}$, and $c_{3}$ from simulations, \\ all others are zero\end{tabular} & \begin{tabular}[c]{@{}l@{}}$c_{1}$ and $c_{2}$ from simulations,\\ all others are zero\end{tabular} & \begin{tabular}[c]{@{}l@{}}$c_{1}$ and $c_{2}$ from simulations, \\ all others from FGPA\end{tabular} \\ \hline
$\mu=0$               & 0.081\%                                                                                                        & 0.088\%                                                                                             & 0.149\%                                                                                               \\
$\mu=1/\sqrt{3}$      & 0.066\%                                                                                                        & 0.070\%                                                                                             & 0.093\%                                                                                               \\
$\mu=1$               & 0.053\%                                                                                                        & 0.054\%                                                                                             & 0.058\%                                                                                               \\ \hline
\end{tabular}
\caption{The BAO peak shift for three different $\mu$ values and three different choices of bias coefficients.}
\label{shift_percents}
\end{table}

\section{\label{sec:conclusion}Discussion}

In the era of precision cosmology it is imperative that researchers have a grasp of systematics that affect our data. In this work we explored the impact of one systematic -- namely, early universe streaming velocity -- on the BAO signature in the Lyman-$\alpha$ forest. To do this we introduced a completely generalized second-order perturbative expansion for an arbitrary cosmological tracer in redshift space. Applying this expansion to the Lyman-$\alpha$ forest auto-power spectrum, we found shifts in the transverse BAO scale of  $0.081$--$0.149\%$ and shifts in the radial BAO scale of $0.053$--$0.058\%$. This can be compared with the expected aggregate DESI precision of Lyman-$\alpha$ forest (+ quasar) BAO of 0.46\% \cite{DESI}. The range of estimated peak shifts in the columns of Table \ref{shift_percents} demonstrates the importance of accurately knowing bias coefficients. 

In future work we will focus on determining all second-order bias coefficients $c_3 - c_8$, which will lead to a more accurate predicted BAO scale shift. We leave as future work an investigation into biasing terms involving the difference $\delta_{\rm b}-\delta_{\rm m}$ between baryonic and total matter perturbations, and their potential impact on BAO scale measurements. We also leave to future work the treatment of the streaming velocity correction to cross-correlation BAO measurements, such as Lyman-$\alpha$ forest $\times$ quasars, which are a significant contributor to the BOSS and DESI constraining power at $z>2$ \cite{Font-Ribera2014, 2017A&A...608A.130D}. A plethora of information also exists outside of the BAO region. As a project for the more distant future, we would like to find all bias parameters to completely model the broadband 1-loop Lyman-$\alpha$ forest power spectrum in the presence of streaming velocity. This would involve using the galaxy redshift space power spectrum model presented in \cite{Desjacques2018jcap} with added streaming velocity contributions and fitting it to simulations. This work should be regarded as our first step toward this endeavor. Having this broadband model and the bispectrum complete with all bias coefficients will allow researchers to better extract cosmological results (e.g. $\Omega_{m}$, $\sigma_{8}$, and $\Sigma_{\nu}m_{\nu}$) from DESI data.

We have ignored astrophysical effects such as patchy reionization \cite{Montero-Camacho2019} and fluctuations in the UV background which can impact the distribution of {H\,\sc{i}} on scales of $\mathcal{O}(10-100)$ Mpc \cite{2014PhRvD..89h3010P, 2014MNRAS.442..187G}. These effects are not inherently coupled to BAO physics, but their impact on the large-scale 3D power spectrum of the Lyman-$\alpha$ forest is significant and may affect broadband fitting. Another source of potential bias arises from the fact that our biasing model applies to total matter, making no distinction between baryonic and dark matter perturbations; this may be important since the BAO feature is somewhat different in total versus baryonic matter, even at low redshift (e.g., \cite{2007ApJ...664..660E}). There are additional effects included in the streaming velocity analysis of \cite{Schmidt2016} which we ignored for reasons outlined in Section \ref{sec:lyabias} but are nonetheless potentially significant sources of systematics.

Here we have focused our attention on using Lyman-$\alpha$ absorption as a cosmological tracer, however {H\,\sc{i}} intensity mapping offers a complementary probe of the Universe's matter distribution over a similar redshift range. Our formalism applies equally well to the either case, but there are important physical differences between them which will give rise to differences in bias parameters. For instance, {H\,\sc{i}} used in intensity mapping studies lies primarily in damped Lyman-$\alpha$ systems---dense gas clouds found inside of galaxies \cite{2010ARA&A..48..127M}. The 21 cm emission may therefore be affected by galaxy intrinsic alignments, which would add a source of anisotropy beyond standard RSD effects to the observed signal. In that case, the relationship between intensity mapping biases $c_i$ and galaxy biases $b_i$ given in Table \ref{galaxymap} will differ. On the other hand, galaxies zero out transmitted flux independent of their orientations. Thus the Lyman-$\alpha$ forest bias coefficients contain no information about galaxy intrinsic alignments. As another example, optical depth effects are more important in the Lyman-$\alpha$ forest because the Lyman-$\alpha$ transition has a larger cross section than the 21 cm transition. Perhaps the most apparent difference in bias parameters between 21 cm intensity mapping and the Lyman-$\alpha$ forest lies in their respective signs. Since galaxies are positively biased tracers of the matter field, so is the integrated 21 cm emission signal. This is in contrast to the Lyman-$\alpha$ forest flux fluctuation which is a negatively biased tracer of the matter field.

%Currently at its pre-survey stage, the DESI instrument has demonstrated an impressive ability to obtain highly resolved flux spectra of its targets. This success has given the community confidence that the instrument can achieve goals outlined in
The DESI instrument is currently being commissioned and is planned to begin its survey this year. The statistical power of DESI, both for BAOs and broadband applications of the Lyman-$\alpha$ forest, is impressive \cite{DESI}, but that hinges on us having the correct tools to analyze, and framework in which to interpret, the data. Here we have offered a tool for Lyman-$\alpha$ surveys like DESI and the 4MOST Cosmology Redshift Survey (4MOST CRS) \cite{2019Msngr.175...50R}, and one which can be readily adapted to 21 cm surveys such as the Canadian Hydrogen Intensity Mapping Experiment (CHIME) \cite{2014SPIE.9145E..22B}. 

\section*{Acknowledgements}
We thank Joe McEwen and Jonathan Blazek for providing the scripts used to calculate the BAO peak shift. We thank Paulo Montero-Camacho, Andreu Font-Ribera, Jonathan Blazek, and Fabian Schmidt for useful comments on the draft. JG thanks Emmanuele Castorina for fruitful discussions during his visit to Ohio State. JG also thanks Florian Beutler for his assistance in replicating results from \cite{Beutler} and Xiao Fang for general help with FAST-PT.

We acknowledge support from Simons Foundation award 60052667; US Department of Energy award DE-SC0019083; and NASA award 15-WFIRST15-0008.

\appendix
\section{\label{sec: florian}Comparison of galaxy power spectra terms}

Results we provide in Section \ref{sec:galaxybias}---while in some cases appearing different at first glance---are identical to those in \cite{Beutler}. As an example case, we compare Eq. (\ref{comparison_ex}) presented above to the analogous equations in \cite{Beutler}, Eqs. (A24)-(A26)\footnote{There is a minor discrepancy between the two works---to wit, there is an additional factor of $k^{2}$ multiplying Eqs. (A25) and (A26). We believe this to be a typo and will disregard the factor in our comparison.}. 

\begin{figure}[t]
    \centering
    \includegraphics[width=0.7\textwidth]{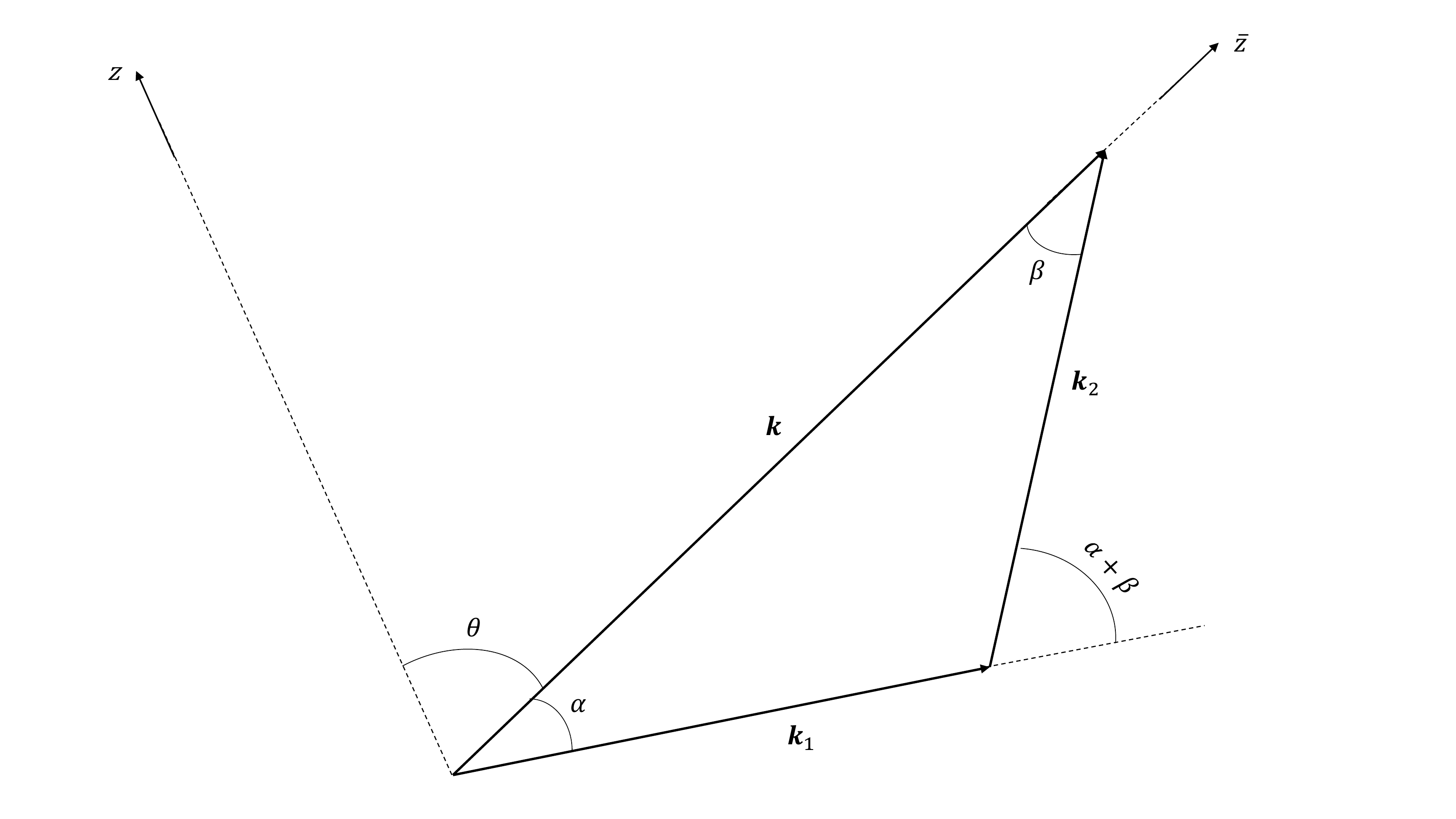}
    \caption{Diagram showing the relationship between different wave vectors. The angles shown are related to $\mu's$ in the following way: $\mu=\cos \theta$, $\mu_{1}=\cos \alpha$, $\mu_{2}=\cos \beta$, and $\mu_{12}= \cos \,(\alpha+\beta)$.}
    \label{pt_triangle}
\end{figure}

Recall that the full power spectrum  Eq. (\ref{Pg}) contains an overall factor of $f^{2}\mu^{2}$ in front of $I_{1}$ and a factor of $f^{2}\mu^{4}$ in front of $I_{2}$. This fact will be necessary for making the final comparison. Focus now on the kernel of $I_{1}$. Start with Eq. (A25) which is the piece of $P_{v_{z}^{2}|v^{2}}$ with a prefactor of $\mu^{0}$. Written in notation consistent with Fig. \ref{pt_triangle} and our conventions, the integral is
\begin{equation}
 \begin{aligned}
     I_{1}(k) = &-\int\frac{d^{3}\mathbf{k}_{1}}{(2\pi)^{3}} \frac{k^{2}(1-\mu_{1})(k_{1}-k\mu_{1})}{k_{2}^{3}}T_{v}(k_{1})T_{v}(k_{2}) \\ &\times P_{\textrm{lin}}(k_{1})P_{\textrm{lin}}(k_{2}).
 \end{aligned}
\end{equation}
 Under symmetrization we have
\begin{equation}
 \begin{aligned}
     \frac{k^{2}(1-\mu_{1})(k_{1}-k\mu_{1})}{k_{2}^{3}} &\rightarrow \frac{k^{2}}{2k_{2}^{3}} \textrm{sin}^{2}\alpha(k_{1}-k\,\textrm{cos}\,\alpha) \\ &+ \frac{k^{2}}{2k_{1}^{3}} \textrm{sin}^{2}\beta(k_{2}-k\,\textrm{cos}\,\beta)
 \end{aligned}    
\end{equation}
which by the law of sines reduces to 
\begin{equation}
    \frac{1}{2}\textrm{sin}^{2}(\alpha+\beta)\left[\frac{k_{1}-k\,\textrm{cos}\,\alpha}{k_{2}}+\frac{k_{2}-k\,\textrm{cos}\,\beta}{k_{1}}\right].
\end{equation}
Applying the law of cosines gives the kernel
\begin{equation}\label{I1}
    -\frac{k^{2}}{k_{1}k_{2}}\textrm{cos}(\alpha+\beta)\,\textrm{sin}\,\alpha\,\textrm{sin}\,\beta.
\end{equation}

We follow a similar line of steps with Eq. (A26). The integral is
\begin{equation}
 \begin{aligned}
    I_{2}(k)= &-\int\frac{d^{3}\mathbf{k}_{1}}{(2\pi)^{3}}k^{2}\left[\frac{2k^{2}\mu_{1}^{2}-3kk_{1}\mu_{1}^{3}-kk_{1}\mu_{1}}{k_{1}k_{2}^{3}}\right. \\ &+ \left. \frac{3k_{1}^{2}\mu_{1}^{2}-k_{1}^{2}}{k_{1}k_{2}^{3}}\right] T_{v}(k_{1})T_{v}(k_{2}) P_{\textrm{lin}}(k_{1})P_{\textrm{lin}}(k_{2}).
 \end{aligned}
\end{equation}
Turning attention to the kernel of $I_{2}$ and repeatedly applying the law of sines gives
\begin{equation}\label{I2}
 \begin{aligned}
    &k^{2}\left[\frac{2\,\textrm{sin}^{2}(\alpha+\beta)\,\textrm{cos}^{2}\,\alpha}{k_{2}^{2}\,\textrm{sin}\,\alpha\, \textrm{sin}\, \beta} -\frac{3\,\textrm{sin}(\alpha+\beta)\,\textrm{cos}^{3}\,\alpha}{k_{2}^{2}\,\textrm{sin}\,\alpha} \right. \\ &- \left. \frac{\textrm{sin}(\alpha+\beta)\,\textrm{cos}\,\alpha - 3\,\textrm{cos}^{2}\,\alpha\,\textrm{sin}\,\beta + \textrm{sin}\,\beta}{k_{2}^{2}\,\textrm{sin}\,\alpha}\right] \\ &= k^{2}\left[\frac{2\,\textrm{sin}^{2}(\alpha+\beta)\,\textrm{cos}^{2}\,\alpha + \textrm{sin}^{2}\,\beta\,(3\,\textrm{cos}^{2}\,\alpha -1)}{k_{1}k_{2}\,\textrm{sin}^{2}\,\alpha} \right. \\ &- \left. \frac{\textrm{cos}\,\alpha\,\textrm{sin}\,\beta\,\textrm{sin}(\alpha+\beta)\,(3\,\textrm{cos}^{2}\,\alpha +1)}{k_{1}k_{2}\,\textrm{sin}^{2}\,\alpha} \right] \\ &= k^{2} \left[\frac{\textrm{cos}(\alpha+\beta)\,(\textrm{sin}\,\alpha\,\textrm{sin}\,\beta+ 2\,\textrm{cos}\,\alpha\,\textrm{cos}\,\beta)}{k_{1}k_{2}}\right].
 \end{aligned}
\end{equation}

Now we focus on the kernel of Eq. (\ref{comparison_ex}). Complete with the factor of 2 from Eq. (\ref{Pg}), the kernel is
\begin{widetext}
\begin{equation}
 \begin{aligned}
    &2k^{2}\left[\frac{\textrm{cos}^{2}(\alpha+\beta)}{3k_{1}k_{2}} + \left(\frac{3}{2}\mu^{2}-\frac{1}{2}\right)\frac{\textrm{cos}(\alpha+\beta)}{k_{1}k_{2}}\left(\textrm{cos}\,\alpha\,\textrm{cos}\,\beta-\frac{1}{3}\textrm{cos}(\alpha+\beta)\right)\right] \\ &= 2k^{2}\frac{\textrm{cos}(\alpha+\beta)}{4k_{1}k_{2}}\left[3\mu^{2}\,\textrm{cos}(\alpha-\beta)+\mu^{2}\,\textrm{cos}(\alpha+\beta)-\textrm{cos}(\alpha-\beta)+\textrm{cos}(\alpha+\beta)\right] \\ &= \frac{k^{2}\mu^{2}}{k_{1}k_{2}}\textrm{cos}(\alpha+\beta)\,(\textrm{sin}\,\alpha\,\textrm{sin}\,\beta+ 2\,\textrm{cos}\,\alpha\,\textrm{cos}\,\beta) -\frac{k^{2}}{k_{1}k_{2}}\textrm{cos}(\alpha+\beta)\,\textrm{sin}\,\alpha\,\textrm{sin}\,\beta,
 \end{aligned}
\end{equation}
\end{widetext}
which agrees with results from Eqs. (\ref{I1}) and (\ref{I2}) after all prefactors are included.

\section{\label{sec:group theory}Proof of universality of Equation \ref{deltaF}}

\begin{figure}
    \centering
    \includegraphics[width=0.8\textwidth]{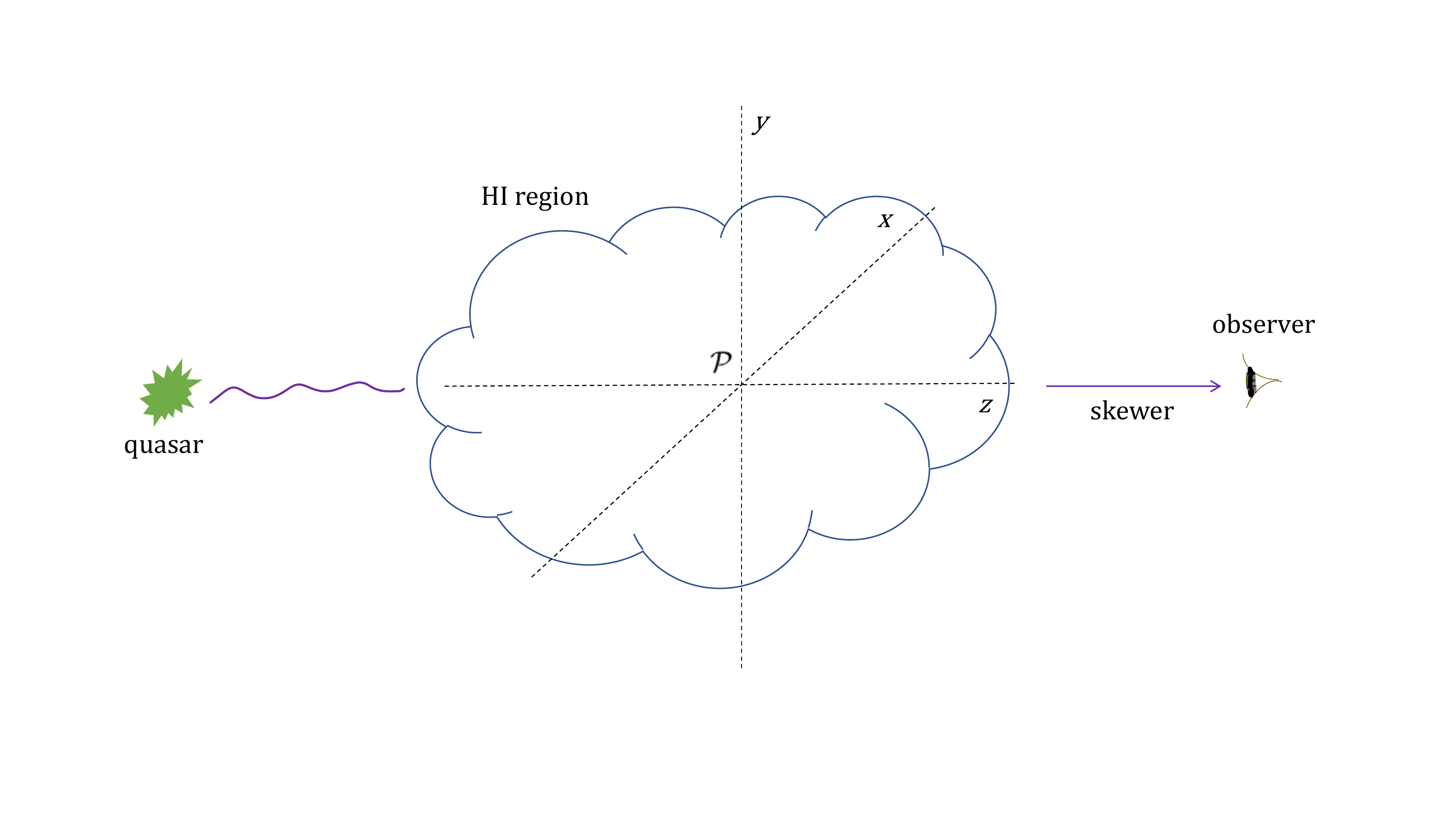}
    \caption{Cartoon showing the relative spatial orientation between a quasar, a region of neutral hydrogen, and the observer. The origin at point $\mathcal{P}$ will remain fixed for all transformations.}
    \label{HIcloud}
\end{figure}

\begin{figure}
    \centering
    \includegraphics[width=0.8\textwidth]{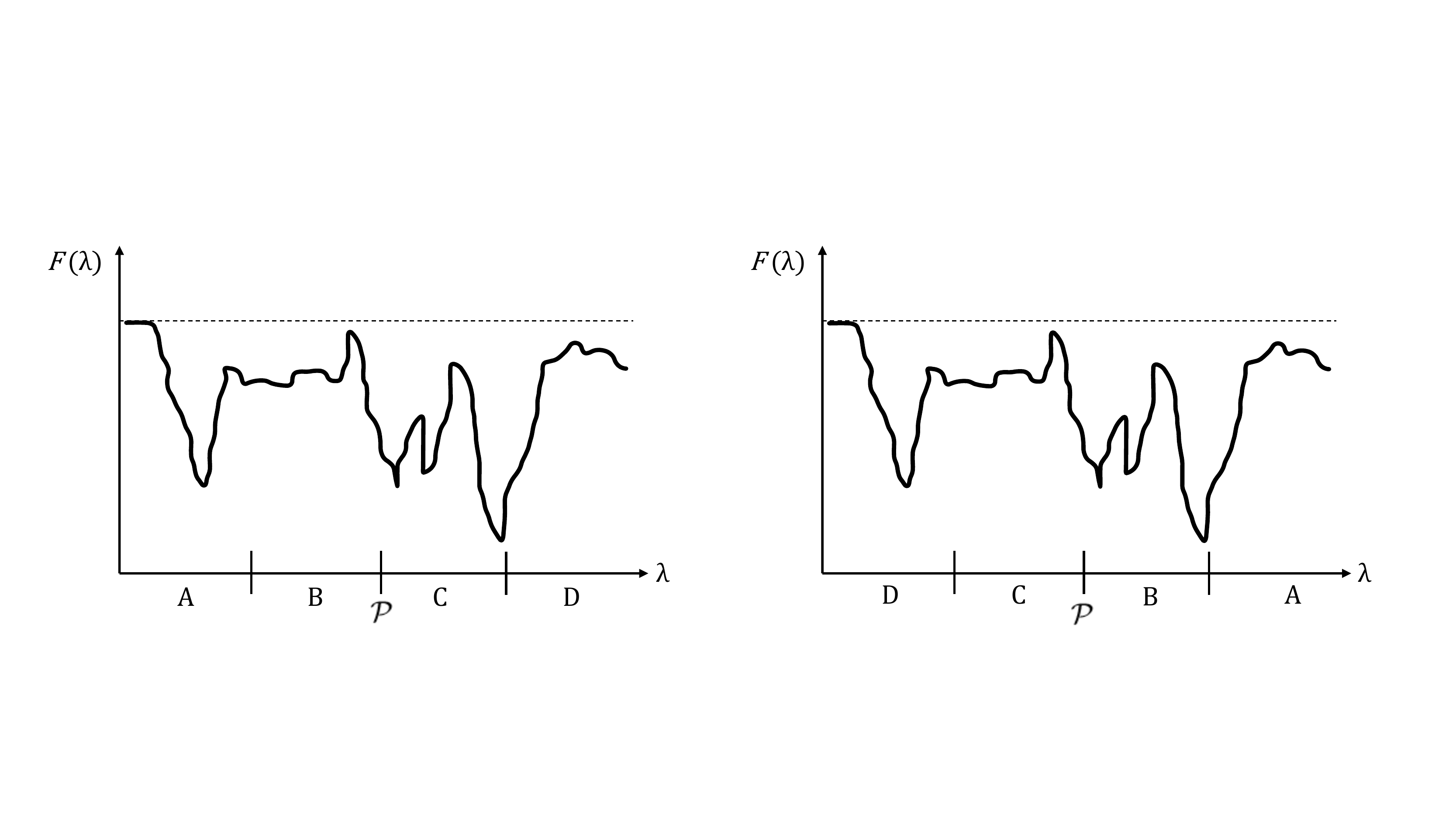}
    \caption{Lyman-$\alpha$ transmitted flux as a function of wavelength. Labels A---D indicate different spatial positions along the $z$-axis. The left plot describes an {H\,\sc{i}} region in its default orientation while the right plot shows the same region after a given transformation occurs which sends $z \rightarrow -z$ but leaves the origin $\mathcal{P}$ fixed.}
    \label{sample_flux}
\end{figure}

A perturbative expansion for $\delta_{F}$ is necessary to calculate the streaming velocity-induced BAO scale shift. Previous literature such as \cite{Arinyo2015} provides an analytic model applicable at linear order in $\delta$ but which relies on simulation and model-fitting to obtain a functional form of the nonlinear power spectrum. Streaming velocity terms (i.e. those with $b_{v}$, $b_{1v}$, or $b_{sv}$) contribute at second- and third-order in $\delta$; since we want correlations with streaming velocity terms that go up to fourth-order in $\delta$, we need the {\em non-streaming velocity} terms in $\delta_{F}$ up to second order in $\delta$.

We want to find the most general perturbative expansion possible for $\delta_{F}$ by constraining what terms it can have. To do this, we will write down exact and approximate symmetries of the Lyman-$\alpha$ system then use these to build a group theoretic framework. In the proceeding discussion we assume that we're looking at a snippet of the {H\,\sc{i}} region sufficiently small such that $\Delta \lambda/\lambda \ll 1$ (i.e. the entire snippet is at approximately the same redshift) but large enough that $\Delta \lambda/\lambda \gg \sqrt{k_{B}T/m_{H}}/c$ (i.e., large compared to the thermal broadening and Jeans length so that the snippet contains many Lyman-$\alpha$ clouds). Clouds in our snippet should not evolve significantly during the time it takes light to pass through them. 

We start with an {H\,\sc{i}} region oriented as in Figure \ref{HIcloud}. It has a transmitted flux spectrum shown in the left-hand plot of Figure \ref{sample_flux} (hereafter, ``\ref{sample_flux}L''). We are interested in the symmetry group under which $\delta_F$ at a particular point ${\cal P}$ remains fixed. The following are all possible symmetry operations:
\begin{enumerate}[(i)]
    \item \label{item:E} Do nothing. This is the identity operation $\hat{E}$.
    \item \label{item:sigmah} Reflect the snippet across the $xy$-plane. The resulting flux spectrum looks like the right-hand plot of Figure \ref{sample_flux} (hereafter, ``\ref{sample_flux}R''). This is the symmetry operation $\hat{\sigma}_{h}$.
    \item \label{item:sigmav} Reflect the snippet across any plane containing the $z$-axis. There are an infinite number of such planes each making a unique angle with the $x$-axis. This infinite set of symmetry operations forms the conjugacy class $\infty\hat{\sigma}_{v}$. The flux spectrum we obtain is that of Figure~\ref{sample_flux}L.
    \item Perform (\ref{item:sigmah}) and (\ref{item:sigmav}) in either order. The combination of these actions makes it clear that the snippet has a center of symmetry. This corresponds to the symmetry operation $\hat{i}$.
    \item \label{item:c2} Consider a parcel of gas in the snippet with position vector $\mathbf{r}=(r_{x},r_{y},r_{z})$. We can rotate this by an angle $\pi$ around the axis $(\cos\varphi,\sin\varphi,0)$, which is in the plane of the sky at position angle $\varphi$ (range: $0\le\varphi<\pi$). The matrix describing this rotation is
    \begin{equation}
    \mathbf{R}=
    \begin{bmatrix}
    \cos 2\varphi & \sin 2\varphi & 0 \\
    \sin 2\varphi & -\cos 2\varphi & 0 \\
    0 & 0 & -1
    \end{bmatrix}.
    \end{equation}
    \comment{which we want to rotate by some angle $\gamma$ about an arbitrary axis $\mathbf{u}=(u_{x},u_{y},u_{z})$ which passes through the origin and lies in the $xy$-plane. The matrix describing this rotation is 
    \begin{equation}
    \mathbf{R}(\gamma)=
    \begin{bmatrix}
    (1-\cos\gamma)\,u_{x}^{2}+\cos\gamma & (1-\cos\gamma)\,u_{x}u_{y} - \sin\gamma\,u_{z} & (1-\cos\gamma)\,u_{x}u_{z} + \sin\gamma\,u_{y}\\
    (1-\cos\gamma)\,u_{x}u_{y} + \sin\gamma\,u_{z} & (1-\cos\gamma)\,u_{y}^{2}+\cos\gamma & (1-\cos\gamma)\,u_{y}u_{z} - \sin\gamma\,u_{x}\\
    (1-\cos\gamma)\,u_{x}u_{z} - \sin\gamma\,u_{y} & (1-\cos\gamma)\,u_{y}u_{z} + \sin\gamma\,u_{x} & (1-\cos\gamma)\,u_{z}^{2}+\cos\gamma
    \end{bmatrix}.
    \end{equation}
    
    If we performed this rotation by an angle other than $0^{\circ}$ or $180^{\circ}$ then parcels initially on the $z$-axis may be moved off it and a parcels initially off the $z$-axis may be moved onto it. In those cases the resulting flux spectrum would be different from Figure \ref{sample_flux}, meaning we would have not performed a symmetry transformation. $\mathbf{R}(\gamma=0^{\circ})=\textrm{diag }(1,1,1)$ which is the case covered in (\ref{item:E}). Therefore we look at the case where $\gamma=180^{\circ}$. Note that we can only measure flux along our line of sight, so we restrict ourselves to parcels for which $v_{x}=v_{y}=0$. Since $\mathbf{u}$ is in the $xy$-plane, $u_{z}=0$.}
    The net result of all these restrictions is the transformation $\mathbf{R}(0,0,r_z)=(0,0,-r_z)$, i.e., we start with the transmitted flux in Figure \ref{sample_flux}L and end with Figure \ref{sample_flux}R. There are an infinite number of position angles $\varphi$ that can be chosen. We call this collection of symmetry operations the conjugacy class $\infty\hat{C}_{2}$.
    \item \label{item:2cinf} Rotate the snippet about the $z$-axis by an angle $\varphi$. We can rotate clockwise or counterclockwise by any infinite number of angles. These two symmetry operations are the conjugacy class $2\hat{C}^{\varphi}_{\infty}$.
    \item \label{item:2sinf} Perform (\ref{item:2cinf}) and (\ref{item:sigmah}) in either order (they commute). This is the conjugacy class $2\hat{S}^{\varphi}_{\infty}$.
\end{enumerate}
We have demonstrated that the Lyman-$\alpha$ system possesses the conjugacy classes of the $D_{\infty h}$ point group. The observable quantity in our analysis is the Lyman-$\alpha$ transmitted flux at ${\cal P}$, which is identical in either Figure~\ref{sample_flux}L or Figure~\ref{sample_flux}R. This is true for $F$ so it necessarily holds for $\delta_{F}$. Therefore $\delta_{F}$ must belong to the irreducible representation $\Sigma_{g}^{+}$.

These symmetries are general properties of \textit{any} of the commonly considered tracers in redshift space, whether they are continuous (e.g. Lyman-$\alpha$ forest flux and 21 cm flux) or discrete (e.g., galaxies and voids). Herein lies the true power of Eq. (\ref{deltaF}): given the proper choice of bias coefficients we can find any $P_{11}$- or $P_{22}$-type correlation between tracers, as well as some $P_{13}$-type correlations. We note that in principle it is possible to build a more complete expansion that covers all $P_{13}$-type correlations, but the number of correlations becomes intractable.  

How does one build the non-streaming velocity components of $\delta_{F}$? We begin by considering all unique fields in our problem, up to second order:
\begin{equation}\label{first-order}
    \textrm{First order}: \delta, s_{ij} 
\end{equation}
and
\begin{equation}\label{second-order}
    \textrm{Second order}: -\frac{1}{aH}\theta - \delta, t_{ij}. 
\end{equation}
The local velocity field $\mathbf{v}(\mathbf{r})$ and local gravitational potential $\phi(\mathbf{r})$ are not included because homogeneous changes in them---and in the gravitational force---should not be observable \cite{mcdonald}. Next, decompose each field into a direct sum of its irreducible representations. For the moment let's focus on the two first-order fields. 

Matter density $\rho$ is a scalar so $\delta = \rho/\overline{\rho}-1$ must also be a scalar; it decomposes into $\Sigma_{g}^{+}.$ The tidal tensor $s_{ij}$ is traceless and symmetric, meaning that only five of its nine components are independent. Which five components we choose is arbitrary---indeed one could rotate from our choice of components to any other---but only one choice gives a matrix representation of $s_{ij}$ in block-diagonal form (i.e. a matrix built from an orthogonal basis) which is the most useful format for finding its decomposition. We determine this basis with the aid of Table \ref{character_table}, the $D_{\infty h}$ character table. In its last column are six quadratic basis functions (read: the basis functions for a symmetric rank-2 tensor) which give the desired result. For instance, the appearance of $x^{2}+y^{2}$ in the first row implies that $s_{xx}+s_{yy}$ transforms under a representation of $\Sigma_{g}^{+}$. From Table \ref{character_table} it is clear that $s_{ij}$ decomposes into $\Sigma_{g}^{+} \oplus \Pi_{g} \oplus \Delta_{g}$. 

\begin{table}
\caption{$D_{\infty h}$ character table. Conjugacy classes are given across the top row and irreducible representations are given down the leftmost column. The last two columns list basis representations according to their transformation properties. Parentheses indicate that the entries separated by a comma must be considered together. This table is adapted from those appearing in \cite{Ferraro&Ziomek,dinfh}.}
\begin{ruledtabular}
\begin{tabular}{llllllllllll}
 $D_{\infty h}$ & $\hat{E}$ & $2\hat{C}_{\infty}^{\varphi}$ & \ldots & $\infty\hat{\sigma}_{v}$ & $\hat{i}$ & $\hat{\sigma}_{h}$ & $2\hat{S}_{\infty}^{\varphi}$ & \ldots & $\infty\hat{C}_{2}$ & linear functions \& rotations & quadratic functions\\
 $\Sigma^{+}_{g}$ & $+1$ & $+1$ & \ldots & $+1$ & $+1$ & $+1$ & $+1$ & \ldots & $+1$ & $-$ & $x^{2}+y^{2}, z^{2}$\\
 $\Sigma^{-}_{g}$ & $+1$ & $+1$ & \ldots & $-1$ & $+1$ & $+1$ & $+1$ & \ldots & $-1$ & $R_{z}$ & $-$\\
 $\Pi_{g}$ & $+2$ & $+2\cos\varphi$ & \ldots & $0$ & $+2$ & $-2$ & $-2\cos\varphi$ &\ldots & $0$ & $(R_{x}, R_{y})$ & $(xz, yz)$ \\
 $\Delta_{g}$ & $+2$ & $+2\cos 2\varphi$ & \ldots & $0$ & $+2$ & $+2$ & $+2\cos 2\varphi$ &\ldots & $0$ & $-$ & $(x^{2}-y^{2}, xy)$\\
 $\Phi_{g}$ & $+2$ & $+2\cos 3\varphi$ & \ldots & $0$ & $+2$ & $-2$ & $-2\cos 3\varphi$ &\ldots & $0$ & $-$ & $-$\\
 $\ldots$ & $\ldots$ & $\ldots$ & $\ldots$ & $\ldots$ & $\ldots$ & $\ldots$ & $\ldots$ & $\ldots$ & $\ldots$ & $-$ & $-$ \\
 $E_{ng}$ & $+2$ & $+2\cos n\varphi$ & \ldots & $0$ & $+2$ & $(-1)^{n}2$ & $(-1)^{n}2\cos n\varphi$ &\ldots & $0$ & $-$ & $-$\\
 $\ldots$ & $\ldots$ & $\ldots$ & $\ldots$ & $\ldots$ & $\ldots$ & $\ldots$ & $\ldots$ & $\ldots$ & $\ldots$ & $-$ & $-$ \\
 $\Sigma^{+}_{u}$ & $+1$ & $+1$ & \ldots & $+1$ & $-1$ & $-1$ & $-1$ & \ldots & $-1$ & $z$ & $-$\\
 $\Sigma^{-}_{u}$ & $+1$ & $+1$ & \ldots & $-1$ & $-1$ & $-1$ & $-1$ & \ldots & $+1$ & $-$ & $-$\\
 $\Pi_{u}$ & $+2$ & $+2\cos\varphi$ & \ldots & $0$ & $-2$ & $+2$ & $+2\cos\varphi$ &\ldots & $0$ & $(x, y)$ & $-$\\
 $\Delta_{u}$ & $+2$ & $+2\cos 2\varphi$ & \ldots & $0$ & $-2$ & $-2$ & $-2\cos 2\varphi$ &\ldots & $0$ & $-$ & $-$\\
 $\Phi_{u}$ & $+2$ & $+2\cos 3\varphi$ & \ldots & $0$ & $-2$ & $+2$ & $+2\cos 3\varphi$ &\ldots & $0$ & $-$ & $-$ \\
 $\ldots$ & $\ldots$ & $\ldots$ & $\ldots$ & $\ldots$ & $\ldots$ & $\ldots$ & $\ldots$ & $\ldots$ & $\ldots$ & $-$ & $-$ \\
 $E_{nu}$ & $+2$ & $+2\cos n\varphi$ & \ldots & $0$ & $-2$ & $(-1)^{n+1}2$ & $(-1)^{n+1}2\cos n\varphi$ &\ldots & $0$ & $-$ & $-$ \\
 $\ldots$ & $\ldots$ & $\ldots$ & $\ldots$ & $\ldots$ & $\ldots$ & $\ldots$ & $\ldots$ & $\ldots$ & $\ldots$ & $-$ & $-$ 
\end{tabular}
\end{ruledtabular}
\label{character_table}
\end{table}

Knowing the decomposition into irreducible representations, we now turn our attention to finding the linear combination of terms describing the first-order flux overdensity. 
\begin{equation}\label{eq:deltaF1}
    \delta_{F}^{(1)} = 
    \begin{bmatrix}
    a_{1}\,|\, a_{2}\,|\, a_{3} & a_{4}\,|\, a_{5} & a_{6}
    \end{bmatrix}
    \begin{bmatrix}
    \delta \\ \hline s_{zz} \\ \hline s_{xz} \\ s_{yz} \\ \hline s_{xx}-s_{yy} \\ s_{xy}
    \end{bmatrix},
\end{equation}
where bars are used both to separate components according to their irreducible representation and to separate the two fields. Symmetries of the Lyman-$\alpha$ system are such that it doesn't matter whether we first perform a transformation in (\ref{item:E})--(\ref{item:2sinf}) on the initial fields then act with a coefficient matrix on the resultant fields, or first act with the coefficient matrix on the initial fields then perform a transformation in (\ref{item:E})--(\ref{item:2sinf}) on the resultant fields; both cases result in $\delta_{F}^{(1)}$. We write the group theoretic version of this statement as
\begin{equation}\label{eq:schur}
    \Gamma_{\Sigma_{g}^{+}}(g) = f\Gamma_{a}(g)\mathbf{x} = \Gamma_{b}(g)f\mathbf{x},
\end{equation}
where $\Gamma_{i}(g)$ is the irreducible representation $i$ for all group elements $g$, $f$ is a linear mapping between the vector spaces of irreducible representations $a$ and $b$, and $\mathbf{x}$ is any block of the vertical vector in Eq. (\ref{eq:deltaF1}) we're considering. Suppose $a \neq b$. Schur's lemma applied to the second equality in Eq. (\ref{eq:schur}) tells us that either $f=0$ or $f$ is an equivalence between $a$ and $b$. However since $a$ and $b$ are irreducible, there cannot be an equivalence between them, so $f=0$. The first equality in Eq. (\ref{eq:schur}) implies no term is contributed to the expansion for $\delta_{F}^{(1)}$. Alternatively, suppose $a=b$. This implies $a=\Sigma_{g}^{+}$, the representation for which all transformations are given by the identity matrix. Therefore the map $f$ must go from one vector space to itself (i.e. $f: V \rightarrow V $). By Schur's lemma, $f$ must be equal to a constant times the identity matrix. Taken together, the preceding results mean that $a_{3}\textrm{\textemdash}a_{6}$ in Eq. (\ref{eq:deltaF1}) are all zero.

We now turn our attention to the second-order fields and write
\begin{equation}\label{eq:deltaF2}
    \delta_{F}^{(2)} = 
    \begin{bmatrix}
    a_{1}'\,|\, a_{2}'\,|\, a_{3}' & a_{4}'\,|\, a_{5}' & a_{6}'
    \end{bmatrix}
    \begin{bmatrix}
    -\theta/aH-\delta \\ \hline t_{zz} \\ \hline t_{xz} \\ t_{yz} \\ \hline t_{xx}-t_{yy} \\ t_{xy}
    \end{bmatrix}.
\end{equation}
All arguments previously given to find the linear combination of terms that build $\delta_{F}^{(1)}$ apply to building $\delta_{F}^{(2)}$ as well. Therefore $a_{3}'\textrm{\textemdash}a_{6}'$ in Eq. (\ref{eq:deltaF2}) are all zero.

Other second-order contributions are built from products of first-order elements. Take these products and decompose them into irreducible representations:
\begin{equation}\label{eq:d2decomp}
    \delta^{2}: \Sigma_{g}^{+} \otimes \Sigma_{g}^{+} = \Sigma_{g}^{+}\,,
\end{equation}
\begin{equation}\label{eq:dsdecomp}
    \delta s_{ij}: \Sigma_{g}^{+} \otimes (\Sigma_{g}^{+} \oplus \Pi_{g} \oplus \Delta_{g}) = \Sigma_{g}^{+} \oplus \Pi_{g} \oplus \Delta_{g}\,,
\end{equation}
and
\begin{eqnarray}\label{eq:s2decomp}
    s_{ij}s_{kl}: (\Sigma_{g}^{+} \oplus \Pi_{g} \oplus \Delta_{g}) \otimes (\Sigma_{g}^{+} \oplus \Pi_{g} \oplus \Delta_{g}) &=& 3\Sigma_{g}^{+} \oplus 2[\Sigma_{g}^{-}] \oplus 2\Pi_{g} \oplus [2\Pi_g] \oplus 2\Delta_{g} \oplus [\Delta_g] \oplus \Phi_{g}\oplus [\Phi_{g}] \oplus E_{4g}
    \nonumber \\
    &\rightarrow &
    3\Sigma_{g}^{+} \oplus 2\Pi_{g} \oplus 2\Delta_{g}  \oplus \Phi_{g} \oplus E_{4g}
    \,.
\end{eqnarray}
The direct sums and products above are calculated using the characters in Table~\ref{character_table}. (Some representations appear in brackets in Eq.~\ref{eq:s2decomp} because they arise from the antisymmetric part of the direct product; since $s_{ij}$ and $s_{kl}$ are the {\em same} tidal field, we will {\em not} have these terms. Alternatively the symmetrical part of the product of the dimension 5 representation $\Sigma_{g}^{+} \oplus \Pi_{g} \oplus \Delta_{g}$ with itself has dimension $5\times 6/2 = 15$ instead of $5^2=25$.) As before, Schur's lemma tells us that only the pieces of terms in $\delta^{2}$, $\delta s_{ij}$, and $s_{ij}s_{kl}$ which transform as $\Sigma_{g}^{+}$ may have a nonzero coefficient. Two of these are obvious: $\delta^{2}$ and $\delta s_{zz}$. Eq. (\ref{eq:s2decomp}) shows that there are three independent terms in the expansion of $s^{2}$ which may have nonzero coefficients. Note that $s^{2}=s_{ij}s_{ij}$ is a rank-4 tensor. Just as quadratic basis functions in Table \ref{character_table} describe various components of the rank-2 tensor $s_{ij}$, products of these same functions should describe pieces of $s^{2}$. However not every possible product will contribute. Products of irreducible representations which can be decomposed into direct sums involving $\Sigma_{g}^{+}$ are $\Sigma_{g}^{+} \otimes \Sigma_{g}^{+}$, $\Pi_{g} \otimes \Pi_{g}$, and $\Delta_{g} \otimes \Delta_{g}$. These correspond to basis functions of the form $s_{zz}^{2}$, $s_{xz}^{2}+s_{yz}^{2}$, and $(s_{xx}-s_{yy})^{2}+s_{xy}^{2}$, respectively\footnote{For simplicity we've simply added the squares of the elements for each two-tuple in the last column of Table \ref{character_table}, but the correct linear combination to build quartic basis functions is more elaborate. Do not get perturbed when, for example, we give different coefficients to $(s_{xx}-s_{yy})^{2}$ and $s_{xy}^{2}$ in Eq. (\ref{eq:s2expand}).}. The full expansion for $s^{2}$ written in a format consistent with a quartic basis is 
\begin{equation}\label{eq:s2expand}
\begin{aligned}
    s^{2} &= s_{xx}^{2}+s_{yy}^{2}+s_{zz}^{2}+2(s_{xy}^{2}+s_{xz}^{2}+s_{yz}^{2})\\
    &= s_{zz}^{2} +2(s_{xz}^{2}+s_{yz}^{2})+\frac{1}{2}(s_{xx}-s_{yy})^{2}+\frac{1}{2}(s_{xx}+s_{yy})^{2}+2s_{xy}^{2} \\
    &= \frac{3}{2}s_{zz}^{2} + 2(s_{xz}^{2}+s_{yz}^{2}) + \frac{1}{2}(s_{xx}-s_{yy})^{2} + 2s_{xy}^{2}.
\end{aligned}    
\end{equation}
Without loss of generality we choose $a_{1}''s^{2}$, $\frac{3}{2}a_{2}''s_{zz}^{2}$, and $2a_{3}''(s_{xz}^{2}+s_{yz}^{2})$ as the three independent terms (with coefficients) in Eq. (\ref{eq:s2expand}).

We make note of Eq. (\ref{eq:mcdonald}) which allows us to absorb the $-\theta/aH-\delta$ contribution from Eq. (\ref{eq:deltaF2}) into $\delta^{2}$ and $s^{2}$ contributions. Taking the results of this appendix and redefining constants, we obtain
\begin{equation}
    \delta_{F,\textrm{no bv}}(\mathbf{s}) = c_{1}\delta(\mathbf{r})+c_{2}s_{zz}(\mathbf{r})+c_{3}\delta^{2}(\mathbf{r})+c_{4}s^{2}(\mathbf{r}) +c_{5}\delta(\mathbf{r})s_{zz}(\mathbf{r})+c_{6}t_{zz}(\mathbf{r})+c_{7}s_{zz}^{2}(\mathbf{r})+c_{8}[s_{xz}^{2}(\mathbf{r})+s_{yz}^{2}(\mathbf{r})].
\end{equation}

\bibliographystyle{apsrev4-2.bst}
\bibliography{references.bib}

\providecommand{\noopsort}[1]{}\providecommand{\singleletter}[1]{#1}%
\begin{thebibliography}{72}%
\makeatletter
\providecommand \@ifxundefined [1]{%
 \@ifx{#1\undefined}
}%
\providecommand \@ifnum [1]{%
 \ifnum #1\expandafter \@firstoftwo
 \else \expandafter \@secondoftwo
 \fi
}%
\providecommand \@ifx [1]{%
 \ifx #1\expandafter \@firstoftwo
 \else \expandafter \@secondoftwo
 \fi
}%
\providecommand \natexlab [1]{#1}%
\providecommand \enquote  [1]{``#1''}%
\providecommand \bibnamefont  [1]{#1}%
\providecommand \bibfnamefont [1]{#1}%
\providecommand \citenamefont [1]{#1}%
\providecommand \href@noop [0]{\@secondoftwo}%
\providecommand \href [0]{\begingroup \@sanitize@url \@href}%
\providecommand \@href[1]{\@@startlink{#1}\@@href}%
\providecommand \@@href[1]{\endgroup#1\@@endlink}%
\providecommand \@sanitize@url [0]{\catcode `\\12\catcode `\$12\catcode
  `\&12\catcode `\#12\catcode `\^12\catcode `\_12\catcode `\%12\relax}%
\providecommand \@@startlink[1]{}%
\providecommand \@@endlink[0]{}%
\providecommand \url  [0]{\begingroup\@sanitize@url \@url }%
\providecommand \@url [1]{\endgroup\@href {#1}{\urlprefix }}%
\providecommand \urlprefix  [0]{URL }%
\providecommand \Eprint [0]{\href }%
\providecommand \doibase [0]{https://doi.org/}%
\providecommand \selectlanguage [0]{\@gobble}%
\providecommand \bibinfo  [0]{\@secondoftwo}%
\providecommand \bibfield  [0]{\@secondoftwo}%
\providecommand \translation [1]{[#1]}%
\providecommand \BibitemOpen [0]{}%
\providecommand \bibitemStop [0]{}%
\providecommand \bibitemNoStop [0]{.\EOS\space}%
\providecommand \EOS [0]{\spacefactor3000\relax}%
\providecommand \BibitemShut  [1]{\csname bibitem#1\endcsname}%
\let\auto@bib@innerbib\@empty
%</preamble>
\bibitem [{\citenamefont {{Planck Collaboration}}\ \emph
  {et~al.}(2018)\citenamefont {{Planck Collaboration}}, \citenamefont
  {{Aghanim}}, \citenamefont {{Akrami}}, \citenamefont {{Ashdown}},
  \citenamefont {{Aumont}}, \citenamefont {{Baccigalupi}}, \citenamefont
  {{Ballardini}} \emph {et~al.}}]{Planck18}%
  \BibitemOpen
  \bibfield  {author} {\bibinfo {author} {\bibnamefont {{Planck
  Collaboration}}}, \bibinfo {author} {\bibfnamefont {N.}~\bibnamefont
  {{Aghanim}}}, \bibinfo {author} {\bibfnamefont {Y.}~\bibnamefont {{Akrami}}},
  \bibinfo {author} {\bibfnamefont {M.}~\bibnamefont {{Ashdown}}}, \bibinfo
  {author} {\bibfnamefont {J.}~\bibnamefont {{Aumont}}}, \bibinfo {author}
  {\bibfnamefont {C.}~\bibnamefont {{Baccigalupi}}}, \bibinfo {author}
  {\bibfnamefont {M.}~\bibnamefont {{Ballardini}}}, \emph {et~al.},\
  }\href@noop {} {\bibfield  {journal} {\bibinfo  {journal} {ArXiv e-prints}\
  ,\ \bibinfo {eid} {arXiv:1807.06209}} (\bibinfo {year} {2018})},\ \Eprint
  {https://arxiv.org/abs/1807.06209} {arXiv:1807.06209 [astro-ph.CO]}
  \BibitemShut {NoStop}%
\bibitem [{\citenamefont {Riess}\ \emph {et~al.}(1998)\citenamefont {Riess},
  \citenamefont {Filippenko}, \citenamefont {Challis}, \citenamefont
  {Clocchiatti}, \citenamefont {Diercks}, \citenamefont {Garnavich},
  \citenamefont {Gilliland} \emph {et~al.}}]{Riess98}%
  \BibitemOpen
  \bibfield  {author} {\bibinfo {author} {\bibfnamefont {A.~G.}\ \bibnamefont
  {Riess}}, \bibinfo {author} {\bibfnamefont {A.~V.}\ \bibnamefont
  {Filippenko}}, \bibinfo {author} {\bibfnamefont {P.}~\bibnamefont {Challis}},
  \bibinfo {author} {\bibfnamefont {A.}~\bibnamefont {Clocchiatti}}, \bibinfo
  {author} {\bibfnamefont {A.}~\bibnamefont {Diercks}}, \bibinfo {author}
  {\bibfnamefont {P.~M.}\ \bibnamefont {Garnavich}}, \bibinfo {author}
  {\bibfnamefont {R.~L.}\ \bibnamefont {Gilliland}}, \emph {et~al.},\
  }\href@noop {} {\bibfield  {journal} {\bibinfo  {journal} {Astron. J.}\
  }\textbf {\bibinfo {volume} {116}},\ \bibinfo {pages} {1009} (\bibinfo {year}
  {1998})},\ \Eprint {https://arxiv.org/abs/astro-ph/9805201}
  {astro-ph/9805201} \BibitemShut {NoStop}%
\bibitem [{\citenamefont {Perlmutter}\ \emph {et~al.}(1999)\citenamefont
  {Perlmutter}, \citenamefont {Aldering}, \citenamefont {Goldhaber},
  \citenamefont {Knop}, \citenamefont {Nugent}, \citenamefont {Castro},
  \citenamefont {Deustua} \emph {et~al.}}]{Perlmutter99}%
  \BibitemOpen
  \bibfield  {author} {\bibinfo {author} {\bibfnamefont {S.}~\bibnamefont
  {Perlmutter}}, \bibinfo {author} {\bibfnamefont {G.}~\bibnamefont
  {Aldering}}, \bibinfo {author} {\bibfnamefont {G.}~\bibnamefont {Goldhaber}},
  \bibinfo {author} {\bibfnamefont {R.~A.}\ \bibnamefont {Knop}}, \bibinfo
  {author} {\bibfnamefont {P.}~\bibnamefont {Nugent}}, \bibinfo {author}
  {\bibfnamefont {P.~G.}\ \bibnamefont {Castro}}, \bibinfo {author}
  {\bibfnamefont {S.}~\bibnamefont {Deustua}}, \emph {et~al.},\ }\href@noop {}
  {\bibfield  {journal} {\bibinfo  {journal} {Astrophys. J.}\ }\textbf
  {\bibinfo {volume} {517}},\ \bibinfo {pages} {565} (\bibinfo {year}
  {1999})},\ \Eprint {https://arxiv.org/abs/astro-ph/9812133}
  {astro-ph/9812133} \BibitemShut {NoStop}%
\bibitem [{\citenamefont {Weinberg}\ \emph {et~al.}(2013)\citenamefont
  {Weinberg}, \citenamefont {Mortonson}, \citenamefont {Eisenstein},
  \citenamefont {Hirata}, \citenamefont {Riess},\ and\ \citenamefont
  {Rozo}}]{obsprobes}%
  \BibitemOpen
  \bibfield  {author} {\bibinfo {author} {\bibfnamefont {D.~H.}\ \bibnamefont
  {Weinberg}}, \bibinfo {author} {\bibfnamefont {M.~J.}\ \bibnamefont
  {Mortonson}}, \bibinfo {author} {\bibfnamefont {D.~J.}\ \bibnamefont
  {Eisenstein}}, \bibinfo {author} {\bibfnamefont {C.}~\bibnamefont {Hirata}},
  \bibinfo {author} {\bibfnamefont {A.~G.}\ \bibnamefont {Riess}},\ and\
  \bibinfo {author} {\bibfnamefont {E.}~\bibnamefont {Rozo}},\ }\href@noop {}
  {\bibfield  {journal} {\bibinfo  {journal} {Phys. Rep.}\ }\textbf {\bibinfo
  {volume} {530}},\ \bibinfo {pages} {87} (\bibinfo {year} {2013})},\ \Eprint
  {https://arxiv.org/abs/1201.2434} {arXiv:1201.2434} \BibitemShut {NoStop}%
\bibitem [{\citenamefont {Ryden}(2003)}]{Ryden}%
  \BibitemOpen
  \bibfield  {author} {\bibinfo {author} {\bibfnamefont {B.~S.}\ \bibnamefont
  {Ryden}},\ }\href@noop {} {\emph {\bibinfo {title} {Introduction to
  Cosmology}}}\ (\bibinfo  {publisher} {Addison Wesley},\ \bibinfo {year}
  {2003})\BibitemShut {NoStop}%
\bibitem [{\citenamefont {{Percival}}\ \emph {et~al.}(2001)\citenamefont
  {{Percival}}, \citenamefont {{Baugh}}, \citenamefont {{Bland-Hawthorn}},
  \citenamefont {{Bridges}}, \citenamefont {{Cannon}}, \citenamefont {{Cole}},
  \citenamefont {{Colless}}, \citenamefont {{Collins}}, \citenamefont
  {{Couch}},\ and\ \citenamefont {{Dalton}}}]{Percival2001}%
  \BibitemOpen
  \bibfield  {author} {\bibinfo {author} {\bibfnamefont {W.~J.}\ \bibnamefont
  {{Percival}}}, \bibinfo {author} {\bibfnamefont {C.~M.}\ \bibnamefont
  {{Baugh}}}, \bibinfo {author} {\bibfnamefont {J.}~\bibnamefont
  {{Bland-Hawthorn}}}, \bibinfo {author} {\bibfnamefont {T.}~\bibnamefont
  {{Bridges}}}, \bibinfo {author} {\bibfnamefont {R.}~\bibnamefont {{Cannon}}},
  \bibinfo {author} {\bibfnamefont {S.}~\bibnamefont {{Cole}}}, \bibinfo
  {author} {\bibfnamefont {M.}~\bibnamefont {{Colless}}}, \bibinfo {author}
  {\bibfnamefont {C.}~\bibnamefont {{Collins}}}, \bibinfo {author}
  {\bibfnamefont {W.}~\bibnamefont {{Couch}}},\ and\ \bibinfo {author}
  {\bibfnamefont {G.}~\bibnamefont {{Dalton}}},\ }\href
  {https://doi.org/10.1046/j.1365-8711.2001.04827.x} {\bibfield  {journal}
  {\bibinfo  {journal} {Mon. Not. R. Astron. Soc.}\ }\textbf {\bibinfo {volume}
  {327}},\ \bibinfo {pages} {1297} (\bibinfo {year} {2001})},\ \Eprint
  {https://arxiv.org/abs/astro-ph/0105252} {arXiv:astro-ph/0105252 [astro-ph]}
  \BibitemShut {NoStop}%
\bibitem [{\citenamefont {{Blake}}\ and\ \citenamefont
  {{Glazebrook}}(2003)}]{Blake&Glazebrook}%
  \BibitemOpen
  \bibfield  {author} {\bibinfo {author} {\bibfnamefont {C.}~\bibnamefont
  {{Blake}}}\ and\ \bibinfo {author} {\bibfnamefont {K.}~\bibnamefont
  {{Glazebrook}}},\ }\href {https://doi.org/10.1086/376983} {\bibfield
  {journal} {\bibinfo  {journal} {\apj}\ }\textbf {\bibinfo {volume} {594}},\
  \bibinfo {pages} {665} (\bibinfo {year} {2003})},\ \Eprint
  {https://arxiv.org/abs/astro-ph/0301632} {arXiv:astro-ph/0301632 [astro-ph]}
  \BibitemShut {NoStop}%
\bibitem [{\citenamefont {{Hu}}\ and\ \citenamefont
  {{Haiman}}(2003)}]{Hu&Haiman}%
  \BibitemOpen
  \bibfield  {author} {\bibinfo {author} {\bibfnamefont {W.}~\bibnamefont
  {{Hu}}}\ and\ \bibinfo {author} {\bibfnamefont {Z.}~\bibnamefont
  {{Haiman}}},\ }\href {https://doi.org/10.1103/PhysRevD.68.063004} {\bibfield
  {journal} {\bibinfo  {journal} {\prd}\ }\textbf {\bibinfo {volume} {68}},\
  \bibinfo {eid} {063004} (\bibinfo {year} {2003})},\ \Eprint
  {https://arxiv.org/abs/astro-ph/0306053} {arXiv:astro-ph/0306053 [astro-ph]}
  \BibitemShut {NoStop}%
\bibitem [{\citenamefont {{Linder}}(2003)}]{Linder}%
  \BibitemOpen
  \bibfield  {author} {\bibinfo {author} {\bibfnamefont {E.~V.}\ \bibnamefont
  {{Linder}}},\ }\href {https://doi.org/10.1103/PhysRevD.68.083504} {\bibfield
  {journal} {\bibinfo  {journal} {\prd}\ }\textbf {\bibinfo {volume} {68}},\
  \bibinfo {eid} {083504} (\bibinfo {year} {2003})},\ \Eprint
  {https://arxiv.org/abs/astro-ph/0304001} {arXiv:astro-ph/0304001 [astro-ph]}
  \BibitemShut {NoStop}%
\bibitem [{\citenamefont {{Seo}}\ and\ \citenamefont
  {{Eisenstein}}(2003)}]{Seo&Eisenstein}%
  \BibitemOpen
  \bibfield  {author} {\bibinfo {author} {\bibfnamefont {H.-J.}\ \bibnamefont
  {{Seo}}}\ and\ \bibinfo {author} {\bibfnamefont {D.~J.}\ \bibnamefont
  {{Eisenstein}}},\ }\href {https://doi.org/10.1086/379122} {\bibfield
  {journal} {\bibinfo  {journal} {\apj}\ }\textbf {\bibinfo {volume} {598}},\
  \bibinfo {pages} {720} (\bibinfo {year} {2003})},\ \Eprint
  {https://arxiv.org/abs/astro-ph/0307460} {arXiv:astro-ph/0307460 [astro-ph]}
  \BibitemShut {NoStop}%
\bibitem [{\citenamefont {{Cole}}\ \emph {et~al.}(2005)\citenamefont {{Cole}},
  \citenamefont {{Percival}}, \citenamefont {{Peacock}}, \citenamefont
  {{Norberg}}, \citenamefont {{Baugh}}, \citenamefont {{Frenk}}, \citenamefont
  {{Baldry}}, \citenamefont {{Bland-Hawthorn}}, \citenamefont {{Bridges}},\
  and\ \citenamefont {{Cannon}}}]{Cole2005}%
  \BibitemOpen
  \bibfield  {author} {\bibinfo {author} {\bibfnamefont {S.}~\bibnamefont
  {{Cole}}}, \bibinfo {author} {\bibfnamefont {W.~J.}\ \bibnamefont
  {{Percival}}}, \bibinfo {author} {\bibfnamefont {J.~A.}\ \bibnamefont
  {{Peacock}}}, \bibinfo {author} {\bibfnamefont {P.}~\bibnamefont
  {{Norberg}}}, \bibinfo {author} {\bibfnamefont {C.~M.}\ \bibnamefont
  {{Baugh}}}, \bibinfo {author} {\bibfnamefont {C.~S.}\ \bibnamefont
  {{Frenk}}}, \bibinfo {author} {\bibfnamefont {I.}~\bibnamefont {{Baldry}}},
  \bibinfo {author} {\bibfnamefont {J.}~\bibnamefont {{Bland-Hawthorn}}},
  \bibinfo {author} {\bibfnamefont {T.}~\bibnamefont {{Bridges}}},\ and\
  \bibinfo {author} {\bibfnamefont {R.}~\bibnamefont {{Cannon}}},\ }\href
  {https://doi.org/10.1111/j.1365-2966.2005.09318.x} {\bibfield  {journal}
  {\bibinfo  {journal} {Mon. Not. R. Astron. Soc.}\ }\textbf {\bibinfo {volume}
  {362}},\ \bibinfo {pages} {505} (\bibinfo {year} {2005})},\ \Eprint
  {https://arxiv.org/abs/astro-ph/0501174} {arXiv:astro-ph/0501174 [astro-ph]}
  \BibitemShut {NoStop}%
\bibitem [{\citenamefont {{Eisenstein}}\ \emph {et~al.}(2005)\citenamefont
  {{Eisenstein}}, \citenamefont {{Zehavi}}, \citenamefont {{Hogg}},
  \citenamefont {{Scoccimarro}}, \citenamefont {{Blanton}}, \citenamefont
  {{Nichol}}, \citenamefont {{Scranton}}, \citenamefont {{Seo}}, \citenamefont
  {{Tegmark}},\ and\ \citenamefont {{Zheng}}}]{Eisenstein2005}%
  \BibitemOpen
  \bibfield  {author} {\bibinfo {author} {\bibfnamefont {D.~J.}\ \bibnamefont
  {{Eisenstein}}}, \bibinfo {author} {\bibfnamefont {I.}~\bibnamefont
  {{Zehavi}}}, \bibinfo {author} {\bibfnamefont {D.~W.}\ \bibnamefont
  {{Hogg}}}, \bibinfo {author} {\bibfnamefont {R.}~\bibnamefont
  {{Scoccimarro}}}, \bibinfo {author} {\bibfnamefont {M.~R.}\ \bibnamefont
  {{Blanton}}}, \bibinfo {author} {\bibfnamefont {R.~C.}\ \bibnamefont
  {{Nichol}}}, \bibinfo {author} {\bibfnamefont {R.}~\bibnamefont
  {{Scranton}}}, \bibinfo {author} {\bibfnamefont {H.-J.}\ \bibnamefont
  {{Seo}}}, \bibinfo {author} {\bibfnamefont {M.}~\bibnamefont {{Tegmark}}},\
  and\ \bibinfo {author} {\bibfnamefont {Z.}~\bibnamefont {{Zheng}}},\ }\href
  {https://doi.org/10.1086/466512} {\bibfield  {journal} {\bibinfo  {journal}
  {\apj}\ }\textbf {\bibinfo {volume} {633}},\ \bibinfo {pages} {560} (\bibinfo
  {year} {2005})},\ \Eprint {https://arxiv.org/abs/astro-ph/0501171}
  {arXiv:astro-ph/0501171 [astro-ph]} \BibitemShut {NoStop}%
\bibitem [{\citenamefont {{Beutler}}\ \emph {et~al.}(2011)\citenamefont
  {{Beutler}}, \citenamefont {{Blake}}, \citenamefont {{Colless}},
  \citenamefont {{Jones}}, \citenamefont {{Staveley-Smith}}, \citenamefont
  {{Campbell}}, \citenamefont {{Parker}}, \citenamefont {{Saunders}},\ and\
  \citenamefont {{Watson}}}]{Beutler2011}%
  \BibitemOpen
  \bibfield  {author} {\bibinfo {author} {\bibfnamefont {F.}~\bibnamefont
  {{Beutler}}}, \bibinfo {author} {\bibfnamefont {C.}~\bibnamefont {{Blake}}},
  \bibinfo {author} {\bibfnamefont {M.}~\bibnamefont {{Colless}}}, \bibinfo
  {author} {\bibfnamefont {D.~H.}\ \bibnamefont {{Jones}}}, \bibinfo {author}
  {\bibfnamefont {L.}~\bibnamefont {{Staveley-Smith}}}, \bibinfo {author}
  {\bibfnamefont {L.}~\bibnamefont {{Campbell}}}, \bibinfo {author}
  {\bibfnamefont {Q.}~\bibnamefont {{Parker}}}, \bibinfo {author}
  {\bibfnamefont {W.}~\bibnamefont {{Saunders}}},\ and\ \bibinfo {author}
  {\bibfnamefont {F.}~\bibnamefont {{Watson}}},\ }\href
  {https://doi.org/10.1111/j.1365-2966.2011.19250.x} {\bibfield  {journal}
  {\bibinfo  {journal} {Mon. Not. R. Astron. Soc.}\ }\textbf {\bibinfo {volume}
  {416}},\ \bibinfo {pages} {3017} (\bibinfo {year} {2011})},\ \Eprint
  {https://arxiv.org/abs/1106.3366} {arXiv:1106.3366 [astro-ph.CO]}
  \BibitemShut {NoStop}%
\bibitem [{\citenamefont {{Blake}}\ \emph {et~al.}(2011)\citenamefont
  {{Blake}}, \citenamefont {{Kazin}}, \citenamefont {{Beutler}}, \citenamefont
  {{Davis}}, \citenamefont {{Parkinson}}, \citenamefont {{Brough}},
  \citenamefont {{Colless}}, \citenamefont {{Contreras}}, \citenamefont
  {{Couch}},\ and\ \citenamefont {{Croom}}}]{Blake2011}%
  \BibitemOpen
  \bibfield  {author} {\bibinfo {author} {\bibfnamefont {C.}~\bibnamefont
  {{Blake}}}, \bibinfo {author} {\bibfnamefont {E.~A.}\ \bibnamefont
  {{Kazin}}}, \bibinfo {author} {\bibfnamefont {F.}~\bibnamefont {{Beutler}}},
  \bibinfo {author} {\bibfnamefont {T.~M.}\ \bibnamefont {{Davis}}}, \bibinfo
  {author} {\bibfnamefont {D.}~\bibnamefont {{Parkinson}}}, \bibinfo {author}
  {\bibfnamefont {S.}~\bibnamefont {{Brough}}}, \bibinfo {author}
  {\bibfnamefont {M.}~\bibnamefont {{Colless}}}, \bibinfo {author}
  {\bibfnamefont {C.}~\bibnamefont {{Contreras}}}, \bibinfo {author}
  {\bibfnamefont {W.}~\bibnamefont {{Couch}}},\ and\ \bibinfo {author}
  {\bibfnamefont {S.}~\bibnamefont {{Croom}}},\ }\href
  {https://doi.org/10.1111/j.1365-2966.2011.19592.x} {\bibfield  {journal}
  {\bibinfo  {journal} {Mon. Not. R. Astron. Soc.}\ }\textbf {\bibinfo {volume}
  {418}},\ \bibinfo {pages} {1707} (\bibinfo {year} {2011})},\ \Eprint
  {https://arxiv.org/abs/1108.2635} {arXiv:1108.2635 [astro-ph.CO]}
  \BibitemShut {NoStop}%
\bibitem [{\citenamefont {{Beutler}}\ \emph
  {et~al.}(2017{\natexlab{a}})\citenamefont {{Beutler}}, \citenamefont {{Seo}},
  \citenamefont {{Ross}}, \citenamefont {{McDonald}}, \citenamefont {{Saito}},
  \citenamefont {{Bolton}}, \citenamefont {{Brownstein}}, \citenamefont
  {{Chuang}}, \citenamefont {{Cuesta}},\ and\ \citenamefont
  {{Eisenstein}}}]{Beutler2017BOSS}%
  \BibitemOpen
  \bibfield  {author} {\bibinfo {author} {\bibfnamefont {F.}~\bibnamefont
  {{Beutler}}}, \bibinfo {author} {\bibfnamefont {H.-J.}\ \bibnamefont
  {{Seo}}}, \bibinfo {author} {\bibfnamefont {A.~J.}\ \bibnamefont {{Ross}}},
  \bibinfo {author} {\bibfnamefont {P.}~\bibnamefont {{McDonald}}}, \bibinfo
  {author} {\bibfnamefont {S.}~\bibnamefont {{Saito}}}, \bibinfo {author}
  {\bibfnamefont {A.~S.}\ \bibnamefont {{Bolton}}}, \bibinfo {author}
  {\bibfnamefont {J.~R.}\ \bibnamefont {{Brownstein}}}, \bibinfo {author}
  {\bibfnamefont {C.-H.}\ \bibnamefont {{Chuang}}}, \bibinfo {author}
  {\bibfnamefont {A.~J.}\ \bibnamefont {{Cuesta}}},\ and\ \bibinfo {author}
  {\bibfnamefont {D.~J.}\ \bibnamefont {{Eisenstein}}},\ }\href
  {https://doi.org/10.1093/mnras/stw2373} {\bibfield  {journal} {\bibinfo
  {journal} {Mon. Not. R. Astron. Soc.}\ }\textbf {\bibinfo {volume} {464}},\
  \bibinfo {pages} {3409} (\bibinfo {year} {2017}{\natexlab{a}})},\ \Eprint
  {https://arxiv.org/abs/1607.03149} {arXiv:1607.03149 [astro-ph.CO]}
  \BibitemShut {NoStop}%
\bibitem [{\citenamefont {{Ross}}\ \emph {et~al.}(2017)\citenamefont {{Ross}},
  \citenamefont {{Beutler}}, \citenamefont {{Chuang}}, \citenamefont
  {{Pellejero-Ibanez}}, \citenamefont {{Seo}}, \citenamefont
  {{Vargas-Maga{\~n}a}}, \citenamefont {{Cuesta}}, \citenamefont {{Percival}},
  \citenamefont {{Burden}},\ and\ \citenamefont {{S{\'a}nchez}}}]{Ross2017}%
  \BibitemOpen
  \bibfield  {author} {\bibinfo {author} {\bibfnamefont {A.~J.}\ \bibnamefont
  {{Ross}}}, \bibinfo {author} {\bibfnamefont {F.}~\bibnamefont {{Beutler}}},
  \bibinfo {author} {\bibfnamefont {C.-H.}\ \bibnamefont {{Chuang}}}, \bibinfo
  {author} {\bibfnamefont {M.}~\bibnamefont {{Pellejero-Ibanez}}}, \bibinfo
  {author} {\bibfnamefont {H.-J.}\ \bibnamefont {{Seo}}}, \bibinfo {author}
  {\bibfnamefont {M.}~\bibnamefont {{Vargas-Maga{\~n}a}}}, \bibinfo {author}
  {\bibfnamefont {A.~J.}\ \bibnamefont {{Cuesta}}}, \bibinfo {author}
  {\bibfnamefont {W.~J.}\ \bibnamefont {{Percival}}}, \bibinfo {author}
  {\bibfnamefont {A.}~\bibnamefont {{Burden}}},\ and\ \bibinfo {author}
  {\bibfnamefont {A.~G.}\ \bibnamefont {{S{\'a}nchez}}},\ }\href
  {https://doi.org/10.1093/mnras/stw2372} {\bibfield  {journal} {\bibinfo
  {journal} {Mon. Not. R. Astron. Soc.}\ }\textbf {\bibinfo {volume} {464}},\
  \bibinfo {pages} {1168} (\bibinfo {year} {2017})},\ \Eprint
  {https://arxiv.org/abs/1607.03145} {arXiv:1607.03145 [astro-ph.CO]}
  \BibitemShut {NoStop}%
\bibitem [{\citenamefont {{Alam}}\ \emph {et~al.}(2017)\citenamefont {{Alam}},
  \citenamefont {{Ata}}, \citenamefont {{Bailey}}, \citenamefont {{Beutler}},
  \citenamefont {{Bizyaev}}, \citenamefont {{Blazek}}, \citenamefont
  {{Bolton}}, \citenamefont {{Brownstein}}, \citenamefont {{Burden}},\ and\
  \citenamefont {{Chuang}}}]{Alam2017}%
  \BibitemOpen
  \bibfield  {author} {\bibinfo {author} {\bibfnamefont {S.}~\bibnamefont
  {{Alam}}}, \bibinfo {author} {\bibfnamefont {M.}~\bibnamefont {{Ata}}},
  \bibinfo {author} {\bibfnamefont {S.}~\bibnamefont {{Bailey}}}, \bibinfo
  {author} {\bibfnamefont {F.}~\bibnamefont {{Beutler}}}, \bibinfo {author}
  {\bibfnamefont {D.}~\bibnamefont {{Bizyaev}}}, \bibinfo {author}
  {\bibfnamefont {J.~A.}\ \bibnamefont {{Blazek}}}, \bibinfo {author}
  {\bibfnamefont {A.~S.}\ \bibnamefont {{Bolton}}}, \bibinfo {author}
  {\bibfnamefont {J.~R.}\ \bibnamefont {{Brownstein}}}, \bibinfo {author}
  {\bibfnamefont {A.}~\bibnamefont {{Burden}}},\ and\ \bibinfo {author}
  {\bibfnamefont {C.-H.}\ \bibnamefont {{Chuang}}},\ }\href
  {https://doi.org/10.1093/mnras/stx721} {\bibfield  {journal} {\bibinfo
  {journal} {Mon. Not. R. Astron. Soc.}\ }\textbf {\bibinfo {volume} {470}},\
  \bibinfo {pages} {2617} (\bibinfo {year} {2017})},\ \Eprint
  {https://arxiv.org/abs/1607.03155} {arXiv:1607.03155 [astro-ph.CO]}
  \BibitemShut {NoStop}%
\bibitem [{\citenamefont {{Chen}}\ \emph {et~al.}(2019)\citenamefont {{Chen}},
  \citenamefont {{Castorina}},\ and\ \citenamefont {{White}}}]{Chen2019}%
  \BibitemOpen
  \bibfield  {author} {\bibinfo {author} {\bibfnamefont {S.-F.}\ \bibnamefont
  {{Chen}}}, \bibinfo {author} {\bibfnamefont {E.}~\bibnamefont
  {{Castorina}}},\ and\ \bibinfo {author} {\bibfnamefont {M.}~\bibnamefont
  {{White}}},\ }\href {https://doi.org/10.1088/1475-7516/2019/06/006}
  {\bibfield  {journal} {\bibinfo  {journal} {J. Cosmol. Astropart. Phys.}\
  }\textbf {\bibinfo {volume} {2019}}\bibfield  {number} {\bibinfo  {number} {
  (6)},\ \bibinfo {eid} {006}},\ }\Eprint {https://arxiv.org/abs/1903.00437}
  {arXiv:1903.00437 [astro-ph.CO]} \BibitemShut {NoStop}%
\bibitem [{\citenamefont {{Bahcall}}\ and\ \citenamefont
  {{Salpeter}}(1965)}]{1965ApJ...142.1677B}%
  \BibitemOpen
  \bibfield  {author} {\bibinfo {author} {\bibfnamefont {J.~N.}\ \bibnamefont
  {{Bahcall}}}\ and\ \bibinfo {author} {\bibfnamefont {E.~E.}\ \bibnamefont
  {{Salpeter}}},\ }\href {https://doi.org/10.1086/148460} {\bibfield  {journal}
  {\bibinfo  {journal} {\apj}\ }\textbf {\bibinfo {volume} {142}},\ \bibinfo
  {pages} {1677} (\bibinfo {year} {1965})}\BibitemShut {NoStop}%
\bibitem [{\citenamefont {{Croft}}\ \emph {et~al.}(1998)\citenamefont
  {{Croft}}, \citenamefont {{Weinberg}}, \citenamefont {{Katz}},\ and\
  \citenamefont {{Hernquist}}}]{1998ApJ...495...44C}%
  \BibitemOpen
  \bibfield  {author} {\bibinfo {author} {\bibfnamefont {R.~A.~C.}\
  \bibnamefont {{Croft}}}, \bibinfo {author} {\bibfnamefont {D.~H.}\
  \bibnamefont {{Weinberg}}}, \bibinfo {author} {\bibfnamefont
  {N.}~\bibnamefont {{Katz}}},\ and\ \bibinfo {author} {\bibfnamefont
  {L.}~\bibnamefont {{Hernquist}}},\ }\href {https://doi.org/10.1086/305289}
  {\bibfield  {journal} {\bibinfo  {journal} {\apj}\ }\textbf {\bibinfo
  {volume} {495}},\ \bibinfo {pages} {44} (\bibinfo {year} {1998})},\ \Eprint
  {https://arxiv.org/abs/astro-ph/9708018} {arXiv:astro-ph/9708018 [astro-ph]}
  \BibitemShut {NoStop}%
\bibitem [{\citenamefont {{Croft}}\ \emph {et~al.}(1999)\citenamefont
  {{Croft}}, \citenamefont {{Weinberg}}, \citenamefont {{Pettini}},
  \citenamefont {{Hernquist}},\ and\ \citenamefont
  {{Katz}}}]{1999ApJ...520....1C}%
  \BibitemOpen
  \bibfield  {author} {\bibinfo {author} {\bibfnamefont {R.~A.~C.}\
  \bibnamefont {{Croft}}}, \bibinfo {author} {\bibfnamefont {D.~H.}\
  \bibnamefont {{Weinberg}}}, \bibinfo {author} {\bibfnamefont
  {M.}~\bibnamefont {{Pettini}}}, \bibinfo {author} {\bibfnamefont
  {L.}~\bibnamefont {{Hernquist}}},\ and\ \bibinfo {author} {\bibfnamefont
  {N.}~\bibnamefont {{Katz}}},\ }\href {https://doi.org/10.1086/307438}
  {\bibfield  {journal} {\bibinfo  {journal} {\apj}\ }\textbf {\bibinfo
  {volume} {520}},\ \bibinfo {pages} {1} (\bibinfo {year} {1999})},\ \Eprint
  {https://arxiv.org/abs/astro-ph/9809401} {arXiv:astro-ph/9809401 [astro-ph]}
  \BibitemShut {NoStop}%
\bibitem [{\citenamefont {{McDonald}}\ \emph {et~al.}(2000)\citenamefont
  {{McDonald}}, \citenamefont {{Miralda-Escud{\'e}}}, \citenamefont {{Rauch}},
  \citenamefont {{Sargent}}, \citenamefont {{Barlow}}, \citenamefont {{Cen}},\
  and\ \citenamefont {{Ostriker}}}]{2000ApJ...543....1M}%
  \BibitemOpen
  \bibfield  {author} {\bibinfo {author} {\bibfnamefont {P.}~\bibnamefont
  {{McDonald}}}, \bibinfo {author} {\bibfnamefont {J.}~\bibnamefont
  {{Miralda-Escud{\'e}}}}, \bibinfo {author} {\bibfnamefont {M.}~\bibnamefont
  {{Rauch}}}, \bibinfo {author} {\bibfnamefont {W.~L.~W.}\ \bibnamefont
  {{Sargent}}}, \bibinfo {author} {\bibfnamefont {T.~A.}\ \bibnamefont
  {{Barlow}}}, \bibinfo {author} {\bibfnamefont {R.}~\bibnamefont {{Cen}}},\
  and\ \bibinfo {author} {\bibfnamefont {J.~P.}\ \bibnamefont {{Ostriker}}},\
  }\href {https://doi.org/10.1086/317079} {\bibfield  {journal} {\bibinfo
  {journal} {\apj}\ }\textbf {\bibinfo {volume} {543}},\ \bibinfo {pages} {1}
  (\bibinfo {year} {2000})},\ \Eprint {https://arxiv.org/abs/astro-ph/9911196}
  {arXiv:astro-ph/9911196 [astro-ph]} \BibitemShut {NoStop}%
\bibitem [{\citenamefont {{Busca}}\ \emph {et~al.}(2013)\citenamefont
  {{Busca}}, \citenamefont {{Delubac}}, \citenamefont {{Rich}}, \citenamefont
  {{Bailey}}, \citenamefont {{Font-Ribera}}, \citenamefont {{Kirkby}},
  \citenamefont {{Le Goff}}, \citenamefont {{Pieri}}, \citenamefont
  {{Slosar}},\ and\ \citenamefont {{Aubourg}}}]{Busca2013}%
  \BibitemOpen
  \bibfield  {author} {\bibinfo {author} {\bibfnamefont {N.~G.}\ \bibnamefont
  {{Busca}}}, \bibinfo {author} {\bibfnamefont {T.}~\bibnamefont {{Delubac}}},
  \bibinfo {author} {\bibfnamefont {J.}~\bibnamefont {{Rich}}}, \bibinfo
  {author} {\bibfnamefont {S.}~\bibnamefont {{Bailey}}}, \bibinfo {author}
  {\bibfnamefont {A.}~\bibnamefont {{Font-Ribera}}}, \bibinfo {author}
  {\bibfnamefont {D.}~\bibnamefont {{Kirkby}}}, \bibinfo {author}
  {\bibfnamefont {J.~M.}\ \bibnamefont {{Le Goff}}}, \bibinfo {author}
  {\bibfnamefont {M.~M.}\ \bibnamefont {{Pieri}}}, \bibinfo {author}
  {\bibfnamefont {A.}~\bibnamefont {{Slosar}}},\ and\ \bibinfo {author}
  {\bibfnamefont {{\'E}.}~\bibnamefont {{Aubourg}}},\ }\href
  {https://doi.org/10.1051/0004-6361/201220724} {\bibfield  {journal} {\bibinfo
   {journal} {Astron. Astrophys.}\ }\textbf {\bibinfo {volume} {552}},\
  \bibinfo {eid} {A96} (\bibinfo {year} {2013})},\ \Eprint
  {https://arxiv.org/abs/1211.2616} {arXiv:1211.2616 [astro-ph.CO]}
  \BibitemShut {NoStop}%
\bibitem [{\citenamefont {{Slosar}}\ \emph {et~al.}(2013)\citenamefont
  {{Slosar}}, \citenamefont {{Ir{\v{s}}i{\v{c}}}}, \citenamefont {{Kirkby}},
  \citenamefont {{Bailey}}, \citenamefont {{Busca}}, \citenamefont {{Delubac}},
  \citenamefont {{Rich}}, \citenamefont {{Aubourg}}, \citenamefont
  {{Bautista}},\ and\ \citenamefont {{Bhardwaj}}}]{Slosar2013}%
  \BibitemOpen
  \bibfield  {author} {\bibinfo {author} {\bibfnamefont {A.}~\bibnamefont
  {{Slosar}}}, \bibinfo {author} {\bibfnamefont {V.}~\bibnamefont
  {{Ir{\v{s}}i{\v{c}}}}}, \bibinfo {author} {\bibfnamefont {D.}~\bibnamefont
  {{Kirkby}}}, \bibinfo {author} {\bibfnamefont {S.}~\bibnamefont {{Bailey}}},
  \bibinfo {author} {\bibfnamefont {N.~G.}\ \bibnamefont {{Busca}}}, \bibinfo
  {author} {\bibfnamefont {T.}~\bibnamefont {{Delubac}}}, \bibinfo {author}
  {\bibfnamefont {J.}~\bibnamefont {{Rich}}}, \bibinfo {author} {\bibfnamefont
  {{\'E}.}~\bibnamefont {{Aubourg}}}, \bibinfo {author} {\bibfnamefont {J.~E.}\
  \bibnamefont {{Bautista}}},\ and\ \bibinfo {author} {\bibfnamefont
  {V.}~\bibnamefont {{Bhardwaj}}},\ }\href
  {https://doi.org/10.1088/1475-7516/2013/04/026} {\bibfield  {journal}
  {\bibinfo  {journal} {J. Cosmol. Astropart. Phys.}\ }\textbf {\bibinfo
  {volume} {2013}}\bibfield  {number} {\bibinfo  {number} { (4)},\ \bibinfo
  {eid} {026}},\ }\Eprint {https://arxiv.org/abs/1301.3459} {arXiv:1301.3459
  [astro-ph.CO]} \BibitemShut {NoStop}%
\bibitem [{\citenamefont {{Font-Ribera}}\ \emph {et~al.}(2014)\citenamefont
  {{Font-Ribera}}, \citenamefont {{Kirkby}}, \citenamefont {{Busca}},
  \citenamefont {{Miralda-Escud{\'e}}}, \citenamefont {{Ross}}, \citenamefont
  {{Slosar}}, \citenamefont {{Rich}}, \citenamefont {{Aubourg}}, \citenamefont
  {{Bailey}},\ and\ \citenamefont {{Bhardwaj}}}]{Font-Ribera2014}%
  \BibitemOpen
  \bibfield  {author} {\bibinfo {author} {\bibfnamefont {A.}~\bibnamefont
  {{Font-Ribera}}}, \bibinfo {author} {\bibfnamefont {D.}~\bibnamefont
  {{Kirkby}}}, \bibinfo {author} {\bibfnamefont {N.}~\bibnamefont {{Busca}}},
  \bibinfo {author} {\bibfnamefont {J.}~\bibnamefont {{Miralda-Escud{\'e}}}},
  \bibinfo {author} {\bibfnamefont {N.~P.}\ \bibnamefont {{Ross}}}, \bibinfo
  {author} {\bibfnamefont {A.}~\bibnamefont {{Slosar}}}, \bibinfo {author}
  {\bibfnamefont {J.}~\bibnamefont {{Rich}}}, \bibinfo {author} {\bibfnamefont
  {{\'E}.}~\bibnamefont {{Aubourg}}}, \bibinfo {author} {\bibfnamefont
  {S.}~\bibnamefont {{Bailey}}},\ and\ \bibinfo {author} {\bibfnamefont
  {V.}~\bibnamefont {{Bhardwaj}}},\ }\href
  {https://doi.org/10.1088/1475-7516/2014/05/027} {\bibfield  {journal}
  {\bibinfo  {journal} {J. Cosmol. Astropart. Phys.}\ }\textbf {\bibinfo
  {volume} {2014}}\bibfield  {number} {\bibinfo  {number} { (5)},\ \bibinfo
  {eid} {027}},\ }\Eprint {https://arxiv.org/abs/1311.1767} {arXiv:1311.1767
  [astro-ph.CO]} \BibitemShut {NoStop}%
\bibitem [{\citenamefont {{Delubac}}\ \emph {et~al.}(2015)\citenamefont
  {{Delubac}}, \citenamefont {{Bautista}}, \citenamefont {{Busca}},
  \citenamefont {{Rich}}, \citenamefont {{Kirkby}}, \citenamefont {{Bailey}},
  \citenamefont {{Font-Ribera}}, \citenamefont {{Slosar}}, \citenamefont
  {{Lee}},\ and\ \citenamefont {{Pieri}}}]{Delubac2015}%
  \BibitemOpen
  \bibfield  {author} {\bibinfo {author} {\bibfnamefont {T.}~\bibnamefont
  {{Delubac}}}, \bibinfo {author} {\bibfnamefont {J.~E.}\ \bibnamefont
  {{Bautista}}}, \bibinfo {author} {\bibfnamefont {N.~G.}\ \bibnamefont
  {{Busca}}}, \bibinfo {author} {\bibfnamefont {J.}~\bibnamefont {{Rich}}},
  \bibinfo {author} {\bibfnamefont {D.}~\bibnamefont {{Kirkby}}}, \bibinfo
  {author} {\bibfnamefont {S.}~\bibnamefont {{Bailey}}}, \bibinfo {author}
  {\bibfnamefont {A.}~\bibnamefont {{Font-Ribera}}}, \bibinfo {author}
  {\bibfnamefont {A.}~\bibnamefont {{Slosar}}}, \bibinfo {author}
  {\bibfnamefont {K.-G.}\ \bibnamefont {{Lee}}},\ and\ \bibinfo {author}
  {\bibfnamefont {M.~M.}\ \bibnamefont {{Pieri}}},\ }\href
  {https://doi.org/10.1051/0004-6361/201423969} {\bibfield  {journal} {\bibinfo
   {journal} {Astron. Astrophys.}\ }\textbf {\bibinfo {volume} {574}},\
  \bibinfo {eid} {A59} (\bibinfo {year} {2015})},\ \Eprint
  {https://arxiv.org/abs/1404.1801} {arXiv:1404.1801 [astro-ph.CO]}
  \BibitemShut {NoStop}%
\bibitem [{\citenamefont {{Bautista}}\ \emph {et~al.}(2017)\citenamefont
  {{Bautista}}, \citenamefont {{Busca}}, \citenamefont {{Guy}}, \citenamefont
  {{Rich}}, \citenamefont {{Blomqvist}}, \citenamefont {{du Mas des Bourboux}},
  \citenamefont {{Pieri}}, \citenamefont {{Font-Ribera}}, \citenamefont
  {{Bailey}},\ and\ \citenamefont {{Delubac}}}]{Bautista2017}%
  \BibitemOpen
  \bibfield  {author} {\bibinfo {author} {\bibfnamefont {J.~E.}\ \bibnamefont
  {{Bautista}}}, \bibinfo {author} {\bibfnamefont {N.~G.}\ \bibnamefont
  {{Busca}}}, \bibinfo {author} {\bibfnamefont {J.}~\bibnamefont {{Guy}}},
  \bibinfo {author} {\bibfnamefont {J.}~\bibnamefont {{Rich}}}, \bibinfo
  {author} {\bibfnamefont {M.}~\bibnamefont {{Blomqvist}}}, \bibinfo {author}
  {\bibfnamefont {H.}~\bibnamefont {{du Mas des Bourboux}}}, \bibinfo {author}
  {\bibfnamefont {M.~M.}\ \bibnamefont {{Pieri}}}, \bibinfo {author}
  {\bibfnamefont {A.}~\bibnamefont {{Font-Ribera}}}, \bibinfo {author}
  {\bibfnamefont {S.}~\bibnamefont {{Bailey}}},\ and\ \bibinfo {author}
  {\bibfnamefont {T.}~\bibnamefont {{Delubac}}},\ }\href
  {https://doi.org/10.1051/0004-6361/201730533} {\bibfield  {journal} {\bibinfo
   {journal} {Astron. Astrophys.}\ }\textbf {\bibinfo {volume} {603}},\
  \bibinfo {eid} {A12} (\bibinfo {year} {2017})},\ \Eprint
  {https://arxiv.org/abs/1702.00176} {arXiv:1702.00176 [astro-ph.CO]}
  \BibitemShut {NoStop}%
\bibitem [{\citenamefont {{de Sainte Agathe}}\ \emph
  {et~al.}(2019)\citenamefont {{de Sainte Agathe}}, \citenamefont {{Balland}},
  \citenamefont {{du Mas des Bourboux}}, \citenamefont {{Busca}}, \citenamefont
  {{Blomqvist}}, \citenamefont {{Guy}}, \citenamefont {{Rich}}, \citenamefont
  {{Font-Ribera}}, \citenamefont {{Pieri}}, \citenamefont {{Bautista}},
  \citenamefont {{Dawson}}, \citenamefont {{Le Goff}}, \citenamefont {{de la
  Macorra}}, \citenamefont {{Palanque-Delabrouille}}, \citenamefont
  {{Percival}}, \citenamefont {{P{\'e}rez-R{\`a}fols}}, \citenamefont
  {{Schneider}}, \citenamefont {{Slosar}},\ and\ \citenamefont
  {{Y{\`e}che}}}]{deSainteAgathe}%
  \BibitemOpen
  \bibfield  {author} {\bibinfo {author} {\bibfnamefont {V.}~\bibnamefont {{de
  Sainte Agathe}}}, \bibinfo {author} {\bibfnamefont {C.}~\bibnamefont
  {{Balland}}}, \bibinfo {author} {\bibfnamefont {H.}~\bibnamefont {{du Mas des
  Bourboux}}}, \bibinfo {author} {\bibfnamefont {N.~G.}\ \bibnamefont
  {{Busca}}}, \bibinfo {author} {\bibfnamefont {M.}~\bibnamefont
  {{Blomqvist}}}, \bibinfo {author} {\bibfnamefont {J.}~\bibnamefont {{Guy}}},
  \bibinfo {author} {\bibfnamefont {J.}~\bibnamefont {{Rich}}}, \bibinfo
  {author} {\bibfnamefont {A.}~\bibnamefont {{Font-Ribera}}}, \bibinfo {author}
  {\bibfnamefont {M.~M.}\ \bibnamefont {{Pieri}}}, \bibinfo {author}
  {\bibfnamefont {J.~E.}\ \bibnamefont {{Bautista}}}, \bibinfo {author}
  {\bibfnamefont {K.}~\bibnamefont {{Dawson}}}, \bibinfo {author}
  {\bibfnamefont {J.-M.}\ \bibnamefont {{Le Goff}}}, \bibinfo {author}
  {\bibfnamefont {A.}~\bibnamefont {{de la Macorra}}}, \bibinfo {author}
  {\bibfnamefont {N.}~\bibnamefont {{Palanque-Delabrouille}}}, \bibinfo
  {author} {\bibfnamefont {W.~J.}\ \bibnamefont {{Percival}}}, \bibinfo
  {author} {\bibfnamefont {I.}~\bibnamefont {{P{\'e}rez-R{\`a}fols}}}, \bibinfo
  {author} {\bibfnamefont {D.~P.}\ \bibnamefont {{Schneider}}}, \bibinfo
  {author} {\bibfnamefont {A.}~\bibnamefont {{Slosar}}},\ and\ \bibinfo
  {author} {\bibfnamefont {C.}~\bibnamefont {{Y{\`e}che}}},\ }\href
  {https://doi.org/10.1051/0004-6361/201935638} {\bibfield  {journal} {\bibinfo
   {journal} {Astron. Astrophys.}\ }\textbf {\bibinfo {volume} {629}},\
  \bibinfo {eid} {A85} (\bibinfo {year} {2019})},\ \Eprint
  {https://arxiv.org/abs/1904.03400} {arXiv:1904.03400 [astro-ph.CO]}
  \BibitemShut {NoStop}%
\bibitem [{\citenamefont {{Hirata}}(2018)}]{hirataLya}%
  \BibitemOpen
  \bibfield  {author} {\bibinfo {author} {\bibfnamefont {C.~M.}\ \bibnamefont
  {{Hirata}}},\ }\href {https://doi.org/10.1093/mnras/stx2854} {\bibfield
  {journal} {\bibinfo  {journal} {Mon. Not. R. Astron. Soc.}\ }\textbf
  {\bibinfo {volume} {474}},\ \bibinfo {pages} {2173} (\bibinfo {year}
  {2018})}\BibitemShut {NoStop}%
\bibitem [{\citenamefont {{DESI Collaboration}}\ \emph
  {et~al.}(2016)\citenamefont {{DESI Collaboration}}, \citenamefont
  {{Aghamousa}}, \citenamefont {{Aguilar}}, \citenamefont {{Ahlen}},
  \citenamefont {{Alam}}, \citenamefont {{Allen}}, \citenamefont {{Allende
  Prieto}}, \citenamefont {{Annis}}, \citenamefont {{Bailey}},\ and\
  \citenamefont {{Balland}}}]{DESI}%
  \BibitemOpen
  \bibfield  {author} {\bibinfo {author} {\bibnamefont {{DESI Collaboration}}},
  \bibinfo {author} {\bibfnamefont {A.}~\bibnamefont {{Aghamousa}}}, \bibinfo
  {author} {\bibfnamefont {J.}~\bibnamefont {{Aguilar}}}, \bibinfo {author}
  {\bibfnamefont {S.}~\bibnamefont {{Ahlen}}}, \bibinfo {author} {\bibfnamefont
  {S.}~\bibnamefont {{Alam}}}, \bibinfo {author} {\bibfnamefont {L.~E.}\
  \bibnamefont {{Allen}}}, \bibinfo {author} {\bibfnamefont {C.}~\bibnamefont
  {{Allende Prieto}}}, \bibinfo {author} {\bibfnamefont {J.}~\bibnamefont
  {{Annis}}}, \bibinfo {author} {\bibfnamefont {S.}~\bibnamefont {{Bailey}}},\
  and\ \bibinfo {author} {\bibfnamefont {C.}~\bibnamefont {{Balland}}},\
  }\href@noop {} {\bibfield  {journal} {\bibinfo  {journal} {arXiv e-prints}\
  ,\ \bibinfo {eid} {arXiv:1611.00036}} (\bibinfo {year} {2016})},\ \Eprint
  {https://arxiv.org/abs/1611.00036} {arXiv:1611.00036 [astro-ph.IM]}
  \BibitemShut {NoStop}%
\bibitem [{\citenamefont {Silk}(1968)}]{Silk68}%
  \BibitemOpen
  \bibfield  {author} {\bibinfo {author} {\bibfnamefont {J.}~\bibnamefont
  {Silk}},\ }\href@noop {} {\bibfield  {journal} {\bibinfo  {journal}
  {Astrophys. J.}\ }\textbf {\bibinfo {volume} {151}},\ \bibinfo {pages} {459}
  (\bibinfo {year} {1968})}\BibitemShut {NoStop}%
\bibitem [{\citenamefont {{Tseliakhovich}}\ and\ \citenamefont
  {{Hirata}}(2010)}]{Tse10}%
  \BibitemOpen
  \bibfield  {author} {\bibinfo {author} {\bibfnamefont {D.}~\bibnamefont
  {{Tseliakhovich}}}\ and\ \bibinfo {author} {\bibfnamefont {C.}~\bibnamefont
  {{Hirata}}},\ }\href {https://doi.org/10.1103/PhysRevD.82.083520} {\bibfield
  {journal} {\bibinfo  {journal} {\prd}\ }\textbf {\bibinfo {volume} {82}},\
  \bibinfo {eid} {083520} (\bibinfo {year} {2010})},\ \Eprint
  {https://arxiv.org/abs/1005.2416} {arXiv:1005.2416 [astro-ph.CO]}
  \BibitemShut {NoStop}%
\bibitem [{\citenamefont {{Blazek}}\ \emph {et~al.}(2016)\citenamefont
  {{Blazek}}, \citenamefont {{McEwen}},\ and\ \citenamefont
  {{Hirata}}}]{blazek}%
  \BibitemOpen
  \bibfield  {author} {\bibinfo {author} {\bibfnamefont {J.~A.}\ \bibnamefont
  {{Blazek}}}, \bibinfo {author} {\bibfnamefont {J.~E.}\ \bibnamefont
  {{McEwen}}},\ and\ \bibinfo {author} {\bibfnamefont {C.~M.}\ \bibnamefont
  {{Hirata}}},\ }\href {https://doi.org/10.1103/PhysRevLett.116.121303}
  {\bibfield  {journal} {\bibinfo  {journal} {\prl}\ }\textbf {\bibinfo
  {volume} {116}},\ \bibinfo {eid} {121303} (\bibinfo {year}
  {2016})}\BibitemShut {NoStop}%
\bibitem [{\citenamefont {{Tseliakhovich}}\ \emph {et~al.}(2011)\citenamefont
  {{Tseliakhovich}}, \citenamefont {{Barkana}},\ and\ \citenamefont
  {{Hirata}}}]{Tse_Bar_Hir2011}%
  \BibitemOpen
  \bibfield  {author} {\bibinfo {author} {\bibfnamefont {D.}~\bibnamefont
  {{Tseliakhovich}}}, \bibinfo {author} {\bibfnamefont {R.}~\bibnamefont
  {{Barkana}}},\ and\ \bibinfo {author} {\bibfnamefont {C.~M.}\ \bibnamefont
  {{Hirata}}},\ }\href {https://doi.org/10.1111/j.1365-2966.2011.19541.x}
  {\bibfield  {journal} {\bibinfo  {journal} {Mon. Not. R. Astron. Soc.}\
  }\textbf {\bibinfo {volume} {418}},\ \bibinfo {pages} {906} (\bibinfo {year}
  {2011})},\ \Eprint {https://arxiv.org/abs/1012.2574} {arXiv:1012.2574
  [astro-ph.CO]} \BibitemShut {NoStop}%
\bibitem [{\citenamefont {{Dalal}}\ \emph {et~al.}(2010)\citenamefont
  {{Dalal}}, \citenamefont {{Pen}},\ and\ \citenamefont {{Seljak}}}]{Dalal}%
  \BibitemOpen
  \bibfield  {author} {\bibinfo {author} {\bibfnamefont {N.}~\bibnamefont
  {{Dalal}}}, \bibinfo {author} {\bibfnamefont {U.-L.}\ \bibnamefont {{Pen}}},\
  and\ \bibinfo {author} {\bibfnamefont {U.}~\bibnamefont {{Seljak}}},\ }\href
  {https://doi.org/10.1088/1475-7516/2010/11/007} {\bibfield  {journal}
  {\bibinfo  {journal} {J. Cosmol. Astropart. Phys.}\ }\textbf {\bibinfo
  {volume} {2010}}\bibfield  {number} {\bibinfo  {number} { (11)},\ \bibinfo
  {eid} {007}},\ }\Eprint {https://arxiv.org/abs/1009.4704} {arXiv:1009.4704
  [astro-ph.CO]} \BibitemShut {NoStop}%
\bibitem [{\citenamefont {{Yoo}}\ \emph {et~al.}(2011)\citenamefont {{Yoo}},
  \citenamefont {{Dalal}},\ and\ \citenamefont {{Seljak}}}]{Yoo11}%
  \BibitemOpen
  \bibfield  {author} {\bibinfo {author} {\bibfnamefont {J.}~\bibnamefont
  {{Yoo}}}, \bibinfo {author} {\bibfnamefont {N.}~\bibnamefont {{Dalal}}},\
  and\ \bibinfo {author} {\bibfnamefont {U.}~\bibnamefont {{Seljak}}},\ }\href
  {https://doi.org/10.1088/1475-7516/2011/07/018} {\bibfield  {journal}
  {\bibinfo  {journal} {J. Cosmol. Astropart. Phys.}\ }\textbf {\bibinfo
  {volume} {2011}}\bibfield  {number} {\bibinfo  {number} { (7)},\ \bibinfo
  {eid} {018}},\ }\Eprint {https://arxiv.org/abs/1105.3732} {arXiv:1105.3732
  [astro-ph.CO]} \BibitemShut {NoStop}%
\bibitem [{\citenamefont {{Yoo}}\ and\ \citenamefont {{Seljak}}(2013)}]{Yoo13}%
  \BibitemOpen
  \bibfield  {author} {\bibinfo {author} {\bibfnamefont {J.}~\bibnamefont
  {{Yoo}}}\ and\ \bibinfo {author} {\bibfnamefont {U.}~\bibnamefont
  {{Seljak}}},\ }\href {https://doi.org/10.1103/PhysRevD.88.103520} {\bibfield
  {journal} {\bibinfo  {journal} {\prd}\ }\textbf {\bibinfo {volume} {88}},\
  \bibinfo {eid} {103520} (\bibinfo {year} {2013})},\ \Eprint
  {https://arxiv.org/abs/1308.1401} {arXiv:1308.1401 [astro-ph.CO]}
  \BibitemShut {NoStop}%
\bibitem [{\citenamefont {{Slepian}}\ and\ \citenamefont
  {{Eisenstein}}(2015)}]{Slepian}%
  \BibitemOpen
  \bibfield  {author} {\bibinfo {author} {\bibfnamefont {Z.}~\bibnamefont
  {{Slepian}}}\ and\ \bibinfo {author} {\bibfnamefont {D.~J.}\ \bibnamefont
  {{Eisenstein}}},\ }\href {https://doi.org/10.1093/mnras/stu2627} {\bibfield
  {journal} {\bibinfo  {journal} {Mon. Not. R. Astron. Soc.}\ }\textbf
  {\bibinfo {volume} {448}},\ \bibinfo {pages} {9} (\bibinfo {year} {2015})},\
  \Eprint {https://arxiv.org/abs/1411.4052} {arXiv:1411.4052 [astro-ph.CO]}
  \BibitemShut {NoStop}%
\bibitem [{\citenamefont {{Arinyo-i-Prats}}\ \emph {et~al.}(2015)\citenamefont
  {{Arinyo-i-Prats}}, \citenamefont {{Miralda-Escud{\'e}}}, \citenamefont
  {{Viel}},\ and\ \citenamefont {{Cen}}}]{Arinyo2015}%
  \BibitemOpen
  \bibfield  {author} {\bibinfo {author} {\bibfnamefont {A.}~\bibnamefont
  {{Arinyo-i-Prats}}}, \bibinfo {author} {\bibfnamefont {J.}~\bibnamefont
  {{Miralda-Escud{\'e}}}}, \bibinfo {author} {\bibfnamefont {M.}~\bibnamefont
  {{Viel}}},\ and\ \bibinfo {author} {\bibfnamefont {R.}~\bibnamefont
  {{Cen}}},\ }\href {https://doi.org/10.1088/1475-7516/2015/12/017} {\bibfield
  {journal} {\bibinfo  {journal} {J. Cosmol. Astropart. Phys.}\ }\textbf
  {\bibinfo {volume} {2015}}\bibfield  {number} {\bibinfo  {number} { (12)},\
  \bibinfo {eid} {017}},\ }\Eprint {https://arxiv.org/abs/1506.04519}
  {arXiv:1506.04519 [astro-ph.CO]} \BibitemShut {NoStop}%
\bibitem [{\citenamefont {{Beutler}}\ \emph
  {et~al.}(2017{\natexlab{b}})\citenamefont {{Beutler}}, \citenamefont
  {{Seljak}},\ and\ \citenamefont {{Vlah}}}]{Beutler}%
  \BibitemOpen
  \bibfield  {author} {\bibinfo {author} {\bibfnamefont {F.}~\bibnamefont
  {{Beutler}}}, \bibinfo {author} {\bibfnamefont {U.}~\bibnamefont
  {{Seljak}}},\ and\ \bibinfo {author} {\bibfnamefont {Z.}~\bibnamefont
  {{Vlah}}},\ }\href {https://doi.org/10.1093/mnras/stx1196} {\bibfield
  {journal} {\bibinfo  {journal} {Mon. Not. R. Astron. Soc.}\ }\textbf
  {\bibinfo {volume} {470}},\ \bibinfo {pages} {2723} (\bibinfo {year}
  {2017}{\natexlab{b}})}\BibitemShut {NoStop}%
\bibitem [{\citenamefont {{Bernardeau}}\ \emph {et~al.}(2002)\citenamefont
  {{Bernardeau}}, \citenamefont {{Colombi}}, \citenamefont {{Gazta{\~n}aga}},\
  and\ \citenamefont {{Scoccimarro}}}]{PT}%
  \BibitemOpen
  \bibfield  {author} {\bibinfo {author} {\bibfnamefont {F.}~\bibnamefont
  {{Bernardeau}}}, \bibinfo {author} {\bibfnamefont {S.}~\bibnamefont
  {{Colombi}}}, \bibinfo {author} {\bibfnamefont {E.}~\bibnamefont
  {{Gazta{\~n}aga}}},\ and\ \bibinfo {author} {\bibfnamefont {R.}~\bibnamefont
  {{Scoccimarro}}},\ }\href {https://doi.org/10.1016/S0370-1573(02)00135-7}
  {\bibfield  {journal} {\bibinfo  {journal} {Phys. Rep.}\ }\textbf {\bibinfo
  {volume} {367}},\ \bibinfo {pages} {1} (\bibinfo {year} {2002})},\ \Eprint
  {https://arxiv.org/abs/astro-ph/0112551} {arXiv:astro-ph/0112551 [astro-ph]}
  \BibitemShut {NoStop}%
\bibitem [{\citenamefont {{McDonald}}\ and\ \citenamefont
  {{Roy}}(2009)}]{mcdonald}%
  \BibitemOpen
  \bibfield  {author} {\bibinfo {author} {\bibfnamefont {P.}~\bibnamefont
  {{McDonald}}}\ and\ \bibinfo {author} {\bibfnamefont {A.}~\bibnamefont
  {{Roy}}},\ }\href {https://doi.org/10.1088/1475-7516/2009/08/020} {\bibfield
  {journal} {\bibinfo  {journal} {J. Cosmol. Astropart. Phys.}\ }\textbf
  {\bibinfo {volume} {2009}}\bibfield  {number} {\bibinfo  {number} { (8)},\
  \bibinfo {eid} {020}},\ }\Eprint {https://arxiv.org/abs/0902.0991}
  {arXiv:0902.0991 [astro-ph.CO]} \BibitemShut {NoStop}%
\bibitem [{\citenamefont {{Kaiser}}(1987)}]{Kaiser}%
  \BibitemOpen
  \bibfield  {author} {\bibinfo {author} {\bibfnamefont {N.}~\bibnamefont
  {{Kaiser}}},\ }\href {https://doi.org/10.1093/mnras/227.1.1} {\bibfield
  {journal} {\bibinfo  {journal} {Mon. Not. R. Astron. Soc.}\ }\textbf
  {\bibinfo {volume} {227}},\ \bibinfo {pages} {1} (\bibinfo {year}
  {1987})}\BibitemShut {NoStop}%
\bibitem [{\citenamefont {{Rauch}}(1998)}]{1998ARA&A..36..267R}%
  \BibitemOpen
  \bibfield  {author} {\bibinfo {author} {\bibfnamefont {M.}~\bibnamefont
  {{Rauch}}},\ }\href {https://doi.org/10.1146/annurev.astro.36.1.267}
  {\bibfield  {journal} {\bibinfo  {journal} {\araa}\ }\textbf {\bibinfo
  {volume} {36}},\ \bibinfo {pages} {267} (\bibinfo {year} {1998})},\ \Eprint
  {https://arxiv.org/abs/astro-ph/9806286} {arXiv:astro-ph/9806286 [astro-ph]}
  \BibitemShut {NoStop}%
\bibitem [{\citenamefont {{Seljak}}(2012)}]{seljak}%
  \BibitemOpen
  \bibfield  {author} {\bibinfo {author} {\bibfnamefont {U.}~\bibnamefont
  {{Seljak}}},\ }\href {https://doi.org/10.1088/1475-7516/2012/03/004}
  {\bibfield  {journal} {\bibinfo  {journal} {J. Cosmol. Astropart. Phys.}\
  }\textbf {\bibinfo {volume} {2012}}\bibfield  {number} {\bibinfo  {number} {
  (3)},\ \bibinfo {eid} {004}},\ }\Eprint {https://arxiv.org/abs/1201.0594}
  {arXiv:1201.0594 [astro-ph.CO]} \BibitemShut {NoStop}%
\bibitem [{\citenamefont {{Desjacques}}\ \emph
  {et~al.}(2018{\natexlab{a}})\citenamefont {{Desjacques}}, \citenamefont
  {{Jeong}},\ and\ \citenamefont {{Schmidt}}}]{Desjacques2018}%
  \BibitemOpen
  \bibfield  {author} {\bibinfo {author} {\bibfnamefont {V.}~\bibnamefont
  {{Desjacques}}}, \bibinfo {author} {\bibfnamefont {D.}~\bibnamefont
  {{Jeong}}},\ and\ \bibinfo {author} {\bibfnamefont {F.}~\bibnamefont
  {{Schmidt}}},\ }\href {https://doi.org/10.1016/j.physrep.2017.12.002}
  {\bibfield  {journal} {\bibinfo  {journal} {Phys. Rep.}\ }\textbf {\bibinfo
  {volume} {733}},\ \bibinfo {pages} {1} (\bibinfo {year}
  {2018}{\natexlab{a}})},\ \Eprint {https://arxiv.org/abs/1611.09787}
  {arXiv:1611.09787 [astro-ph.CO]} \BibitemShut {NoStop}%
\bibitem [{\citenamefont {{Desjacques}}\ \emph
  {et~al.}(2018{\natexlab{b}})\citenamefont {{Desjacques}}, \citenamefont
  {{Jeong}},\ and\ \citenamefont {{Schmidt}}}]{Desjacques2018jcap}%
  \BibitemOpen
  \bibfield  {author} {\bibinfo {author} {\bibfnamefont {V.}~\bibnamefont
  {{Desjacques}}}, \bibinfo {author} {\bibfnamefont {D.}~\bibnamefont
  {{Jeong}}},\ and\ \bibinfo {author} {\bibfnamefont {F.}~\bibnamefont
  {{Schmidt}}},\ }\href {https://doi.org/10.1088/1475-7516/2018/12/035}
  {\bibfield  {journal} {\bibinfo  {journal} {J. Cosmol. Astropart. Phys.}\
  }\textbf {\bibinfo {volume} {2018}}\bibfield  {number} {\bibinfo  {number} {
  (12)},\ \bibinfo {eid} {035}},\ }\Eprint {https://arxiv.org/abs/1806.04015}
  {arXiv:1806.04015 [astro-ph.CO]} \BibitemShut {NoStop}%
\bibitem [{\citenamefont {{Schmidt}}(2016)}]{Schmidt2016}%
  \BibitemOpen
  \bibfield  {author} {\bibinfo {author} {\bibfnamefont {F.}~\bibnamefont
  {{Schmidt}}},\ }\href {https://doi.org/10.1103/PhysRevD.94.063508} {\bibfield
   {journal} {\bibinfo  {journal} {\prd}\ }\textbf {\bibinfo {volume} {94}},\
  \bibinfo {eid} {063508} (\bibinfo {year} {2016})},\ \Eprint
  {https://arxiv.org/abs/1602.09059} {arXiv:1602.09059 [astro-ph.CO]}
  \BibitemShut {NoStop}%
\bibitem [{\citenamefont {{Gnedin}}\ and\ \citenamefont
  {{Hui}}(1998)}]{1998MNRAS.296...44G}%
  \BibitemOpen
  \bibfield  {author} {\bibinfo {author} {\bibfnamefont {N.~Y.}\ \bibnamefont
  {{Gnedin}}}\ and\ \bibinfo {author} {\bibfnamefont {L.}~\bibnamefont
  {{Hui}}},\ }\href {https://doi.org/10.1046/j.1365-8711.1998.01249.x}
  {\bibfield  {journal} {\bibinfo  {journal} {Mon. Not. R. Astron. Soc.}\
  }\textbf {\bibinfo {volume} {296}},\ \bibinfo {pages} {44} (\bibinfo {year}
  {1998})},\ \Eprint {https://arxiv.org/abs/astro-ph/9706219}
  {arXiv:astro-ph/9706219 [astro-ph]} \BibitemShut {NoStop}%
\bibitem [{\citenamefont {{McQuinn}}\ and\ \citenamefont
  {{White}}(2011)}]{2011MNRAS.415.2257M}%
  \BibitemOpen
  \bibfield  {author} {\bibinfo {author} {\bibfnamefont {M.}~\bibnamefont
  {{McQuinn}}}\ and\ \bibinfo {author} {\bibfnamefont {M.}~\bibnamefont
  {{White}}},\ }\href {https://doi.org/10.1111/j.1365-2966.2011.18855.x}
  {\bibfield  {journal} {\bibinfo  {journal} {Mon. Not. R. Astron. Soc.}\
  }\textbf {\bibinfo {volume} {415}},\ \bibinfo {pages} {2257} (\bibinfo {year}
  {2011})},\ \Eprint {https://arxiv.org/abs/1102.1752} {arXiv:1102.1752
  [astro-ph.CO]} \BibitemShut {NoStop}%
\bibitem [{\citenamefont {{Eisenstein}}\ \emph {et~al.}(2007)\citenamefont
  {{Eisenstein}}, \citenamefont {{Seo}},\ and\ \citenamefont
  {{White}}}]{2007ApJ...664..660E}%
  \BibitemOpen
  \bibfield  {author} {\bibinfo {author} {\bibfnamefont {D.~J.}\ \bibnamefont
  {{Eisenstein}}}, \bibinfo {author} {\bibfnamefont {H.-J.}\ \bibnamefont
  {{Seo}}},\ and\ \bibinfo {author} {\bibfnamefont {M.}~\bibnamefont
  {{White}}},\ }\href {https://doi.org/10.1086/518755} {\bibfield  {journal}
  {\bibinfo  {journal} {\apj}\ }\textbf {\bibinfo {volume} {664}},\ \bibinfo
  {pages} {660} (\bibinfo {year} {2007})},\ \Eprint
  {https://arxiv.org/abs/astro-ph/0604361} {arXiv:astro-ph/0604361 [astro-ph]}
  \BibitemShut {NoStop}%
\bibitem [{\citenamefont {{McEwen}}\ \emph {et~al.}(2016)\citenamefont
  {{McEwen}}, \citenamefont {{Fang}}, \citenamefont {{Hirata}},\ and\
  \citenamefont {{Blazek}}}]{fastpt1}%
  \BibitemOpen
  \bibfield  {author} {\bibinfo {author} {\bibfnamefont {J.~E.}\ \bibnamefont
  {{McEwen}}}, \bibinfo {author} {\bibfnamefont {X.}~\bibnamefont {{Fang}}},
  \bibinfo {author} {\bibfnamefont {C.~M.}\ \bibnamefont {{Hirata}}},\ and\
  \bibinfo {author} {\bibfnamefont {J.~A.}\ \bibnamefont {{Blazek}}},\ }\href
  {https://doi.org/10.1088/1475-7516/2016/09/015} {\bibfield  {journal}
  {\bibinfo  {journal} {J. Cosmol. Astropart. Phys.}\ }\textbf {\bibinfo
  {volume} {2016}}\bibfield  {number} {\bibinfo  {number} { (9)},\ \bibinfo
  {eid} {015}},\ }\Eprint {https://arxiv.org/abs/1603.04826} {arXiv:1603.04826
  [astro-ph.CO]} \BibitemShut {NoStop}%
\bibitem [{\citenamefont {{Fang}}\ \emph {et~al.}(2017)\citenamefont {{Fang}},
  \citenamefont {{Blazek}}, \citenamefont {{McEwen}},\ and\ \citenamefont
  {{Hirata}}}]{fastpt2}%
  \BibitemOpen
  \bibfield  {author} {\bibinfo {author} {\bibfnamefont {X.}~\bibnamefont
  {{Fang}}}, \bibinfo {author} {\bibfnamefont {J.~A.}\ \bibnamefont
  {{Blazek}}}, \bibinfo {author} {\bibfnamefont {J.~E.}\ \bibnamefont
  {{McEwen}}},\ and\ \bibinfo {author} {\bibfnamefont {C.~M.}\ \bibnamefont
  {{Hirata}}},\ }\href {https://doi.org/10.1088/1475-7516/2017/02/030}
  {\bibfield  {journal} {\bibinfo  {journal} {J. Cosmol. Astropart. Phys.}\
  }\textbf {\bibinfo {volume} {2017}}\bibfield  {number} {\bibinfo  {number} {
  (2)},\ \bibinfo {eid} {030}},\ }\Eprint {https://arxiv.org/abs/1609.05978}
  {arXiv:1609.05978 [astro-ph.CO]} \BibitemShut {NoStop}%
\bibitem [{\citenamefont {Thornton}\ and\ \citenamefont
  {Marion}(2004)}]{Thornton&Marion}%
  \BibitemOpen
  \bibfield  {author} {\bibinfo {author} {\bibfnamefont {S.~T.}\ \bibnamefont
  {Thornton}}\ and\ \bibinfo {author} {\bibfnamefont {J.~B.}\ \bibnamefont
  {Marion}},\ }\href@noop {} {\emph {\bibinfo {title} {Classical Dynamics of
  Particles and Systems}}},\ \bibinfo {edition} {5th}\ ed.\ (\bibinfo
  {publisher} {Brooks/Cole---Thomson Learning},\ \bibinfo {year}
  {2004})\BibitemShut {NoStop}%
\bibitem [{BHP(2016)}]{BHPToolkit}%
  \BibitemOpen
  \href@noop {} {\bibinfo {title} {{Black Hole Perturbation Toolkit}}},\
  \bibinfo {howpublished} {(\href{http://bhptoolkit.org/}{bhptoolkit.org})}
  (\bibinfo {year} {2016})\BibitemShut {NoStop}%
\bibitem [{\citenamefont {{Springel}}(2005)}]{Gadget2}%
  \BibitemOpen
  \bibfield  {author} {\bibinfo {author} {\bibfnamefont {V.}~\bibnamefont
  {{Springel}}},\ }\href {https://doi.org/10.1111/j.1365-2966.2005.09655.x}
  {\bibfield  {journal} {\bibinfo  {journal} {Mon. Not. R. Astron. Soc.}\
  }\textbf {\bibinfo {volume} {364}},\ \bibinfo {pages} {1105} (\bibinfo {year}
  {2005})},\ \Eprint {https://arxiv.org/abs/astro-ph/0505010}
  {arXiv:astro-ph/0505010 [astro-ph]} \BibitemShut {NoStop}%
\bibitem [{\citenamefont {{Gunn}}\ and\ \citenamefont
  {{Peterson}}(1965)}]{FGPA}%
  \BibitemOpen
  \bibfield  {author} {\bibinfo {author} {\bibfnamefont {J.~E.}\ \bibnamefont
  {{Gunn}}}\ and\ \bibinfo {author} {\bibfnamefont {B.~A.}\ \bibnamefont
  {{Peterson}}},\ }\href {https://doi.org/10.1086/148444} {\bibfield  {journal}
  {\bibinfo  {journal} {\apj}\ }\textbf {\bibinfo {volume} {142}},\ \bibinfo
  {pages} {1633} (\bibinfo {year} {1965})}\BibitemShut {NoStop}%
\bibitem [{\citenamefont {{Weinberg}}\ \emph {et~al.}(1997)\citenamefont
  {{Weinberg}}, \citenamefont {{Hernsquit}}, \citenamefont {{Katz}},
  \citenamefont {{Croft}},\ and\ \citenamefont
  {{Miralda-Escud{\'e}}}}]{FGPAexplained}%
  \BibitemOpen
  \bibfield  {author} {\bibinfo {author} {\bibfnamefont {D.~H.}\ \bibnamefont
  {{Weinberg}}}, \bibinfo {author} {\bibfnamefont {L.}~\bibnamefont
  {{Hernsquit}}}, \bibinfo {author} {\bibfnamefont {N.}~\bibnamefont {{Katz}}},
  \bibinfo {author} {\bibfnamefont {R.}~\bibnamefont {{Croft}}},\ and\ \bibinfo
  {author} {\bibfnamefont {J.}~\bibnamefont {{Miralda-Escud{\'e}}}},\ }in\
  \href@noop {} {\emph {\bibinfo {booktitle} {Structure and Evolution of the
  Intergalactic Medium from QSO Absorption Line System}}},\ \bibinfo {editor}
  {edited by\ \bibinfo {editor} {\bibfnamefont {P.}~\bibnamefont
  {{Petitjean}}}\ and\ \bibinfo {editor} {\bibfnamefont {S.}~\bibnamefont
  {{Charlot}}}}\ (\bibinfo {year} {1997})\ p.\ \bibinfo {pages} {133},\ \Eprint
  {https://arxiv.org/abs/astro-ph/9709303} {arXiv:astro-ph/9709303 [astro-ph]}
  \BibitemShut {NoStop}%
\bibitem [{\citenamefont {{Tie}}\ \emph {et~al.}(2019)\citenamefont {{Tie}},
  \citenamefont {{Weinberg}}, \citenamefont {{Martini}}, \citenamefont {{Zhu}},
  \citenamefont {{Peirani}}, \citenamefont {{Suarez}},\ and\ \citenamefont
  {{Colombi}}}]{Tie}%
  \BibitemOpen
  \bibfield  {author} {\bibinfo {author} {\bibfnamefont {S.~S.}\ \bibnamefont
  {{Tie}}}, \bibinfo {author} {\bibfnamefont {D.~H.}\ \bibnamefont
  {{Weinberg}}}, \bibinfo {author} {\bibfnamefont {P.}~\bibnamefont
  {{Martini}}}, \bibinfo {author} {\bibfnamefont {W.}~\bibnamefont {{Zhu}}},
  \bibinfo {author} {\bibfnamefont {S.}~\bibnamefont {{Peirani}}}, \bibinfo
  {author} {\bibfnamefont {T.}~\bibnamefont {{Suarez}}},\ and\ \bibinfo
  {author} {\bibfnamefont {S.}~\bibnamefont {{Colombi}}},\ }\href
  {https://doi.org/10.1093/mnras/stz1632} {\bibfield  {journal} {\bibinfo
  {journal} {Mon. Not. R. Astron. Soc.}\ }\textbf {\bibinfo {volume} {487}},\
  \bibinfo {pages} {5346} (\bibinfo {year} {2019})}\BibitemShut {NoStop}%
\bibitem [{\citenamefont {{Seo}}\ \emph {et~al.}(2008)\citenamefont {{Seo}},
  \citenamefont {{Siegel}}, \citenamefont {{Eisenstein}},\ and\ \citenamefont
  {{White}}}]{Seo2008}%
  \BibitemOpen
  \bibfield  {author} {\bibinfo {author} {\bibfnamefont {H.-J.}\ \bibnamefont
  {{Seo}}}, \bibinfo {author} {\bibfnamefont {E.~R.}\ \bibnamefont {{Siegel}}},
  \bibinfo {author} {\bibfnamefont {D.~J.}\ \bibnamefont {{Eisenstein}}},\ and\
  \bibinfo {author} {\bibfnamefont {M.}~\bibnamefont {{White}}},\ }\href
  {https://doi.org/10.1086/589921} {\bibfield  {journal} {\bibinfo  {journal}
  {\apj}\ }\textbf {\bibinfo {volume} {686}},\ \bibinfo {pages} {13} (\bibinfo
  {year} {2008})},\ \Eprint {https://arxiv.org/abs/0805.0117} {arXiv:0805.0117
  [astro-ph]} \BibitemShut {NoStop}%
\bibitem [{\citenamefont {{Eisenstein}}\ and\ \citenamefont
  {{Hu}}(1998)}]{EH1998}%
  \BibitemOpen
  \bibfield  {author} {\bibinfo {author} {\bibfnamefont {D.~J.}\ \bibnamefont
  {{Eisenstein}}}\ and\ \bibinfo {author} {\bibfnamefont {W.}~\bibnamefont
  {{Hu}}},\ }\href {https://doi.org/10.1086/305424} {\bibfield  {journal}
  {\bibinfo  {journal} {\apj}\ }\textbf {\bibinfo {volume} {496}},\ \bibinfo
  {pages} {605} (\bibinfo {year} {1998})},\ \Eprint
  {https://arxiv.org/abs/astro-ph/9709112} {arXiv:astro-ph/9709112 [astro-ph]}
  \BibitemShut {NoStop}%
\bibitem [{\citenamefont {{Nelder}}\ and\ \citenamefont
  {{Mead}}(1965)}]{NelderMead}%
  \BibitemOpen
  \bibfield  {author} {\bibinfo {author} {\bibfnamefont {J.~A.}\ \bibnamefont
  {{Nelder}}}\ and\ \bibinfo {author} {\bibfnamefont {R.}~\bibnamefont
  {{Mead}}},\ }\href {https://doi.org/10.1093/comjnl/7.4.308} {\bibfield
  {journal} {\bibinfo  {journal} {The Computer Journal}\ }\textbf {\bibinfo
  {volume} {7}},\ \bibinfo {pages} {308} (\bibinfo {year} {1965})}\BibitemShut
  {NoStop}%
\bibitem [{\citenamefont {{Schlegel}}\ \emph {et~al.}(2019)\citenamefont
  {{Schlegel}}, \citenamefont {{Kollmeier}},\ and\ \citenamefont
  {{Ferraro}}}]{2019BAAS...51g.229S}%
  \BibitemOpen
  \bibfield  {author} {\bibinfo {author} {\bibfnamefont {D.}~\bibnamefont
  {{Schlegel}}}, \bibinfo {author} {\bibfnamefont {J.~A.}\ \bibnamefont
  {{Kollmeier}}},\ and\ \bibinfo {author} {\bibfnamefont {S.}~\bibnamefont
  {{Ferraro}}},\ }in\ \href@noop {} {\emph {\bibinfo {booktitle} {Bulletin of
  the American Astronomical Society}}},\ Vol.~\bibinfo {volume} {51}\ (\bibinfo
  {year} {2019})\ p.\ \bibinfo {pages} {229},\ \Eprint
  {https://arxiv.org/abs/1907.11171} {arXiv:1907.11171 [astro-ph.IM]}
  \BibitemShut {NoStop}%
\bibitem [{\citenamefont {{du Mas des Bourboux}}\ \emph
  {et~al.}(2017)\citenamefont {{du Mas des Bourboux}}, \citenamefont {{Le
  Goff}}, \citenamefont {{Blomqvist}}, \citenamefont {{Busca}}, \citenamefont
  {{Guy}}, \citenamefont {{Rich}}, \citenamefont {{Y{\`e}che}}, \citenamefont
  {{Bautista}}, \citenamefont {{Burtin}}, \citenamefont {{Dawson}},
  \citenamefont {{Eisenstein}}, \citenamefont {{Font-Ribera}}, \citenamefont
  {{Kirkby}}, \citenamefont {{Miralda-Escud{\'e}}}, \citenamefont
  {{Noterdaeme}}, \citenamefont {{Palanque-Delabrouille}}, \citenamefont
  {{P{\^a}ris}}, \citenamefont {{Petitjean}}, \citenamefont
  {{P{\'e}rez-R{\`a}fols}}, \citenamefont {{Pieri}}, \citenamefont {{Ross}},
  \citenamefont {{Schlegel}}, \citenamefont {{Schneider}}, \citenamefont
  {{Slosar}}, \citenamefont {{Weinberg}},\ and\ \citenamefont
  {{Zarrouk}}}]{2017A&A...608A.130D}%
  \BibitemOpen
  \bibfield  {author} {\bibinfo {author} {\bibfnamefont {H.}~\bibnamefont {{du
  Mas des Bourboux}}}, \bibinfo {author} {\bibfnamefont {J.-M.}\ \bibnamefont
  {{Le Goff}}}, \bibinfo {author} {\bibfnamefont {M.}~\bibnamefont
  {{Blomqvist}}}, \bibinfo {author} {\bibfnamefont {N.~G.}\ \bibnamefont
  {{Busca}}}, \bibinfo {author} {\bibfnamefont {J.}~\bibnamefont {{Guy}}},
  \bibinfo {author} {\bibfnamefont {J.}~\bibnamefont {{Rich}}}, \bibinfo
  {author} {\bibfnamefont {C.}~\bibnamefont {{Y{\`e}che}}}, \bibinfo {author}
  {\bibfnamefont {J.~E.}\ \bibnamefont {{Bautista}}}, \bibinfo {author}
  {\bibfnamefont {{\'E}.}~\bibnamefont {{Burtin}}}, \bibinfo {author}
  {\bibfnamefont {K.~S.}\ \bibnamefont {{Dawson}}}, \bibinfo {author}
  {\bibfnamefont {D.~J.}\ \bibnamefont {{Eisenstein}}}, \bibinfo {author}
  {\bibfnamefont {A.}~\bibnamefont {{Font-Ribera}}}, \bibinfo {author}
  {\bibfnamefont {D.}~\bibnamefont {{Kirkby}}}, \bibinfo {author}
  {\bibfnamefont {J.}~\bibnamefont {{Miralda-Escud{\'e}}}}, \bibinfo {author}
  {\bibfnamefont {P.}~\bibnamefont {{Noterdaeme}}}, \bibinfo {author}
  {\bibfnamefont {N.}~\bibnamefont {{Palanque-Delabrouille}}}, \bibinfo
  {author} {\bibfnamefont {I.}~\bibnamefont {{P{\^a}ris}}}, \bibinfo {author}
  {\bibfnamefont {P.}~\bibnamefont {{Petitjean}}}, \bibinfo {author}
  {\bibfnamefont {I.}~\bibnamefont {{P{\'e}rez-R{\`a}fols}}}, \bibinfo {author}
  {\bibfnamefont {M.~M.}\ \bibnamefont {{Pieri}}}, \bibinfo {author}
  {\bibfnamefont {N.~P.}\ \bibnamefont {{Ross}}}, \bibinfo {author}
  {\bibfnamefont {D.~J.}\ \bibnamefont {{Schlegel}}}, \bibinfo {author}
  {\bibfnamefont {D.~P.}\ \bibnamefont {{Schneider}}}, \bibinfo {author}
  {\bibfnamefont {A.}~\bibnamefont {{Slosar}}}, \bibinfo {author}
  {\bibfnamefont {D.~H.}\ \bibnamefont {{Weinberg}}},\ and\ \bibinfo {author}
  {\bibfnamefont {P.}~\bibnamefont {{Zarrouk}}},\ }\href
  {https://doi.org/10.1051/0004-6361/201731731} {\bibfield  {journal} {\bibinfo
   {journal} {Astron. Astrophys.}\ }\textbf {\bibinfo {volume} {608}},\
  \bibinfo {eid} {A130} (\bibinfo {year} {2017})},\ \Eprint
  {https://arxiv.org/abs/1708.02225} {arXiv:1708.02225 [astro-ph.CO]}
  \BibitemShut {NoStop}%
\bibitem [{\citenamefont {{Montero-Camacho}}\ \emph {et~al.}(2019)\citenamefont
  {{Montero-Camacho}}, \citenamefont {{Hirata}}, \citenamefont {{Martini}},\
  and\ \citenamefont {{Honscheid}}}]{Montero-Camacho2019}%
  \BibitemOpen
  \bibfield  {author} {\bibinfo {author} {\bibfnamefont {P.}~\bibnamefont
  {{Montero-Camacho}}}, \bibinfo {author} {\bibfnamefont {C.~M.}\ \bibnamefont
  {{Hirata}}}, \bibinfo {author} {\bibfnamefont {P.}~\bibnamefont
  {{Martini}}},\ and\ \bibinfo {author} {\bibfnamefont {K.}~\bibnamefont
  {{Honscheid}}},\ }\href {https://doi.org/10.1093/mnras/stz1388} {\bibfield
  {journal} {\bibinfo  {journal} {Mon. Not. R. Astron. Soc.}\ }\textbf
  {\bibinfo {volume} {487}},\ \bibinfo {pages} {1047} (\bibinfo {year}
  {2019})},\ \Eprint {https://arxiv.org/abs/1902.02892} {arXiv:1902.02892
  [astro-ph.CO]} \BibitemShut {NoStop}%
\bibitem [{\citenamefont {{Pontzen}}(2014)}]{2014PhRvD..89h3010P}%
  \BibitemOpen
  \bibfield  {author} {\bibinfo {author} {\bibfnamefont {A.}~\bibnamefont
  {{Pontzen}}},\ }\href {https://doi.org/10.1103/PhysRevD.89.083010} {\bibfield
   {journal} {\bibinfo  {journal} {\prd}\ }\textbf {\bibinfo {volume} {89}},\
  \bibinfo {eid} {083010} (\bibinfo {year} {2014})},\ \Eprint
  {https://arxiv.org/abs/1402.0506} {arXiv:1402.0506 [astro-ph.CO]}
  \BibitemShut {NoStop}%
\bibitem [{\citenamefont {{Gontcho A Gontcho}}\ \emph
  {et~al.}(2014)\citenamefont {{Gontcho A Gontcho}}, \citenamefont
  {{Miralda-Escud{\'e}}},\ and\ \citenamefont {{Busca}}}]{2014MNRAS.442..187G}%
  \BibitemOpen
  \bibfield  {author} {\bibinfo {author} {\bibfnamefont {S.}~\bibnamefont
  {{Gontcho A Gontcho}}}, \bibinfo {author} {\bibfnamefont {J.}~\bibnamefont
  {{Miralda-Escud{\'e}}}},\ and\ \bibinfo {author} {\bibfnamefont {N.~G.}\
  \bibnamefont {{Busca}}},\ }\href {https://doi.org/10.1093/mnras/stu860}
  {\bibfield  {journal} {\bibinfo  {journal} {Mon. Not. R. Astron. Soc.}\
  }\textbf {\bibinfo {volume} {442}},\ \bibinfo {pages} {187} (\bibinfo {year}
  {2014})},\ \Eprint {https://arxiv.org/abs/1404.7425} {arXiv:1404.7425
  [astro-ph.CO]} \BibitemShut {NoStop}%
\bibitem [{\citenamefont {{Morales}}\ and\ \citenamefont
  {{Wyithe}}(2010)}]{2010ARA&A..48..127M}%
  \BibitemOpen
  \bibfield  {author} {\bibinfo {author} {\bibfnamefont {M.~F.}\ \bibnamefont
  {{Morales}}}\ and\ \bibinfo {author} {\bibfnamefont {J.~S.~B.}\ \bibnamefont
  {{Wyithe}}},\ }\href {https://doi.org/10.1146/annurev-astro-081309-130936}
  {\bibfield  {journal} {\bibinfo  {journal} {\araa}\ }\textbf {\bibinfo
  {volume} {48}},\ \bibinfo {pages} {127} (\bibinfo {year} {2010})},\ \Eprint
  {https://arxiv.org/abs/0910.3010} {arXiv:0910.3010 [astro-ph.CO]}
  \BibitemShut {NoStop}%
\bibitem [{\citenamefont {{Richard}}\ \emph {et~al.}(2019)\citenamefont
  {{Richard}}, \citenamefont {{Kneib}}, \citenamefont {{Blake}}, \citenamefont
  {{Raichoor}}, \citenamefont {{Comparat}}, \citenamefont {{Shanks}},
  \citenamefont {{Sorce}}, \citenamefont {{Sahl{\'e}n}}, \citenamefont
  {{Howlett}}, \citenamefont {{Tempel}}, \citenamefont {{McMahon}},
  \citenamefont {{Bilicki}}, \citenamefont {{Roukema}}, \citenamefont
  {{Loveday}}, \citenamefont {{Pryer}}, \citenamefont {{Buchert}},
  \citenamefont {{Zhao}},\ and\ \citenamefont {{CRS
  Team}}}]{2019Msngr.175...50R}%
  \BibitemOpen
  \bibfield  {author} {\bibinfo {author} {\bibfnamefont {J.}~\bibnamefont
  {{Richard}}}, \bibinfo {author} {\bibfnamefont {J.~P.}\ \bibnamefont
  {{Kneib}}}, \bibinfo {author} {\bibfnamefont {C.}~\bibnamefont {{Blake}}},
  \bibinfo {author} {\bibfnamefont {A.}~\bibnamefont {{Raichoor}}}, \bibinfo
  {author} {\bibfnamefont {J.}~\bibnamefont {{Comparat}}}, \bibinfo {author}
  {\bibfnamefont {T.}~\bibnamefont {{Shanks}}}, \bibinfo {author}
  {\bibfnamefont {J.}~\bibnamefont {{Sorce}}}, \bibinfo {author} {\bibfnamefont
  {M.}~\bibnamefont {{Sahl{\'e}n}}}, \bibinfo {author} {\bibfnamefont
  {C.}~\bibnamefont {{Howlett}}}, \bibinfo {author} {\bibfnamefont
  {E.}~\bibnamefont {{Tempel}}}, \bibinfo {author} {\bibfnamefont
  {R.}~\bibnamefont {{McMahon}}}, \bibinfo {author} {\bibfnamefont
  {M.}~\bibnamefont {{Bilicki}}}, \bibinfo {author} {\bibfnamefont
  {B.}~\bibnamefont {{Roukema}}}, \bibinfo {author} {\bibfnamefont
  {J.}~\bibnamefont {{Loveday}}}, \bibinfo {author} {\bibfnamefont
  {D.}~\bibnamefont {{Pryer}}}, \bibinfo {author} {\bibfnamefont
  {T.}~\bibnamefont {{Buchert}}}, \bibinfo {author} {\bibfnamefont
  {C.}~\bibnamefont {{Zhao}}},\ and\ \bibinfo {author} {\bibnamefont {{CRS
  Team}}},\ }\href {https://doi.org/10.18727/0722-6691/5127} {\bibfield
  {journal} {\bibinfo  {journal} {The Messenger}\ }\textbf {\bibinfo {volume}
  {175}},\ \bibinfo {pages} {50} (\bibinfo {year} {2019})},\ \Eprint
  {https://arxiv.org/abs/1903.02474} {arXiv:1903.02474 [astro-ph.CO]}
  \BibitemShut {NoStop}%
\bibitem [{\citenamefont {{Bandura}}\ \emph {et~al.}(2014)\citenamefont
  {{Bandura}}, \citenamefont {{Addison}}, \citenamefont {{Amiri}},
  \citenamefont {{Bond}}, \citenamefont {{Campbell-Wilson}}, \citenamefont
  {{Connor}}, \citenamefont {{Cliche}}, \citenamefont {{Davis}}, \citenamefont
  {{Deng}}, \citenamefont {{Denman}}, \citenamefont {{Dobbs}}, \citenamefont
  {{Fandino}}, \citenamefont {{Gibbs}}, \citenamefont {{Gilbert}},
  \citenamefont {{Halpern}}, \citenamefont {{Hanna}}, \citenamefont {{Hincks}},
  \citenamefont {{Hinshaw}}, \citenamefont {{H{\"o}fer}}, \citenamefont
  {{Klages}}, \citenamefont {{Landecker}}, \citenamefont {{Masui}},
  \citenamefont {{Mena Parra}}, \citenamefont {{Newburgh}}, \citenamefont
  {{Pen}}, \citenamefont {{Peterson}}, \citenamefont {{Recnik}}, \citenamefont
  {{Shaw}}, \citenamefont {{Sigurdson}}, \citenamefont {{Sitwell}},
  \citenamefont {{Smecher}}, \citenamefont {{Smegal}}, \citenamefont
  {{Vanderlinde}},\ and\ \citenamefont {{Wiebe}}}]{2014SPIE.9145E..22B}%
  \BibitemOpen
  \bibfield  {author} {\bibinfo {author} {\bibfnamefont {K.}~\bibnamefont
  {{Bandura}}}, \bibinfo {author} {\bibfnamefont {G.~E.}\ \bibnamefont
  {{Addison}}}, \bibinfo {author} {\bibfnamefont {M.}~\bibnamefont {{Amiri}}},
  \bibinfo {author} {\bibfnamefont {J.~R.}\ \bibnamefont {{Bond}}}, \bibinfo
  {author} {\bibfnamefont {D.}~\bibnamefont {{Campbell-Wilson}}}, \bibinfo
  {author} {\bibfnamefont {L.}~\bibnamefont {{Connor}}}, \bibinfo {author}
  {\bibfnamefont {J.-F.}\ \bibnamefont {{Cliche}}}, \bibinfo {author}
  {\bibfnamefont {G.}~\bibnamefont {{Davis}}}, \bibinfo {author} {\bibfnamefont
  {M.}~\bibnamefont {{Deng}}}, \bibinfo {author} {\bibfnamefont
  {N.}~\bibnamefont {{Denman}}}, \bibinfo {author} {\bibfnamefont
  {M.}~\bibnamefont {{Dobbs}}}, \bibinfo {author} {\bibfnamefont
  {M.}~\bibnamefont {{Fandino}}}, \bibinfo {author} {\bibfnamefont
  {K.}~\bibnamefont {{Gibbs}}}, \bibinfo {author} {\bibfnamefont
  {A.}~\bibnamefont {{Gilbert}}}, \bibinfo {author} {\bibfnamefont
  {M.}~\bibnamefont {{Halpern}}}, \bibinfo {author} {\bibfnamefont
  {D.}~\bibnamefont {{Hanna}}}, \bibinfo {author} {\bibfnamefont {A.~D.}\
  \bibnamefont {{Hincks}}}, \bibinfo {author} {\bibfnamefont {G.}~\bibnamefont
  {{Hinshaw}}}, \bibinfo {author} {\bibfnamefont {C.}~\bibnamefont
  {{H{\"o}fer}}}, \bibinfo {author} {\bibfnamefont {P.}~\bibnamefont
  {{Klages}}}, \bibinfo {author} {\bibfnamefont {T.~L.}\ \bibnamefont
  {{Landecker}}}, \bibinfo {author} {\bibfnamefont {K.}~\bibnamefont
  {{Masui}}}, \bibinfo {author} {\bibfnamefont {J.}~\bibnamefont {{Mena
  Parra}}}, \bibinfo {author} {\bibfnamefont {L.~B.}\ \bibnamefont
  {{Newburgh}}}, \bibinfo {author} {\bibfnamefont {U.-l.}\ \bibnamefont
  {{Pen}}}, \bibinfo {author} {\bibfnamefont {J.~B.}\ \bibnamefont
  {{Peterson}}}, \bibinfo {author} {\bibfnamefont {A.}~\bibnamefont
  {{Recnik}}}, \bibinfo {author} {\bibfnamefont {J.~R.}\ \bibnamefont
  {{Shaw}}}, \bibinfo {author} {\bibfnamefont {K.}~\bibnamefont {{Sigurdson}}},
  \bibinfo {author} {\bibfnamefont {M.}~\bibnamefont {{Sitwell}}}, \bibinfo
  {author} {\bibfnamefont {G.}~\bibnamefont {{Smecher}}}, \bibinfo {author}
  {\bibfnamefont {R.}~\bibnamefont {{Smegal}}}, \bibinfo {author}
  {\bibfnamefont {K.}~\bibnamefont {{Vanderlinde}}},\ and\ \bibinfo {author}
  {\bibfnamefont {D.}~\bibnamefont {{Wiebe}}},\ }in\ \href
  {https://doi.org/10.1117/12.2054950} {\emph {\bibinfo {booktitle}
  {Ground-based and Airborne Telescopes V}}},\ \bibinfo {series} {Society of
  Photo-Optical Instrumentation Engineers (SPIE) Conference Series}, Vol.\
  \bibinfo {volume} {9145}\ (\bibinfo {year} {2014})\ p.\ \bibinfo {pages}
  {914522},\ \Eprint {https://arxiv.org/abs/1406.2288} {arXiv:1406.2288
  [astro-ph.IM]} \BibitemShut {NoStop}%
\bibitem [{\citenamefont {Ferraro}\ and\ \citenamefont
  {Ziomek}(1975)}]{Ferraro&Ziomek}%
  \BibitemOpen
  \bibfield  {author} {\bibinfo {author} {\bibfnamefont {J.~R.}\ \bibnamefont
  {Ferraro}}\ and\ \bibinfo {author} {\bibfnamefont {J.~S.}\ \bibnamefont
  {Ziomek}},\ }\href@noop {} {\emph {\bibinfo {title} {Introductory Group
  Theory and Its Application to Molecular Structure}}},\ \bibinfo {edition}
  {2nd}\ ed.\ (\bibinfo  {publisher} {Springer, New York},\ \bibinfo {year}
  {1975})\BibitemShut {NoStop}%
\bibitem [{\citenamefont {Gelessus}(2018)}]{dinfh}%
  \BibitemOpen
  \bibfield  {author} {\bibinfo {author} {\bibfnamefont {A.}~\bibnamefont
  {Gelessus}},\ }\href@noop {} {\bibinfo {title} {Character table for the point
  group $d_{\infty h}$}},\ \bibinfo {howpublished}
  {\url{http://symmetry.jacobs-university.de/cgi-bin/group.cgi?group=1001&option=4}}
  (\bibinfo {year} {2018}),\ \bibinfo {note} {[Accessed:
  13-January-2020]}\BibitemShut {NoStop}%
\end{thebibliography}%
\end{document}